\documentclass[poms,final,nonblindrev]{poms1_V1} 

\usepackage{enumitem}
\usepackage{subfigure}
\usepackage{varwidth}
\usepackage{color}
\usepackage{array}
\usepackage{setspace}
\usepackage{algpseudocode}
\algdef{SE}{Begin}{End}{\textbf{begin}}{\textbf{end}}
\usepackage{algorithmicx}
\usepackage{multirow}
\usepackage{graphicx}
\usepackage[flushleft]{threeparttable} 
\usepackage{booktabs}
\usepackage[ruled,vlined]{algorithm2e}
\usepackage{bmpsize}
\usepackage{multirow}
\usepackage{epsfig}
\usepackage{amssymb}
\let\theoremstyle\relax
\let\proof\relax 

\usepackage{amsmath}
\usepackage{amsfonts}
\usepackage{amsthm}
\usepackage{mathtools}
\usepackage{multimedia}
\usepackage{tikz}
\usepackage{lipsum}
\usetikzlibrary{positioning}
\usepackage{array,booktabs}
\usepackage{bm}
\usepackage{mathtools}
\usepackage{empheq}
\usepackage{pgfplots}
\usepackage[parfill]{parskip}
\usepackage{txfonts}
\usepackage{fancybox}
\usepackage{anyfontsize} 

\usepackage[utf8]{inputenc}
\usepackage[thinlines,thiklines]{easybmat}
\usepackage{hhline}
\pgfplotsset{compat=1.8}

\makeatletter
\DeclareMathSizes{\f@size}{11}{6}{6}
\makeatother

\def\VRHDW#1#2#3{\vrule height #1 depth #2 width #3}

\makeatletter
\@ifundefined{myeqdl}{%
}{%
}

\@ifundefined{myeqmodel}{%
  \newcommand{\myeqmodel}[1]{\begin{equation}{\begin{aligned} #1 \end{aligned}}\end{equation}}%
}{%
  \renewcommand{\myeqmodel}[1]{\begin{equation}{\begin{aligned} #1 \end{aligned}}\end{equation}}%
}

\@ifundefined{myeqmodeln}{%
  \newcommand{\myeqmodeln}[1]{\begin{equation*}{\begin{array}{cl} #1 \end{array}}\end{equation*}}%
}{%
  \renewcommand{\myeqmodeln}[1]{\begin{equation*}{\begin{array}{cl} #1 \end{array}}\end{equation*}}%
}

\@ifundefined{myeq}{%
  \newcommand{\myeq}[1]{$  {#1} $}%
}{%
  \renewcommand{\myeq}[1]{$  {#1} $}%
}

\@ifundefined{myeql}{%
  \newcommand{\myeql}[1]{\begin{equation}  {#1} \end{equation}}%
}{%
  \renewcommand{\myeql}[1]{\begin{equation}  {#1} \end{equation}}%
}

\@ifundefined{myeqln}{%
  \newcommand{\myeqln}[1]{\begin{equation*}  {#1} \end{equation*}}%
}{%
  \renewcommand{\myeqln}[1]{\begin{equation*}  {#1} \end{equation*}}%
}

\@ifundefined{myalg}{%
}{%
}

\@ifundefined{mycases}{%
}{%
}

\usepackage{xcolor}

\@ifundefined{myproof}{%
  \newcommand{\myproof}[1]{\par\noindent \quad #1 \hfill $\square$ \par}%
}{%
  \renewcommand{\myproof}[1]{\par\noindent \quad #1 \hfill $\square$ \par}%
}

\@ifundefined{mylemmaproof}{%
  \newcommand{\mylemmaproof}[1]{\par\noindent \quad \textbf{Proof.} #1 \hfill $\square$ \par}%
}{%
  \renewcommand{\mylemmaproof}[1]{\par\noindent \quad \textbf{Proof.} #1 \hfill $\square$ \par}%
}

\@ifundefined{myasmp}{%
  \newcommand{\myasmp}[1]{\begin{assumption} {\it #1}\end{assumption}}%
}{%
  \renewcommand{\myasmp}[1]{\begin{assumption} {\it #1}\end{assumption}}%
}

\@ifundefined{mycol}{%
}{%
}

\@ifundefined{myth}{%
  \newcommand{\myth}[1]{\begin{theorem} {\it #1}\end{theorem}}%
}{%
  \renewcommand{\myth}[1]{\begin{theorem} {\it #1}\end{theorem}}%
}

\@ifundefined{mylemma}{%
  \newcommand{\mylemma}[1]{\begin{lemma} {\it #1}\end{lemma}}%
}{%
  \renewcommand{\mylemma}[1]{\begin{lemma} {\it #1}\end{lemma}}%
}

\@ifundefined{myprop}{%
  \newcommand{\myprop}[1]{\begin{proposition} {\it #1}\end{proposition}}%
}{%
  \renewcommand{\myprop}[1]{\begin{proposition} {\it #1}\end{proposition}}%
}

\@ifundefined{mydef}{%
}{%
}
\makeatother

\makeatletter
\@ifundefined{mzcomment}{%
}{%
}
\@ifundefined{mzcommentb}{%
}{%
}
\makeatother

\makeatletter
\@ifundefined{Prob}{\newcommand{\Prob}{\mathbb{P}}}{\renewcommand{\Prob}{\mathbb{P}}}
\makeatother

\makeatletter
\@ifundefined{mylemmaproof}{%
  \newcommand{\mylemmaproof}[1]{\par\noindent \quad \textbf{Proof.} #1 \hfill $\square$ \par}%
}{%
  \renewcommand{\mylemmaproof}[1]{\par\noindent \quad \textbf{Proof.} #1 \hfill $\square$ \par}%
}
\makeatother

\makeatletter
\@ifundefined{lemma}{\newtheorem{lemma}{Lemma}[section]}{}
\makeatother

\makeatletter
\@ifundefined{mylemma}{%
  \newcommand{\mylemma}[1]{\begin{lemma}\itshape #1\end{lemma}}%
}{%
  \renewcommand{\mylemma}[1]{\begin{lemma}\itshape #1\end{lemma}}%
}
\makeatother

\makeatletter
\@ifundefined{theorem}{\newtheorem{theorem}{Theorem}[section]}{}
\makeatother

\makeatletter
\@ifundefined{lemma}{\newtheorem{lemma}[theorem]{Lemma}}{}
\makeatother

\makeatletter
\@ifundefined{proposition}{\newtheorem{proposition}[theorem]{Proposition}}{}
\makeatother

\makeatletter
\@ifundefined{corollary}{\newtheorem{corollary}[theorem]{Corollary}}{}
\makeatother

\makeatletter
\@ifundefined{assumption}{\newtheorem{assumption}[theorem]{Assumption}}{}
\makeatother

\makeatletter
\@ifundefined{definition}{\newtheorem{definition}[theorem]{Definition}}{}
\makeatother

\makeatletter
\@ifundefined{myasmp}{%
  \newcommand{\myasmp}[1]{\begin{assumption}\itshape #1\end{assumption}}%
}{%
  \renewcommand{\myasmp}[1]{\begin{assumption}\itshape #1\end{assumption}}%
}
\makeatother

\makeatletter
\@ifundefined{mycol}{%
}{%
}
\makeatother

\makeatletter
\@ifundefined{myth}{%
  \newcommand{\myth}[1]{\begin{theorem}\itshape #1\end{theorem}}%
}{%
  \renewcommand{\myth}[1]{\begin{theorem}\itshape #1\end{theorem}}%
}
\makeatother

\makeatletter
\@ifundefined{mylemma}{%
  \newcommand{\mylemma}[1]{\begin{lemma}\itshape #1\end{lemma}}%
}{%
  \renewcommand{\mylemma}[1]{\begin{lemma}\itshape #1\end{lemma}}%
}
\makeatother

\makeatletter
\@ifundefined{myprop}{%
  \newcommand{\myprop}[1]{\begin{proposition}\itshape #1\end{proposition}}%
}{%
  \renewcommand{\myprop}[1]{\begin{proposition}\itshape #1\end{proposition}}%
}
\makeatother

\makeatletter
\@ifundefined{mydef}{%
}{%
}
\makeatother

\makeatletter

\@ifundefined{BFa}{}{}
\@ifundefined{BFb}{}{}
\@ifundefined{BFc}{}{}
\@ifundefined{BFd}{\newcommand{\BFd}{\ensuremath{\mathbf{d}}}}{\renewcommand{\BFd}{\ensuremath{\mathbf{d}}}}
\@ifundefined{BFe}{}{}
\@ifundefined{BFf}{}{}
\@ifundefined{BFg}{}{}
\@ifundefined{BFh}{}{}
\@ifundefined{BFi}{}{}
\@ifundefined{BFj}{}{}
\@ifundefined{BFk}{}{}
\@ifundefined{BFl}{}{}
\@ifundefined{BFm}{}{}
\@ifundefined{BFn}{}{}
\@ifundefined{BFo}{}{}
\@ifundefined{BFp}{}{}
\@ifundefined{BFq}{}{}
\@ifundefined{BFr}{}{}
\@ifundefined{BFs}{}{}
\@ifundefined{BFt}{}{}
\@ifundefined{BFu}{}{}
\@ifundefined{BFv}{}{}
\@ifundefined{BFw}{}{}
\@ifundefined{BFx}{}{}
\@ifundefined{BFy}{\newcommand{\BFy}{\ensuremath{\mathbf{y}}}}{\renewcommand{\BFy}{\ensuremath{\mathbf{y}}}}
\@ifundefined{BFz}{}{}

\@ifundefined{E}{\newcommand{\E}{\ensuremath{\mathbb{E}}}}{\renewcommand{\E}{\ensuremath{\mathbb{E}}}}

\@ifundefined{BFA}{}{}
\@ifundefined{BFB}{}{}
\@ifundefined{BFC}{}{}
\@ifundefined{BFD}{}{}
\@ifundefined{BFE}{}{}
\@ifundefined{BFF}{}{}
\@ifundefined{BFG}{}{}
\@ifundefined{BFH}{}{}
\@ifundefined{BFI}{}{}
\@ifundefined{BFJ}{}{}
\@ifundefined{BFK}{}{}
\@ifundefined{BFL}{}{}
\@ifundefined{BFM}{}{}
\@ifundefined{BFN}{}{}
\@ifundefined{BFO}{}{}
\@ifundefined{BFP}{}{}
\@ifundefined{BFQ}{}{}
\@ifundefined{BFR}{}{}
\@ifundefined{BFS}{}{}
\@ifundefined{BFT}{}{}
\@ifundefined{BFU}{}{}
\@ifundefined{BFV}{}{}
\@ifundefined{BFW}{}{}
\@ifundefined{BFX}{}{}
\@ifundefined{BFY}{}{}
\@ifundefined{BFZ}{}{}

\@ifundefined{BFalpha}{}{}
\@ifundefined{BFbeta}{}{}
\@ifundefined{BFgamma}{}{}
\@ifundefined{BFdelta}{}{}
\@ifundefined{BFepsilon}{}{}
\@ifundefined{BFzeta}{}{}
\@ifundefined{BFeta}{}{}
\@ifundefined{BFtheta}{}{}
\@ifundefined{BFiota}{}{}
\@ifundefined{BFkappa}{}{}
\@ifundefined{BFlambda}{}{}
\@ifundefined{BFmu}{}{}
\@ifundefined{BFnu}{}{}
\@ifundefined{BFxi}{}{}
\@ifundefined{BFpi}{}{}
\@ifundefined{BFrho}{}{}
\@ifundefined{BFsigma}{}{}
\@ifundefined{BFtau}{}{}
\@ifundefined{BFupsilon}{}{}
\@ifundefined{BFphi}{}{}
\@ifundefined{BFchi}{}{}
\@ifundefined{BFpsi}{}{}
\@ifundefined{BFomega}{}{}

\@ifundefined{BFvarepsilon}{}{}
\@ifundefined{BFvarphi}{}{}
\@ifundefined{BFvartheta}{}{}
\@ifundefined{BFvarpi}{}{}
\@ifundefined{BFvarrho}{}{}
\@ifundefined{BFvarsigma}{}{}
\@ifundefined{BFvarkappa}{}{}

\@ifundefined{BFGamma}{}{}
\@ifundefined{BFDelta}{}{}
\@ifundefined{BFTheta}{}{}
\@ifundefined{BFLambda}{}{}
\@ifundefined{BFXi}{}{}
\@ifundefined{BFPi}{}{}
\@ifundefined{BFSigma}{}{}
\@ifundefined{BFUpsilon}{}{}
\@ifundefined{BFPhi}{}{}
\@ifundefined{BFPsi}{}{}
\@ifundefined{BFOmega}{}{}
\makeatother

\OneAndAHalfSpacedXI 
\usepackage{graphicx}
\usepackage{subfigure, epsfig}
\usepackage{natbib}

 \bibpunct[, ]{(}{)}{,}{a}{}{,}%
 \def\bibfont{\small}%

\EquationsNumberedThrough   

\begin{document}

\RUNAUTHOR{Mendelson and Zhu}

\RUNTITLE{Learning by Doing: The Case of Online Lending}

\TITLE{Learning by Doing: The Case of Online Lending}

\ARTICLEAUTHORS{%
\AUTHOR{Haim Mendelson}
\AFF{Graduate School of Business, Stanford University, Stanford, CA 94305, \EMAIL{haim@stanford.edu}} 
\AUTHOR{Mingxi Zhu}
\AFF{Scheller College of Business, Georgia Institute of Technology, Atlanta, GA 30332, \EMAIL{mingxi.zhu@scheller.gatech.edu}}}

\ABSTRACT{
\noindent Online lending, a phenomenon which is becoming mainstream due to the migration of consumer finance to the Internet and the adoption of AI-based lending models, is an example of ``learning by doing." This paper studies optimal policies for a direct online lender. This is an instance of a more general problem: how should a decision-maker experiment sequentially in the face of unknown customer (or other) information? Conventional wisdom suggests the decision-maker should take advantage of sequential learning opportunities by conducting multiple small, ``lean" experiments, each building incrementally on the results of earlier ones. Can a single ``grand experiment," uninformed by earlier experiments, do as well? We find that ``lean" incremental experiments are optimal when the interest rate is exogenous. However, when we extend the lender's action space to setting both the interest rate and the loan amount, we find conditions under which a single ``grand experiment" is optimal. In both cases, income variability can benefit the lender by enabling more effective experimentation. We also study the consumer segmentation associated with each strategy and show that the lender cannot achieve more than half the profit obtained under perfect information.
 }
\maketitle

\section{Introduction and Literature Review}

Financial services have been transformed by the Internet and more recently, by the use of AI. Online lending in particular has benefited from the ability to access consumers online and the use of AI to predict their creditworthiness. Along with other financial technologies, online lending has evolved from its early adoption stage to become mainstream, with leading U.S. Fintechs such as LendingClub, Upstart and SoFi becoming major financial institutions. Further, large financial services firms have implemented their own online lending services to avoid disruption \citep{jamieletter}. The increasing use of AI to improve the underwriting process and make it more dynamic, as well as Covid-19, have accelerated this transformation \citep{FutureGlobalFintech2025}. In the early years of online lending, there has been substantial enthusiasm about peer-to-peer lending, where lenders were consumers or consumer-like professionals who lent their own money to other consumers (for surveys of this literature, see \citep{zhao2017p2p}, \citep{basha2021online}). However, as online lending moved to the mainstream, most of it is direct, with financial institutions risking their own balance sheets to underwrite loans. Since 2020, LendingClub, which was the leading peer-to-peer market in the United States, has been shifting its strategy from peer-to-peer lending from consumers, to lending from financial institutions, and on to direct lending. Upstart and SoFi, the other leading pure-play online lenders in the U.S., are direct lenders with upstart increasingly providing online lending services to traditional banks. In China, there were thousands of online peer-to-peer lenders that went out of business or ceased operations by 2021 \citep{chinap2p2019} and today's Chinese online lending market is a pure direct lending market.     

While the optimization of online lending operations is an interesting and important research area, it is also an interesting example of learning by doing: The lender risks its own capital to learn about the creditworthiness of prospective customers, and when the loans are unsecured, the lender will typically stage its loans to reduce its risk. Online lenders of unsecured consumer loans typically offer a series of loan amounts which depend on the consumer's repayment record. This gives rise to a dynamic optimization problem where in each stage, the lender acquires additional information about the borrower's creditworthiness. This problem belongs to the vast literature on experimentation and learning. We aim to develop insights on the specific application we focus on---online consumer lending, and through it---shed some light on the broader area of experimentation and learning. 

We view a risky loan as an experiment. While the lender’s first loan is inherently risky, once a repayment is made, the lender has to choose between exploiting the repayment information without taking additional risks or exploring further by increasing the loan amount. The lender may operate in an exogenously-set interest rate environment; for example, the  the rate may be determined by regulation. In this case, the lender's problem in each period is uni-dimensional---determining the loan amount offered in that period. Alternatively, the lender may determine in each period both the loan amount and the interest rate, turning its experiments into two-dimensional. We identify two extreme experimentation strategies and study how the dimensionality of the action space and the properties of the demand elasticity determine which one is optimal. We contrast the ``lean experimentation" (LE) strategy, where the lender gradually increases the loan amounts and takes small incremental risks in each experiment, to a ``grand experiment" (GE) strategy whereby the lender makes a single high-risk experiment and, if the consumer repays, continues by ``playing it safe" without experimenting any further. While the LE strategy is intuitively appealing and is often recommended, we find broad conditions under which a GE is optimal.    

We model and study the lender's optimal dynamic control problem under both exogenous and endogenous interest rates. We find that with exogenous interest rates, the optimal policy is LE. With endogenous interest (or discount) rates, the structure of the optimal policy depends on the shape of the elasticity of demand with respect to the discount rate. When the demand elasticity is decreasing in the discount rate, the optimal policy is LE, similar to the exogenous interest rate case. When the demand elasticity is increasing or constant, it is optimal to perform a single GE. We further explore the implications of the optimal policies, studying consumer segmentation and the effects of income variability and other variables. We also find that online lending is inherently a high-risk, low-payoff business: for both exogenous and endogenous interest rates, the lender's expected profit is bounded by half of the full-information profit. 

As discussed above, this paper may be viewed from two perspectives: a ``learning by doing" lens and an optimal online lending lens. There is a vast literature on learning by doing, and we will not review most of it here. 
A common approach to modeling learning by doing is the multi-armed bandit (MAB) model (\cite{katehakis1987multi}), where each action (arm) is associated with a random reward following a probability distribution which is unknown a-priori, and in each period the decision-maker has to choose an action so as to maximize the expected total payoff over the decision horizon. Our model cannot be analyzed using the MAB approach as the lending market is more complex than the MAB environment, where the action selected today does not influence the payoffs on any of the arms in future periods. In contrast, in our model the actions taken today greatly influence future payoffs. Although the literature arguing for many small experiments does not appeal to the MAB framework, MAB provides a theoretical foundation for its recommendations. Under the MAB model, the optimal policy is based on the Gittins index (\cite{gittins2011multi}), which trades off for each arm \emph{(i)} its expected return based on experiments conducted so far, against \emph{(ii)} a term which is inversely related to the number of times the arm has been pulled so far. Term \emph{(i)} favors the exploitation of arms that have already proved to be more profitable, whereas term \emph{(ii)} favors the exploration of more uncertain arms. As the process continues, term \emph{(ii)} diminishes compared to \emph{(i)}, resulting in a lower emphasis on experimentation. The UCB algorithm, which is asymptotically optimal for the MAB problem, shares a similar structure (\cite{lai1985asymptotically}, \cite{agrawal1995sample}, \cite{lattimore2020bandit}).
These results suggest an optimal policy which is based on multiple experiments with the importance of experimentation diminishing over time. At a high level, this structure is similar to our LE approach, although it is generated by a different model. 

One special application of the MAB approach is \cite{weitzman1979optimal}, which offers a framework for finding the optimal policy under sequential experiments. An important driver of the number of experiments is the relationship between the cost per experiment and the potential benefit of the experiment. As the cost of experimentation goes down, the expected optimal number of experiments naturally increases. This framework has been widely applied directly or indirectly in the product development literature. Inherited from the theoretical foundations of \cite{weitzman1979optimal}, the recommendation to conduct many small experiments is prevalent in the entrepreneurship, product development and business practitioner literature.  The ``lean startup" approach (\cite{reis2011lean}, \cite{blank2013lean}), which has become dominant in the entrepreneurship area, recommends building a startup by iterating on a series of ``lean", small and inexpensive experiments.   \cite{furr2022upside} argues for making multiple small experiments (10,000 of them in the case of National Geographic), citing Match.com, Thomas Edison and Amazon.com as masters of such lean experimentation. \cite{moogk2012minimum} and \cite{giardino2014early} propose that by building a series of minimum viable products, essentially small product experiments (prototypes), startups can receive fast market feedback and evolve quickly.  \cite{kerr2014entrepreneurship} and \cite{nanda2016financing} suggest that to maximize returns, venture capital firms should run a large number of initial experiments by funding multiple small startups.







The idea that multiple small experiments are an effective way to resolve uncertainty is a common theme in the product development literature, where prototyping and iterative design are often modeled as experiments \citep{ulrich2000product, krishnan2001product, dahanmendelson2001extreme, loch2001parallel, sommer2004selectionism, kornishhutchisonkrupat2017research}. In particular, online experiments and A/B testing significantly reduce the cost of experimentation \citep{dahan2000predictive,dixonvictorinokwortnikverma2017surprise, wootenulrich2017idea,joo2019optimal,koninghasanchatterji2022experimentation}. 
In most cases, firms run the experiments themselves; in others, they delegate them to third parties (\cite{eratkrishnan2012managing}). 
\cite{thomke2003experimentation} shows that due to the declining costs of experimentation, product development benefits from a process which is based on many small experiments, and \cite{thomke2020building} argues that companies should opt for a culture that favors such experiments: ``It's actually less risky to run a large number of experiments than a small one." He cites examples from Booking.com, Expedia, Microsoft and Netflix that engage in this practice.  \cite{erat2008sequential} model costly sequential tests of correlated product designs, selecting the next design via an index rule and choosing an optimal stopping time. The optimal policy continues only when previous test outcomes fall in a “sweet spot” between two critical thresholds. \cite{thomke2001sequential} address the optimal strategy of a firm that repeatedly tests to find bugs and fix them, as in the software development domain (a ``bug" may reflect technical, customer understanding or production issues). The firm has to find and repair all the bugs by the end of the development period. \cite{thomke2001sequential} show that the optimal number of tests depends on the cost of experimentation, deriving (under certain conditions) an EOQ-like result: the optimal number of tests is given by the square root of the ratio of avoidable cost to the cost of a test. \cite{bhaskaranerat2025optimal} model noisy sequential testing of prototypes with a choice of measurement fidelity, and find that precise measurements are optimal when ex-ante uncertainty is moderate. Their optimal testing policy uses a threshold stopping rule, generalizing \cite{weitzman1979optimal}.
While each of these models has its unique features, they mostly find that an effective product development strategy continues to experiment until the added value of the next experiment is smaller than its fixed cost. Our setting is obviously different, and our results are structurally different as well: the optimal experimentation strategy is either to experiment forever (under LE) or to conduct just a single grand experiment, depending on the dimensionality of the action space and the monotonicity properties of the demand curve, with no options in-between.

Our work is also related to the literature on lending policies. This literature largely comprises macro-finance papers that focus on the effects of  commercial lending policies on the economy, whereas operations-oriented papers focus on other aspects of lending (e.g., dynamic programming for optimizing collection actions and stopping times in consumer credit \citep{dealmeidafilho2010optimizing}, or empirical links between product usage and repayment risk in rent-to-own settings \citep{guajardocohennetessine2019how}). One branch of the macro-finance literature focuses on credit rationing when business demand for commercial loans exceeds the loan supply at prevailing commercial loan rates (\cite{jaffee1990credit}). While this literature considers the determination of loan amounts, the credit rationing problem is only loosely related to our work in both structure and objective, as it aims to study the macroeconomic effects of lending to firms that finance commercial projects (for a review of this literature, see \cite{jaffee1990credit} and \cite{bellier2012lies}; more recent papers are \cite{kremp2013did}, \cite{ambrose2016credit} and \cite{mc2020lending}). A related branch of the literature studies the design of long-term commercial lending contracts and its impact on the firm's valuation and growth under market shocks (\cite{albuquerque2004optimal}, \cite{clementi2006theory}, \cite{biais2007dynamic}). In these studies, the lender (a bank) designs a lending contract that specifies the amount it will lend to businesses in response to macroeconomic market shocks using both the current, observable market shock and distributional information about future shocks; the lender then optimizes its long-term expected profit without modeling a learning environment. Learning is modeled in another branch of the macroeconomic banking literature which focuses on the effect of screening standards for commercial loans on overall lending in the economy. \cite{fishman2024dynamic} build a dynamic model where banks choose whether and to what degree to screen commercial borrowers at a cost. They find that screening reduces the average number of projects financed, which creates negative externalities at the macro level: one bank's tighter standards worsen the borrower pool, incentivizing other banks to also tighten standards, which can amplify economic downturns.  \cite{hu2021theory} aim to explain the phenomenon of ``zombie lending," i.e., making loans with a negative NPV to businesses. They show that when bad news about a business loan don't trigger immediate default, the bank rolls over credit even though the loan has a negative NPV. The reason is that the practice helps the business build its reputation with other lenders, which can benefit the bank. Obviously, our model has different objectives and structure: our borrower is a consumer, not a business; the lender's learning mechanism is entirely different; and we aim to derive structural results on the lender's optimal policy rather than to study the macroeconomic effects of lending practices on the economy.  

The rest of the paper is organized as follows. Section 2 presents our model. Section 3 solves it for the case of exogenous interest rate, and Section 4 studies the endogenous interest rate case. Section 5 examines the effects of income variability and the value of information. Section 6 considers the case where the consumer's income is subject to random shocks, and Section 7 offers our concluding remarks. Proofs are in the Appendix. 

\section{Modeling Approach}
\label{sec_exo_model}

We model the lender's problem as a discrete-time, infinite-horizon, dynamic control problem with the following timeline. A prospective borrower (consumer) has periodic disposable income (or ability to pay) $\theta$, drawn from a known distribution \myeq{F}. We call $\theta$ ``income." In each period $t$, the lender offers the consumer a one-period unsecured loan of $\mu_t$ dollars at interest rate $r_t$. The discount rate corresponding to loan amount $\mu_t$ is $d_t= \frac{1}{1+r_t}$. For technical convenience, we use as our decision variables the discount rate $d_t$ and the repayment amount the borrower would pay at the end of period $t$, $y_t=\mu_t/d_t$.

The realization of $\theta$ is unknown to the lender, who knows only the income distribution $F$. We assume $F$ has a differentiable density $f$ with support $[0,u)$ where $u>0$ may be finite or infinite (our model extends in a straightforward way to distributions with positive support $[l,u)$ for $l>0$). The hazard function corresponding to the income distribution \myeq{F} is $H(x)=\frac{f(x)}{1-F(x)}$. We assume:  

\myasmp{
\label{assumpt1}
$G(x)=xH(x)$ is strictly increasing for all $x\in[0,u)$, $G(0)=0$ and $G(u)\geq1$, 
}
\noindent (when $u$ is infinite, we interpret the condition $G(u)\geq 1$ as $\lim_{x\to\infty}G(x)\geq 1$). Assumption \ref{assumpt1} clearly holds for any distribution $F$ with a (weakly) increasing hazard rate and finite support $[0,u)$, $f(0)<\infty$ and $\lim_{x\to u}f(u)>0$, such as the uniform distribution. Other distributions satisfying Assumption \ref{assumpt1} include the Beta, Gamma and Weibull distributions (see Proofs in Appendix \ref{app_assump1_dist}). 
Not all distributions satisfy Assumption \ref{assumpt1}: for example, the folded Cauchy distribution has a non-monotone $G(x)$.

The prospective borrower observes the offer $(y_t,d_t)$ and decides whether to accept or reject it based on the offered discount rate $d_t$ (or equivalently, the interest rate $r_t$). The acceptance probability $s(d_t)$ is strictly increasing, twice continuously differentiable and strictly concave with \(s(0)=0\), \(s'(d_t)>0\) and \(s''(d_t)<0\); a lower interest rate (higher discount rate) increases the probability of acceptance. Of special interest is the case of constant demand elasticity where $s(d)=d^\alpha, \alpha>0$.

Income \(\theta\) is drawn at \(t=0\) from \(F\) with continuous support \([0,u)\). The lender announces \((d_0,y_0)\); the borrower accepts with probability \(s(d_0)\) and otherwise exits. If accepted, the lender lends $\mu_0=d_0y_0$. Repayment occurs at the end of period \(t\) if \(y_t\le \theta\); otherwise the borrower defaults and exits at \(t+1\). Upon repayment at \(t\), the lender moves to period \(t+1\), announces \((d_{t+1},y_{t+1})\), and the process is repeated. Figure~\ref{fig_timeline} summarizes the sequence of moves.

\begin{figure}[!htb]
\centering
\begin{tikzpicture}
  \draw [dashed, black, ->] (-1,0) -- (13,0)      
        node [above, black] {};              
    \draw [dashed, white, ->] (0,0) -- (0,-3)      
        node [right, black] {};              
        \node at (-2, 0) {$t$};
        \node at (0,0) {$|$};
        \node [below] at (0,-0.3) {}; 
         \node [below] at (0,-1) {\small The consumer's};
         \node [below] at (0,-1.5) {\small  income};
         \node [below] at (0,-2) {\small  $\theta\sim F$ is realized.};
         
          \node at (3,0) {$|$};
         \node [below] at (3,-0.3) {$0$}; 
         \node [below] at (3,-1) {\small The lender};
         \node [below] at (3,-1.5) {\small lends $\mu_0=d_0y_0$};
          \node [below] at (3,-2) {\small to the consumer.};
          \node [below] at (3,-2.7){\small The consumer};
          \node [below] at (3,-3.2){\small accepts the loan};
          \node [below] at (3,-3.7){\small with  probability $s(d_0)$.};

         \node at (7,0) {$|$};
         \node [below] at (7,-0.3) {$1$}; 
         \node [below] at (7,-1) {\small The consumer receives $\theta$ and};
         \node [below] at (7,-1.5) {\small pays back $y_0$, if possible.}; 
                  \node [below] at (7,-2.3) {\small  Receiving the payment,}; 
         \node [below] at (7,-2.8) {\small the lender updates belief};
         \node [below] at (7,-3.3) {\small  and lends $\mu_1=d_1y_1$, with};

         \node [below] at (7,-3.8) {\small   consumer acceptance};
        \node [below] at (7,-4.3) {\small  probability as $s(d_1)$.};
         
         \node at (11,0) {$|$};
         \node [below] at (11, -0.3) {$2$};
         \node [below] at (11, -1) {\small The process};
         \node [below] at (11, -1.5) {\small continues...}; 
\end{tikzpicture}
\caption{Model Timeline.}
\label{fig_timeline}
\end{figure}
\vspace{-10 pt}

In our model, the lender gradually learns what the consumer can repay and dynamically adjusts the loan amounts and discount rates in each period. This setup is a canonical example of costly experimentation in a high-risk environment, where the lender risks losing the entire loan amount in the hope of discovering profitable customers. We first solve the problem where the interest (and discount) rate is exogenously fixed, either by the market or by regulation. We then solve the general problem, where $d_t$ is endogenous (Section \ref{sec_endo}). We compare the two solutions in Section \ref{subsec_discuss_LEGE}.


\section{Exogenous interest rate: Model and Results}  

In this Section, we study the case where the interest rate is exogenously fixed, either by the market or by regulation. In this case, $d_t=d$ is an exogenous constant and the lender has to decide in each period only on $y_t$. Further, since $d$ is fixed, so is the prospective borrower's acceptance probability $s=s(d)$. The lender is risk-neutral and seeks the dynamic policy \myeq{\{y_t, \ t\geq0\}} that maximizes the net present value of its expected payoffs (which we call NPV or expected profit). The lender discounts its cash flows at \(\rho\in(0,1)\). Since \(s\) is constant, we can effectively absorb it into $\rho$ and adjust $d$ accordingly. In this Section, then, $\rho$ means $\rho \cdot s$ and $d$ means $d \cdot s$ (in the endogenous-rate model in Section \ref{sec_endo}, we write \(s(d_t)\) explicitly as it depends on the policy).


We assume $\rho>d$; otherwise the lender would prefer not to lend. Let \myeq{\mathcal{H}_t} denote the lender's information set at time \myeq{t}, which specifies whether the consumer paid back \myeq{y_{i}} in periods \myeq{i< t} up to $t$. The lender's expected payoff (NPV) under policy \myeq{\BFy=\{y_0,y_1,\dots\}\in\mathbb{R}^{\infty}} is 
\myeql{
\label{lender_expected_payoff}
\Pi(\BFy)=\sum^{\infty}_{t=0}\ \rho^{t} \ [\Pi^{t-1}_{i=0}\Prob(\theta\geq y_i|\mathcal{H}_i)] \cdot [\rho\Prob(\theta\geq y_t|\mathcal{H}_t)y_t-d y_t],
} 

\noindent where we define  $\Pi^{-1}_{i=0}\Prob(\theta\geq y_i|\mathcal{H}_i)=1$.

\subsection{Structure of the Optimal Policy}
\label{sec_model_x0_0}

If the consumer has not defaulted by time $t$, the conditional distribution of $\theta$ given history $\mathcal{H}_t$ is $F(\theta|\theta\geq \max_{-1\leq i< t}\ y_i)$, with $y_{-1}$ defined as $0$.  
We define the system state as $x_t= \max_{-1\leq i< t} \ y_i$, the period-$t$ lower-bound on the consumer's income. The lender's decision problem is thus Markovian with decisions depending on $x_t$. The Bellman equation for our dynamic control problem is
\myeql{J(x_t)\ = \ \max_{y_t} \{ \  -dy_t+\rho\Prob(\theta\geq y_t|\theta\geq x_t)(J(x_{t+1})+y_t) \},
\label{eq_J_xt_1}
}
\noindent which satisfies the following Lemma.
\begin{lemma}
\textit{\label{lemma_increasing_control}
If an optimal policy $\{y_t\}$ exists, it is weakly increasing, i.e., $y_{t}\geq y_{t-1}$ for all $t>0$. }
\end{lemma}
Intuitively, when the lender knows that $\theta\geq y_{t-1}$, setting $y_t<y_{t-1}$ is sub-optimal as it could increase the loan amount to $\mu_t = \mu_{t-1} =dy_{t-1}$ knowing that the consumer would pay back $y_t$, which strictly dominates setting $y_t<y_{t-1}$. Further, when the support of $F$ is continuous, Lemma \ref{lemma_increasing_control} implies that if the consumer does not default by the beginning of period \myeq{t}, $x_t=\max_{-1\leq i<t} y_i = y_{t-1}$ and the lender updates the lower bound on income to $y_t$. 
Since \myeq{F} has a continuous positive support on $[0,u)$, it follows that
\myeql{J(x_t)\ = \ \max_{y_t\in[x_t,u)} \ \ \{ x_t-dy_t+\rho\Prob(\theta\geq y_t|\theta\geq x_t)J(y_t)\} \ = \ \max_{y_t\in[x_t,u)} \ \ { x_t-dy_t+\rho\dfrac{1-F(y_t)}{1-F(x_t)}J(y_t) }.
\label{eq_J_xt}
}
As we show in Theorem \ref{them_explore_exploit_1}, an optimal policy exists with the lender increasing the repayment amounts towards an upper limit $\bar{x}$ defined below.  

\begin{lemma}
\textit{\label{lemma_general_barx}
There exists a unique finite solution $0<\bar{x}<u$ to the equation: 
\myeql{
\label{eq_bar_x}
G(\bar{x})=-\frac{(1-\rho)d}{(1-d)\rho}+1.
}
The solution $\bar{x}$ increases in $\rho$ and decreases in $d$.}
\end{lemma} 

We are now ready to state the structure of the optimal policy.

\myth{
\label{them_explore_exploit_1}
 \textbf{Optimal policy}
there exists a unique optimal policy whereby the lender increases the repayment amounts $y_t$ in each period $t$ as long as the consumer repays her loans. The sequence of repayment amounts $\{y_t, t=0,1,2,\ldots\}$ is strictly increasing to the limit \myeq{\bar{x}}.
}


 While we are not aware of empirical work in the context we study, the results of Theorem \ref{them_explore_exploit_1} are consistent with empirical findings on increasing loan amounts in commercial lending (\cite{petersen1994benefits}, \cite{angelini1998availability}, \cite{cenni2015credit}). Theorem \ref{them_explore_exploit_1} shows that the lender follows a classic Lean Experimentation (LE) strategy: it keeps making incremental experiments by gradually increasing the loan amounts, with the increases becoming infinitesimally small. The lender's LE strategy follows a particular structure driven by the threshold $\bar{x}$, which is the indifference point between ceasing experimentation and acquiring new incremental information. To see why, consider two alternative strategies when the lender is in state $x$: \emph{(i)} play it safe and just exploit the information embedded in $x$ by lending $(d \cdot x)$ (with repayment amount $x$) in each period, resulting in a zero probability of default; vs. \emph{(ii)} continue to experiment by increasing the repayment amount to  $(x+\Delta x)$ (lending $(d \cdot (x+\Delta x))$ to further learn about $\theta$; then, if the consumer repays the loan, the lender keeps setting the repayment amounts $y_t$ at $(x+\Delta x)$ in all subsequent periods. The lender's expected going forward NPV under strategy \emph{(i)} is
\myeql{\label{eq_strategy_1}
-d x +  \dfrac{\rho (1-d)}{(1-\rho)}x,
}
where the first term is the current loan amount and the second is the present value of all future interest cash flows. The expected NPV for strategy \emph{(ii)} to a first order in $\Delta x$ is given by
\myeql{\label{eq_strategy_2}
-d(x+\Delta x) + \left(1-H(x)\Delta x\right)\dfrac{\rho (1-d)}{(1-\rho)}(x+\Delta x) 
,}
where the first term is the current loan amount, $(1-H(x)\Delta x)$ is the probability that the consumer repays in the next period, and $\frac{\rho (1-d)}{(1-\rho)}(x+\Delta x)$ is the present value of the future interest cash flows conditional on no default. The lender's first-order incremental NPV from following strategy \emph{(ii)} over \emph{(i)} is thus
\myeql{\frac{\rho (1-d)}{(1-\rho)}\left(1-\frac{(1-\rho)d}{(1-d)\rho} -G(x)\right)\Delta x,
\label{eq_delta_x_2}}
and $\bar{x}$ is the point of indifference between \emph{(i)} and \emph{(ii)}.


\subsection{Implications}

\subsection{Lean Experimentation Strategy}
\label{subsec_lean_exp_strategy}
Our results lend support to the LE strategy structure: the lender keeps experimenting by gradually increasing its loan amounts to identify how much more it can lend. This is best shown by considering the    
case of the uniform income distribution, $F\sim U[0,1]$, where we have a closed-form solution: 

\myth{ 
\label{them_unif}
When $F\sim U[0,1]$, there is a unique solution to the Bellman equation (\ref{eq_J_xt}) with lending threshold \myeql{\bar{x}=\dfrac{(\rho-d)}{2\rho-d-d\rho}\in(0,1).} 
The optimal repayment amounts $y_t$ increase linearly in the state $x$ to the limit $\bar{x}$ with $y_t(x_t)=m x_t+n$, where 
\myeqmodel{
m& = \dfrac{d}{2\rho(1-a)}, \quad n =  \dfrac{(\rho - d + b\rho)}{2\rho(1-a)}, \textup{ and}\\
\label{eq_correct_a}a = & \dfrac{1}{2}\left(1-d\sqrt{\dfrac{\rho-d^2}{\rho d^2}}\right), \quad  b=  \dfrac{1}{2\rho - \rho d - d}\left((\rho -d)\left(d\sqrt{\dfrac{\rho-d^2}{\rho d^2}}+d-1\right)\right),\\
c=& \dfrac{1}{2(1-\rho)(2\rho -d -\rho d)^2}\left((\rho-d)^2\left(1-2d+\rho-(1-\rho)\sqrt{1-\frac{d^2}{\rho}} \right) \right).
}
}

\begin{figure}[htb!]
\centering
{\includegraphics[width=0.45\textwidth]{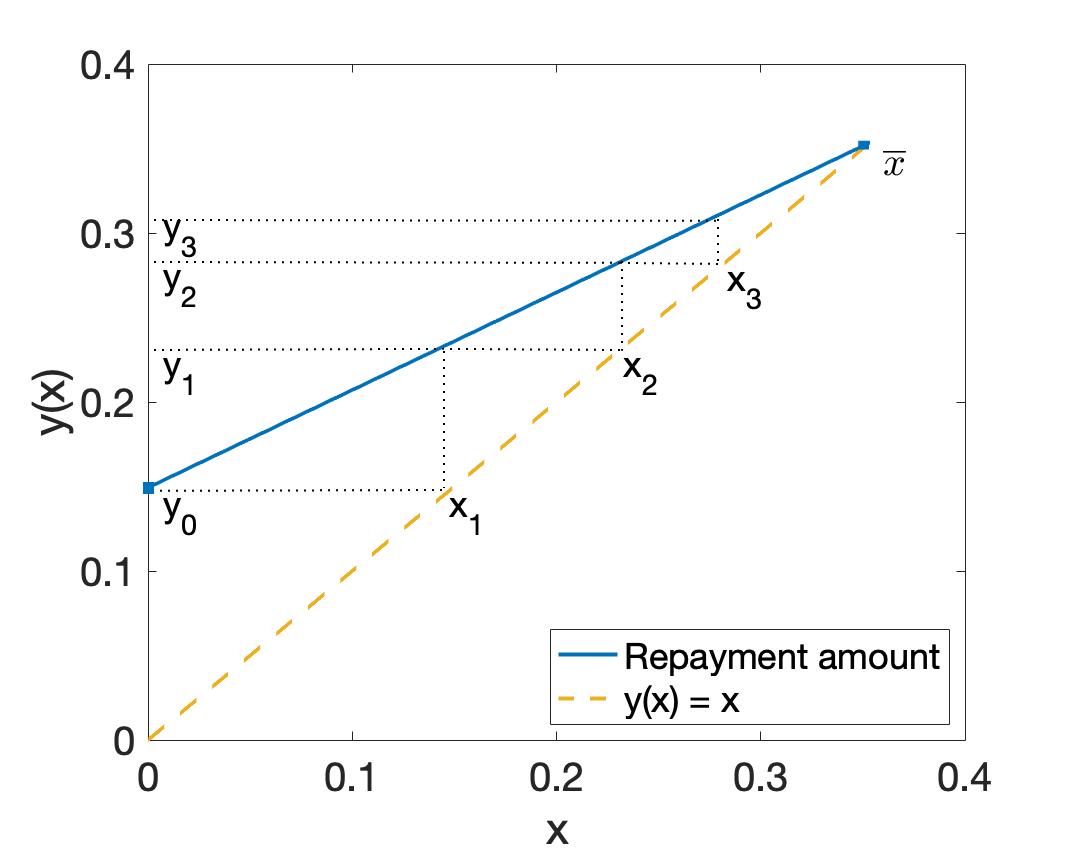}}
\caption{Optimal lending policy $\{y^*_0,y^*_1,y^*_2,\dots\}$ for uniform income distribution with $\rho=0.95$ and $d=0.833 \ (r=20\%)$. The x-axis shows the current state (lower bound on income). On the y-axis, $y_t(x_t)$ is the optimal repayment amount when the current state is $x_t$.  } 
\label{fig_opt_policy_unif}
\end{figure}

\noindent Figure \ref{fig_opt_policy_unif} shows that the optimal policy is classic LE: in each state $x_t$, the lender slightly increases the loan amount to test whether the consumer can repay $y_t(x_t)>x_t$. The amounts at risk, $(y_t(x_t)-x)$, keep decreasing. They are always positive but tend to zero at the limit. 


The results in this Section provide a theoretical foundation for the use of the LE approach commonly recommended in product development and entrepreneurship. LE emphasizes staged learning with incremental experimentation as evidence accumulates, gradually shifting to exploitation as uncertainty narrows. 

We call a consumer \textit{creditworthy} if she will never default under the optimal policy, which happens when $\theta\geq \bar{x}$. An important feature of the LE policy derived in Theorem \ref{them_explore_exploit_1} is that it is extremely conservative: even when the lender encounters a creditworthy consumer, it never lends more than $(d \cdot \bar{x})$. As $y_t$ approaches $\bar{x}$, the lender knows that $\theta\geq \bar{x}$ (i.e., the consumer will repay) almost surely: the probability mass between $y_t$ and $\bar{x}$ becomes infinitesimal whereas the probability that $\theta \geq \bar{x}$ is finite. Nevertheless, the lender never sets the repayment amount at or above $\bar{x}$. This is the result of two factors: $(i)$ a general difference between purely statistical experiments and business experiments, and $(ii)$ the high potential loss associated with lending (losing the entire principal amount) compared to the smaller potential value of earning incremental interest payments. Regarding $(i)$, statistical experiments are designed to estimate the value of an unknown parameter (in this case, $\theta$), whereas business experiments maximize expected net present value, where in our problem the cost of experimentation is endogenous because it is incurred as part of the operation of the business (in this case, lending). For statistical experimentation, binary search would lead to the fastest convergence to the true value of $\theta$. This strategy, however, does not take into account the risk of losing the entire principal amount. As a result, the lender continues to experiment conservatively (because of ($(ii)$) and indefinitely. For a creditworthy borrower, even though the \emph{number} of experiments goes to infinity, the size of these experiments goes to zero. This result changes dramatically when the interest rate in endogenous (Section~\ref{sec_endo}).

\subsubsection{Partitioning the State Space: Segmentation}
\label{subsec_consumer_seg_fi}

From an information economics perspective, experimentation may be viewed as a process of partitioning the state space into subsets to uncover what subset an unknown state variable ($\theta$ in our case) belongs to \citep{marschak1960remarks}. While statistical experimental designs often target point estimation, in the context of a decision problem such as ours it is sufficient to find a partition of the state space into subsets within which the optimal action and optimal payoff remain constant. 
In our problem, we can partition the state space into consumer segments such that all the consumers in the same segment will be offered the same series of loan amounts and will yield the same NPV. Formally, we call a partition $\{C_k, k=1,2,\ldots,\infty\}$ \emph{sufficient} if the optimal decisions and resulting NPVs are the same for all $\theta \in C_k, k=1,2,\ldots$. A sufficient partition also defines a segmentation of consumers. For the policy path \(\{y_t\}\) obtained under the optimal policy, we define
\[
C_k := [y_{k-1},y_k)\quad (k=1,2,\ldots),\qquad C_\infty := [\bar{x},u),\quad\text{with } y_{-1}=0.
\]
Any borrower with \(\theta\in C_k\) repays in periods \(0,\ldots,k-1\) and defaults in period \(k\), and a borrower with \(\theta\in C_\infty\) is creditworthy and never defaults. Hence, within each \(C_k\) (including \(C_\infty\)), \emph{(i)} the per-period cash flows and the default times are identical, and \emph{(ii)} the transitions (continue vs.\ exit) are identical. It follows that the lender's \emph{optimal} action and \emph{optimal} payoff coincide for all $\theta\in C_k$, so the partition (segmentation) \(\{C_k\}_{k\in\mathbb{N}\cup\{\infty\}}\) is sufficient.

We call a consumer a \emph{type-k consumer} when its income is in $C_k$. The NPV of a type-\(k\) consumer comprises \textit{(i)} the NPV of her interest payments in periods \(0, 1, 2, \ldots, k\); and \textit{(ii)} the (negative) present value of the final default loss in period \(k\) (zero when \(k=\infty\)). Component \textit{(i)} is given by
\myeq{\label{eq_PVR}
PV^R_k=\sum^{k-1}_{i=0}\rho^{i}(\rho-d)y_i.
}
Component \textit{(ii)} is given by 
\myeq{
\label{eq_PVI}
PV^I_k=\rho^{k} d y_k.
}
Thus, the overall NPV of a type-\(k\) consumer is
\myeq{\Pi_k = PV^R_k - PV^I_k = \sum^{k-1}_{i=0}\rho^{i}(\rho-d)y_i-\rho^{k} d y_k.}
For a creditworthy consumer (\(k = \infty\)), this reduces to \(\Pi_{\infty}=\sum^{\infty}_{i=0}\rho^{i}(\rho-d)y_i\). The sufficient partition \(\{C_k\}_{k\in\mathbb{N}\cup\{\infty\}}\) makes it unnecessary to identify a consumer's exact $\theta$: it is sufficient to identify which segment of the sufficient partition the consumer is in.

In a business context, managers often use behavioral (or managerial) segmentation which groups states (and the associated consumers) by similar behaviors (e.g., creditworthiness) even though their actual payoffs may differ. It is tempting to segment consumers into two behavioral subsets: \emph{(i)} profitable and creditworthy, vs. \emph{(ii)} unprofitable and non-creditworthy. However, we find that while creditworthy consumers are always profitable, some non-creditworthy consumers may also be profitable. Therefore, a proper behavioral partition of prospective borrowers requires further analysis. 

We create three behavioral consumer segments based on our sufficient partition. First, we split borrowers into creditworthy (\(\theta\ge \bar{x}\), type-\(\infty\)) and non-creditworthy (\(\theta<\bar{x}\)). Among the non-creditworthy, type-\(k\) borrowers have \(\theta\in[y_{k-1},y_k)\), default in period \(k\) after receiving \(d\,y_k\), and their NPV is \(\Pi_k\). This yields the following behavioral groups:
\begin{itemize}
    \item \emph{Unprofitable}: type-\(k\) (for some values of $k$) with \(\Pi_k<0\).
    \item \emph{Profitable (non-creditworthy)}: type-\(k\) (for some values of $k$) with finite \(k\) and \(\Pi_k\ge 0\).
    \item \emph{Creditworthy}: type-\(\infty\) (never default).
\end{itemize}

\begin{figure}[htb!]
\centering
\subfigure[$d=0.67$]{\includegraphics[width=0.4\textwidth]{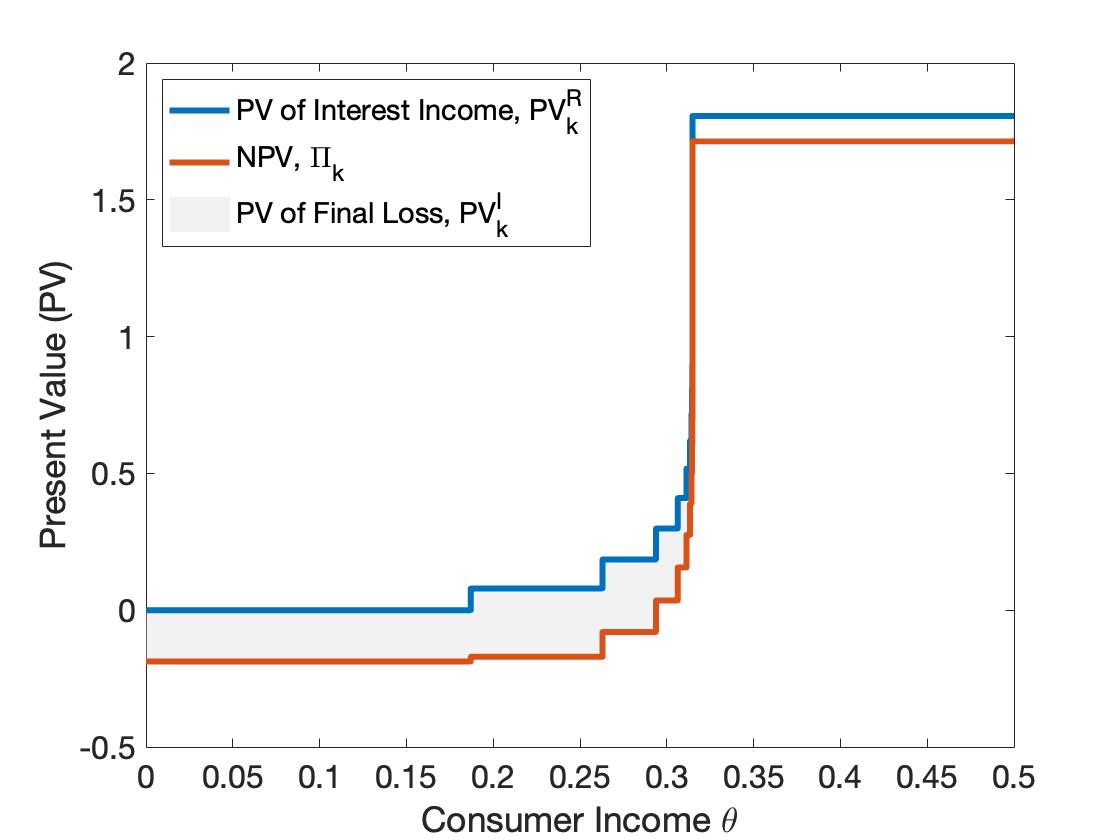}} 
\subfigure[$d=0.83$]{\includegraphics[width=0.4\textwidth]{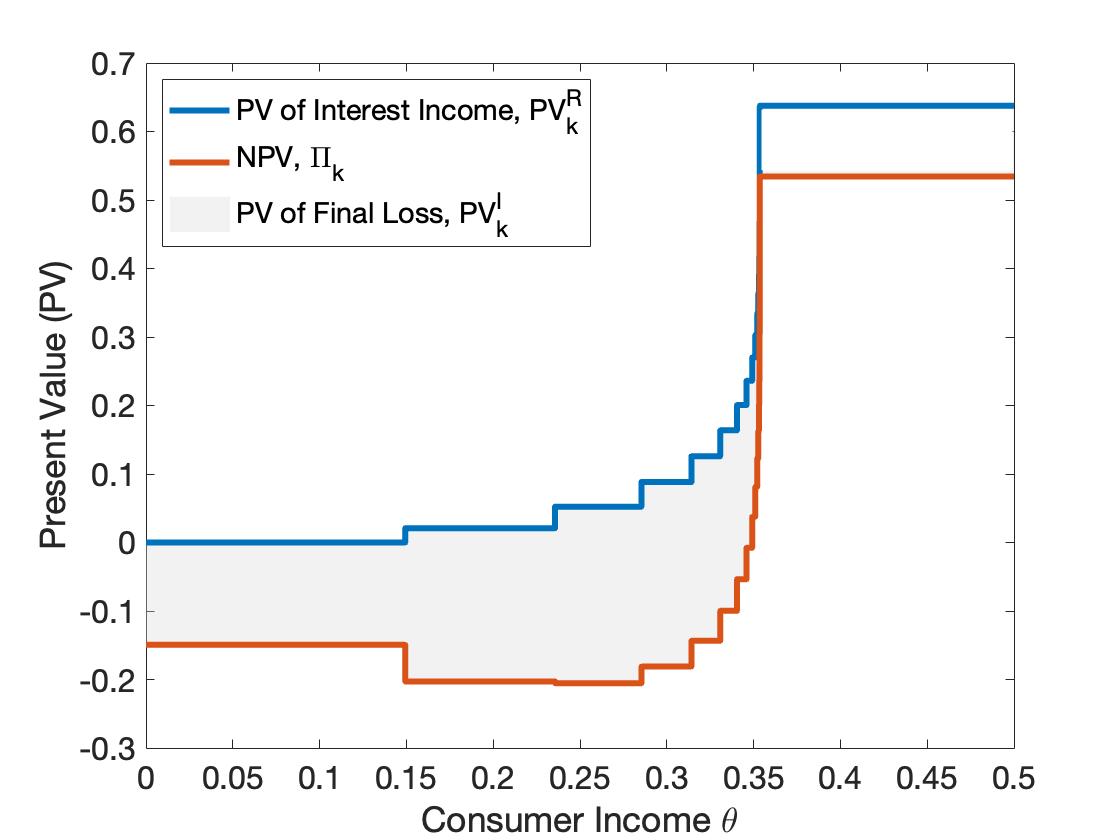}}\\
\caption{Lender NPV and its components as functions of consumer income. Income is uniformly distributed (\(U[0,1]\)) and \(\rho=0.95\). In \emph{(a)}, \(d=0.67\); in \emph{(b)}, \(d=0.83\). }
\label{fig:NPV}
\end{figure}

Figure \ref{fig:NPV} shows how consumers' NPVs (red lines) vary with their income states for the uniform income distribution with $\rho=0.95$ and \emph{(a)} $d=0.67$ or \emph{(b)} $d=0.83$. The NPVs are decomposed into those of the interest payments (in blue) and default loss (shaded). The $k$-th horizontal step in each curve represents consumers in $C_k$; the final horizontal step corresponds to creditworthy consumers in $C_\infty$ who don't default. The lines that seem vertical actually represent numerous shrinking segments obtained as the $y_k$ approach $\bar{x}$. 
As expected, the NPVs are significantly higher when the interest rate is higher \emph{((a)} vs. \emph{(b)}).
NPV is intuitively expected to increase with income, as in Figure \ref{fig:NPV} (a), but this is not always the case (Figure \ref{fig:NPV} \emph{(b)}): with a low interest rate, increases in the loan amounts may not generate enough interest income to compensate for the higher cost of default.\footnote{In Figure \ref{fig:NPV} \emph{(b)}, the curve decreases slightly over the first three steps and increases thereafter.} Nevertheless, it is optimal to make these loans in the hope of identifying more profitable borrowers. In both cases \emph{(a)} and \emph{(b)}, the behavioral partitions of the state space comprise three intervals defined by two cutoff values, $\theta^*$ and $\bar{x}$, such that consumers are unprofitable in $[0,\theta^*)$, profitable but non-creditworthy in $[\theta^*,\bar{x})$, and creditworthy in $[\bar{x},\infty)$. The following Proposition shows that this is a more general phenomenon.    


\myprop{\label{prop_mon_k}
Let income be uniformly distributed and let \(k^*=\min\{k:\Pi_k\geq0\}\). Then, unprofitable consumers default in periods \(k= 1,2,\ldots, k^*-1\); profitable consumers default in periods \(k \ge k^*\); and creditworthy consumers never default. Further, there exists a cutoff value \(\theta^* = y_{k^*}\)
such that
\begin{itemize}
    \item the consumer is unprofitable for \(\theta \in[0,\theta^*)\),
    \item the consumer is profitable (but non-creditworthy) for \(\theta \in[\theta^*,\bar{x})\), and
    \item the consumer is creditworthy for \(\theta\geq \bar{x}\). 
\end{itemize}
\noindent For all \(k\geq k^*\), \(\Pi_{k+1}\geq \Pi_{k}\).
}

The cutoff value \(k^*\) is the minimum number of periods the consumer needs to repay her loans to be profitable. And, although the relationship between income and profitability is not necessarily monotone throughout, it is monotone among profitable (as well as creditworthy) consumers (indeed, the relationship between NPV and $\theta$ is monotone for large enough $\theta$ in both Figures \ref{fig:NPV}\emph{(a)} and \emph{(b)}). The monotonicity of NPV for profitable customers reflects the fact that these consumers only become profitable in later periods. As outlined in Section \ref{subsec_lean_exp_strategy}, in these later stages, the lender makes only small adjustments to the lending amounts, resulting in a stable default loss. Once consumers keep borrowing for enough periods, their cumulative interest payments offset the final default loss.

\section{Endogenous Interest Rate}
\label{sec_endo}


Having analyzed the case of an exogenous interest rate, we now turn to the lender's problem where the interest (equivalently, discount) rate is set by the lender: the lender has to decide in each period \(t\) on the pair \((d_t,y_t)\) of discount rate and repayment amount. Now, the consumer's acceptance rate \(s(d)\) depends on the discount rate $d$ and needs to be modeled explicitly.

In this setting, the lender seeks the dynamic policy \myeq{\{(y_t, d_t) \ t\geq0\}} that maximizes its expected NPV. Let \myeq{\mathcal{H}_t} denote the lender's information set at time \myeq{t}, which specifies whether the consumer paid back \myeq{y_{i}} in periods \myeq{i < t}. The lender's expected payoff under policy \myeq{(\BFy,\BFd)=\{(y_0,d_0),(y_1,d_1),\dots\}\in(\mathbb{R}^{\infty},\mathbb{R}^{\infty})} is 
\myeql{
\label{lender_expected_payoff_dr_1}
\Pi(\BFy,\BFd)=\sum^{\infty}_{t=0}\  \ \rho^{t}[\Pi^{t-1}_{i=0}\Prob(\theta\geq y_i|\mathcal{H}_i)s(d_i)] \cdot [\rho\Prob(\theta\geq y_t|\mathcal{H}_t)y_t-s(d_t)d_ty_t] ,
} 
\noindent where we define  $\Pi^{-1}_{i=0}\Prob(\theta\geq y_i|\mathcal{H}_i)s(d_i)=1$. In this setting, the lender has more decision variables, but it also has the opportunity to perform richer experiments that further partition the state space based on both $d$ and $y$. Can the lender exploit this opportunity?

\subsection{Grand Experiment: The Idea}
\label{subsec_GE_analysis}

Suppose the lender can determine in advance its desired long-term discount rate $d^*$ and repayment amount $y^*$. Then, there is hope that a single ``Grand Experiment" (GE) can be used to partition the state space into two subsets to help the lender determine whether the consumer is creditworthy or not with respect to the long-term repayment amount $y^*$. In essence, the lender will then use the GE to test whether the consumer can repay $y^*$ and if the answer is positive, will set the repayment amount at $y^*$ in each subsequent period without taking any default risk. Thus, the single GE will segment consumers using a sufficient partition into creditworthy and non-creditworthy consumers, and if a consumer is creditworthy, the lender will proceed with the optimal long-term policy. This cannot be done optimally with a uni-dimensional experiment because the GE requires a higher interest rate to compensate the lender for the higher risk it takes when it makes the experiment. Further, as we show below, a GE may or may not achieve this maxim depending on the properties of the elasticity of demand.     

We will now proceed to implement this idea. To do that, we first assume that the lender is constrained to perform a two-step policy that starts with a single experiment which is then followed by a constant policy in all subsequent periods.  In other words, the lender sets \((y_0, d_0)\) to design the GE in period 0, followed by $(y^*, d^*)$ in periods $1,2,3,\ldots$. We first present the following lemma on the existence and uniqueness of $d^*$, which will play a role in both the two-step policy and the optimal dynamic policy. 

\begin{lemma}\label{prop:dstar}
There exits a unique solution $d^*\in(0,\rho)$ that satisfies
\myeql{\label{eq_d_star_dr_1}
\frac{1 - s(d^*)\, d^*}{1 - s(d^*)\, \rho} \;=\; \frac{s(d^*) + s'(d^*)\, d^*}{\rho\, s'(d^*)}\,.
}
\end{lemma}

We are now ready to characterize the optimal two-stage policy, i.e., to identify the parameters \((y_0,d_0)\) of the GE and to determine the subsequent \((y^*,d^*)\), where (we'll show) $d^*$ is given by Lemma \ref{prop:dstar}. As it turns out, the structure of the policy hinges on the demand elasticity, given by
\myeql{\label{eq:demand_elas}
\xi(d)=\dfrac{ds'(d)}{s(d)}.
}
We obtain the solution for the cases of decreasing, increasing or constant elasticity.  

\begin{proposition}\label{prop:grand} When the hazard rate function $\frac{f(x)}{1-F(x)}$ is monotone increasing, the optimal two-stage policy chooses \((y_0,d_0)\) in period \(0\) and then fixes \((y^*,d^*)\) from period \(1\) onward, with $d^*$ given by Lemma \ref{prop:dstar}. The values of $d_0$, $y_0$ and $y^*$ depend on the monotonicity properties of the elasticity function $\xi(d)$ (equation (\ref{eq:demand_elas})):  

(a) when  $\xi(d)$ is decreasing, the pairs \((y_0,d_0)\), $(y^*,d^*)$ satisfy the following system of equations: 
\myeql{\label{eq:grand1} d_0: \ s'(d_0)[\rho(1-F(y_0))V_0-d_0y_0]=s(d_0)y_0}
\myeql{\label{eq:grand2} y_0:\ \rho s(d_0)(1-F(y_0))[y_0-G(y_0)V_0+\rho s(d^*)G(y_0)\frac{1-F(y^*)}{1-F(y_0)}\frac{1-s(d^*)d^*}{1-s(d^*)\rho}y^*]=s(d_0)d_0y_0}
\myeql{\label{eq:grand3} y^* : G(y^*)+\frac{d^*}{\rho}\frac{1-s(d^*)\rho}{1-s(d^*)d^*}\frac{1-F(y_0)}{1-F(y^*)}=1
}
with $V_0:=y_0-s(d^*)d^*y^*+\rho s(d^*)\frac{1-F(y^*)}{1-F(y_0)}\frac{1-s(d^*)d^*}{1-s(d^*)\rho}y^*$.

(b) when $\xi(d)$ is constant or increasing, the pair \((y_0,d_0)\) is uniquely characterized by the following system of equations: 
\myeql{\label{eq:grand_system1}
G(y_0)\;+\;\frac{d_0}{\rho\,(1-F(y_0))}\cdot\frac{1-s(d^*)\,\rho}{1-s(d^*)\,d^*}\;=\;1,}
\myeql{\label{eq:grand_system2}
\frac{(1-F(y_0))\bigl(1-s(d^*)\,d^*\bigr)}{1-s(d^*)\,\rho}
\;=\;\frac{s(d_0)+s'(d_0)\,d_0}{\rho\,s'(d_0)}\,,}
and $y^*=y_0$.
\end{proposition}

When the demand elasticity is constant, i.e., $s(d)=d^\alpha$, we have the following corollary:


Under the two-step policy with constant or increasing elasticity, the lender adjusts the discount rate exactly once at the initial period: it chooses \(d_0<d^*\) at \(t=0\) and then sets \(d_t=d^*\) for all \(t\ge 1\), while holding the repayment amount fixed (i.e., \(y_t=y_0\) for all \(t\)). This means that in this case, the GE indeed identifies a sufficient partition: a consumer who repaid $y_0$ in the GE will be able to repay $y_0$ in all subsequent periods without default. Intuitively, this means that we have achieved our objective in this case, although this still requires a rigorous analysis. However, when the demand elasticity is decreasing, we don't have a sufficient partition of the state space.  We proceed to make these ideas precise in the next section.    

\subsection{Optimal Dynamic Policy}

Let the state of the system be $x_t=\max_{i=0\dots,t-1}\{y_i\}$, the current lower bound on $\theta$. Then, $x_{t+1}=\max(x_t, y_t)$ and the Bellman equation is 
\myeql{J(x_t)\ = \max_{y_t,d_t} \ \{ -s(d_t)d_t y_t+\rho s(d_t)\Prob(\theta\geq y_t|\theta\geq x_t)(J(x_{t+1})+y_t)\} .
\label{eq_J_xt_dr}
}
The counterpart of Lemma \ref{lemma_increasing_control} in this case is: 
\begin{lemma}
\label{lemma_increasing_control_dr_1}
\textit{If an optimal policy $\{y_t, d_t\}$ exists,  then $y_t$ is weakly increasing, i.e., $y_{t}\geq y_{t-1}$ for all $t>0$. }
\end{lemma}

The intuition is similar to that of Lemma \ref{lemma_increasing_control}: when the lender knows that $\theta\geq y_{t-1}$, setting $y_t<y_{t-1}$ is sub-optimal. Thus, the state evolves as $x_t = y_{t-1}$ and the Bellman equation becomes
\myeql{J(x_t)\ = \ \max_{y_t,d_t} \ \ \{ x_t-s(d_t)d_ty_t+\rho s(d_t)\Prob(\theta\geq y_t|\theta\geq x_t)J(y_t) \} .
\label{eq_J_xt_dr_1}
}

\begin{lemma}\textit{
\label{lemma_general_barx_dr_1}
There exists a unique finite solution $0<\bar{x}<u$ to the equation:  
\myeql{
\label{eq_bar_x_dr_1}
G(\bar{x})=-\frac{(1- s(d^*)\rho)}{(1-s(d^*)d^*)}\frac{d^*}{\rho}+1,
}
where $d^*\in(0, \rho)$ is the unique solution of equation (\ref{eq_d_star_dr_1}). }
\end{lemma} 

When the demand elasticity is monotone decreasing, the structure of the optimal lending policy is similar to the one we found in Theorem \ref{them_explore_exploit_1}:
\myth{
\label{them_explore_exploit_dr_1}  
When $\xi(d)$ is monotone decreasing, there exists a unique optimal policy whereby the lender keeps increasing the repayment amounts $y_t$ in each period $t$ as long as the consumer repays her loans. The infinite sequence of repayment amounts $\{y_t, t=0,1,2,\ldots\}$ is strictly increasing to the limit \myeq{\bar{x}}. The optimal discount rates are not necessarily monotone and converge to $d^*$.}

Intuitively, when the demand elasticity is decreasing, the consumer is less likely to balk at higher discount rates. The discount rate generally increases (the interest rate decreases) as the lender accumulates more positive information about the consumer's creditworthiness. This creates an advantage for experimenting later, when the discount rate is higher. Thus, the lender's optimal policy is based on an infinite number of small experiments, as aggressive early experiments will reduce demand by all consumers, leading to the loss of too many creditworthy ones. This is an LE policy similar to the one we obtained with an exogenous interest rate. However, when the demand function has \emph{constant} elasticity, $s(s)=d^{\alpha}$ with $\alpha\in(0,1)$, the optimal policy structure changes dramatically.

\myth{\label{thm_constant_elas} : When the demand elasticity is constant, there exists a unique optimal policy characterized by a single ``grand experiment'': the lender first sets the repayment amount at $y_0=\bar{x}$ and the discount rate at $d_0=(1-F(\bar{x}))d^*$. If the consumer repays $y_0$, the lender fixes the repayment amount at $y_t=\bar{x}$ and the discount rate at $d_t=d^*$ for all subsequent periods.}

Theorem \ref{thm_constant_elas} is remarkable, confirming our intuitive idea for the GE policy. It shows that when the demand elasticity is constant, the GE policy is optimal for the dynamic problem. The GE tests upfront whether the consumer can repay the optimal long-term repayment amount $\bar{x}$. If the answer is positive, the lender has a sufficient partition for lending $\bar{x}$: in each period, it will charge the same interest rate and lend the same amount which together lead to repayment amount $\bar{x}$. The lender takes risk only at the GE and has has no risk in all subsequent periods, as it knows the consumer will repay $\bar{x}$: the partition induced by the GE is sufficient. The lender takes advantage of its ability to control both the discount rate and the loan amount, designing a multidimensional GE that sets a lower loan amount (entailing less risk) and a higher interest rate that together test the optimal repayment amount $\bar{x}$ and partition the state space accordingly. In this way, one test establishes everything the lender needs to know without taking excessive risk. 

We further explore how the (constant) elasticity $\alpha$ shapes the optimal policy. The following proposition states how $d^*$ and $\bar{x}$ change as functions of $\alpha$. 

\begin{proposition}\label{prop_d_increase_in_alpha}
\textit{Under constant elasticity, the optimal long-term discount rate $d^*$ is increasing in $\alpha$, and the optimal lending threshold $\bar{x}$ is decreasing in $\alpha$.}
\end{proposition}

Intuitively, increasing the elasticity $\alpha$ makes acceptance more sensitive to the discount rate.
To retain consumers, the lender then lowers the interest rate so \(d^*\) increases with \(\alpha\). In addition, the lender increases the expected number of consumers it retains by lowering the repayment threshold \(\bar{x}\).

The optimality of the GE strategy does not hinge on the constant elasticity assumption, although the case where demand has increasing elasticity is more complex. 

\myth{\label{thm_increasing_elas} : When both the demand elasticity and the hazard rate function $\frac{f(x)}{1-F(x)}$ are increasing, there exists a unique optimal policy characterized by a single ``grand experiment'':  the lender first sets the  repayment amount at $y_0$ and the discount rate at $d_0$. If the consumer repays $y_0$, the lender fixes the repayment amount at $y_t=y_0$ and the dicount rate at $d_t=d^*$ for all subsequent periods.
    
Here, $y_0>\bar{x}$ and $(y_0,d_0)$ jointly solves
\myeql{G(y_0)+\dfrac{d_0}{\rho(1-F(y_0))}\dfrac{(1-s(d^*)\rho)}{(1-s(d^*)d^*)}=1,\quad\dfrac{(1-F(y_0))(1-s(d^*)d^*)}{(1-s(d^*)\rho)}=\dfrac{s(d_0)+s'(d_0)d_0}{\rho s'(d_0)}.}
}

Under the conditions of Theorem \ref{thm_increasing_elas}, the lender still employs a single GE. However, under strictly increasing elasticity the experiment does not involve repayment amount $\bar{x}$. Instead, the lender tests whether the consumer can repay a $\emph{higher}$ amount $y_0$. The reason is that with increasing demand elasticity, the consumer is less sensitive to a higher interest rate. Taking advantage of that, the lender designs a more aggressive GE with a higher repayment amount than $\bar{x}$. Similar to the constant elasticity case, the first loan has a lower discount rate than $d^*$ and if the consumer repays $y_0$, the lender will not take any risk in subsequent periods by lending $(d^*y_0)$ in each period and setting the discount rate at its optimal level, $d^*$: in this case as well, the GE establishes a sufficient partition of the state space.

\subsection{Implications}
\label{subsec_implications}

\subsubsection{LE vs. GE}
\label{subsec_discuss_LEGE}

Theorems \ref{them_explore_exploit_1} and \ref{thm_constant_elas} show that under our dynamic model, either LE or GE approach may be optimal depending on the properties of the demand function and the scope of the lender's decisions (endogenous vs. exogenous interest rate). Lenders should adopt an LE policy when they don't set the interest rate or when the demand elasticity is decreasing in the discount rate. They should follow GE when the demand elasticity is constant or increasing in the discount rate. The distinction between the two cases lies in the lender's consideration of the risk of losing customers when it charges a high interest rate (low discount rate), which hinges on the shape of the demand function $s(d)$.  The effect of the elasticity function is due to the fact that the lender has to balance its default risk from low-income borrowers against the risk of losing profitable high-income borrowers. As the risk of default declines, so does the interest rate since the lender wishes to retain the customer. With a decreasing demand elasticity, later experiments entail a lower risk of losing customers, which creates an incentive to delay risky experimentation.  

The implications of the LE strategy studied in Section \ref{subsec_lean_exp_strategy} carry over qualitatively to the decreasing demand elasticity case with endogenous rates. We have also seen (Theorems \ref{thm_constant_elas} and \ref{thm_increasing_elas}) that both the constant and increasing elasticity cases lead to qualitatively similar GE results. Hence, we study next the implications of the constant elasticity case, which is more transparent, as a canonical GE case.  



\begin{figure}[htb!]
\centering
{\includegraphics[width=0.55\textwidth]{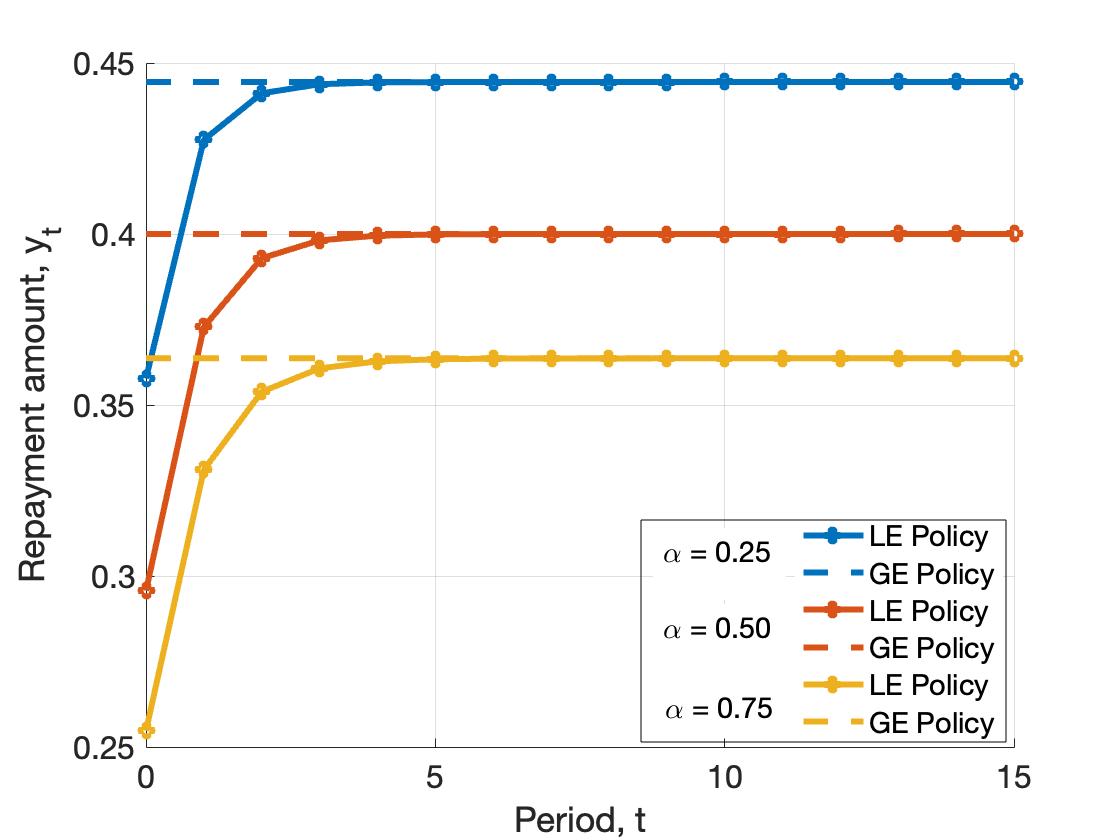}} 
\caption{Optimal repayment amounts ($y$-axis) under LE and GE for the uniform distribution. On the $x$-axis is the time period. We consider $\alpha=0.25,0.5$ and $0.75$, and $\rho=0.95$.} 
\label{fig_opt_policy_unif_compare}
\end{figure}

In Section \ref{subsec_lean_exp_strategy}, we analyzed the LE policy for uniformly-distributed income. We next compare these results to the GE policy under comparable assumptions (except that the discount rate $d$ is now endogenous). In the comparisons, we assume a uniform income distribution and exogenous discount rate $d=d^*$ for LE, so we are comparing the same discount rate in both cases. It follows that the consumer's acceptance probability $s=s(d^*)=(d^*)^{\alpha}$ is also the same in both cases. This implies that the threshold repayment amount $\bar{x}$ under LE is the same as the repayment amounts (in all periods) under GE.
Figure \ref{fig_opt_policy_unif_compare} presents the period-by-period repayment amounts under LE and $\bar{x}$ (the repayment amount in all periods under GE) for different levels of demand elasticity $\alpha=0.25,0.5$ and $0.75$.

Figure \ref{fig_opt_policy_unif_compare} shows that as $\alpha$ increases, the optimal lending policy becomes more conservative (lower repayment amounts) under both LE and GE. This is to be expected as discussed in the context of Proposition~\ref{prop_d_increase_in_alpha}. 
eFurther, the (exogenous rate) LE repayments are always lower than the GE repayment and they converge to the GE level from below. The difference stems from the fact that the two-parameter policies allow the lender to better control its risk. GE jointly optimizes the lending threshold and the discount rate, obtaining a more effective partition of the state space at a lower experimentation cost, whereas the single-parameter LE can adjust only the repayment amount, which makes it more risky and leads to more conservative repayment amounts to mitigate the default risk.

Lenders may adopt the simple two-step GE policy even when it's not optimal. If they do that, how much will they lose relative to the optimal dynamic solution? To answer this question, we consider a range of demand functions $s(d)$ which will give us a range of elasticity functions \(\xi(d)\) encompassing decreasing, constant and increasing elasticities. We choose the demand function $
s(d)\;=\;\frac{\sqrt{d}}{(1-\ln d)^{\,q}}$ parametrized by \(q\in(-0.2,\,0.2)\), where $d\in[0,\rho]$. With this parameterization, $s(d)$ is strictly increasing and concave.  The  demand elasticity is then given by $\xi(d)=\frac{1}{2}+\frac{q}{1-log(d)}$. It is decreasing when \(q<0\), constant when \(q=0\), and increasing when \(q>0\); we know that in the latter two cases, two-step GE is optimal. 

\begin{figure}[htb!]
\centering
{\includegraphics[width=0.45\textwidth]{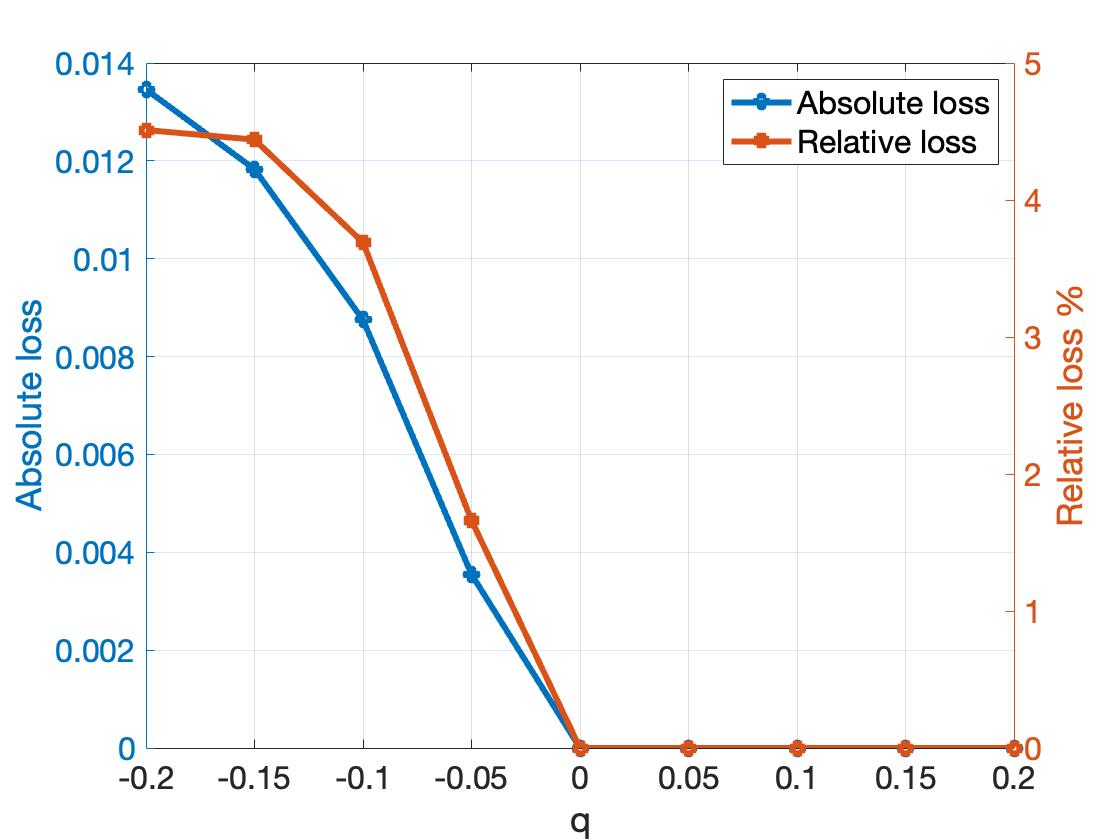}} 
\caption{Comparison between optimal dynamic policy and two-step GE policy, for $s(d)=\frac{\sqrt{d}}{(1-\ln d)^{\,q}}$, uniform income distribution and $\rho=0.95$.} 
\label{fig_GEopt}
\end{figure}

We compute the absolute loss as the difference between the NPV under the optimal dynamic policy and the NPV under the two-step GE policy, and the relative loss as the ratio of absolute loss to NPV under the optimal dynamic policy.  Figure~\ref{fig_GEopt} reports the absolute and relative loss as functions of the parameter \(q\).

Figure \ref{fig_GEopt} shows that when the demand elasticity is constant or increasing $(q \ge 0$), the optimal policy coincides with the two-step GE policy (Theorems \ref{thm_constant_elas} and \ref{thm_increasing_elas}). As $q$ becomes more negative, the absolute and relative losses increase as expected. The loss is small when $q$ is close to zero, and increases to a few percentage points as $q$ decreases. For small $q$, the two-stage approximation may be acceptable; for larger (absolute) values, lenders should opt for an LE policy (Theorem \ref{them_explore_exploit_dr_1}).

\subsubsection{Consumer Segmentation (Sufficient Partition) under GE}
\label{subsec_GE_partition}

Under LE, a sufficient partition of consumer incomes comprises an infinite number of segments which we aggregated into three behavioral segments. Under GE with a constant elasticity, consumer segmentation is much simpler and more intuitive: our sufficient partition of the state space has only two segments, which also correspond to a simple behavioral segmentation into creditworthy (and profitable) vs. non-creditworthy (and unprofitable) consumers:

\begin{itemize}
    \item \textit{Non-creditworthy consumers} with \(\theta< \bar{x}\) default in the first (test) period. The default results in a loss to the lender, i.e., these consumers are unprofitable.
    \item \textit{Creditworthy consumers} with \(\theta\geq \bar{x}\) repay in the first period, and therefore in all subsequent periods, and are profitable.
\end{itemize}

\begin{figure}[htb!]
\centering
{\includegraphics[width=0.55\textwidth]{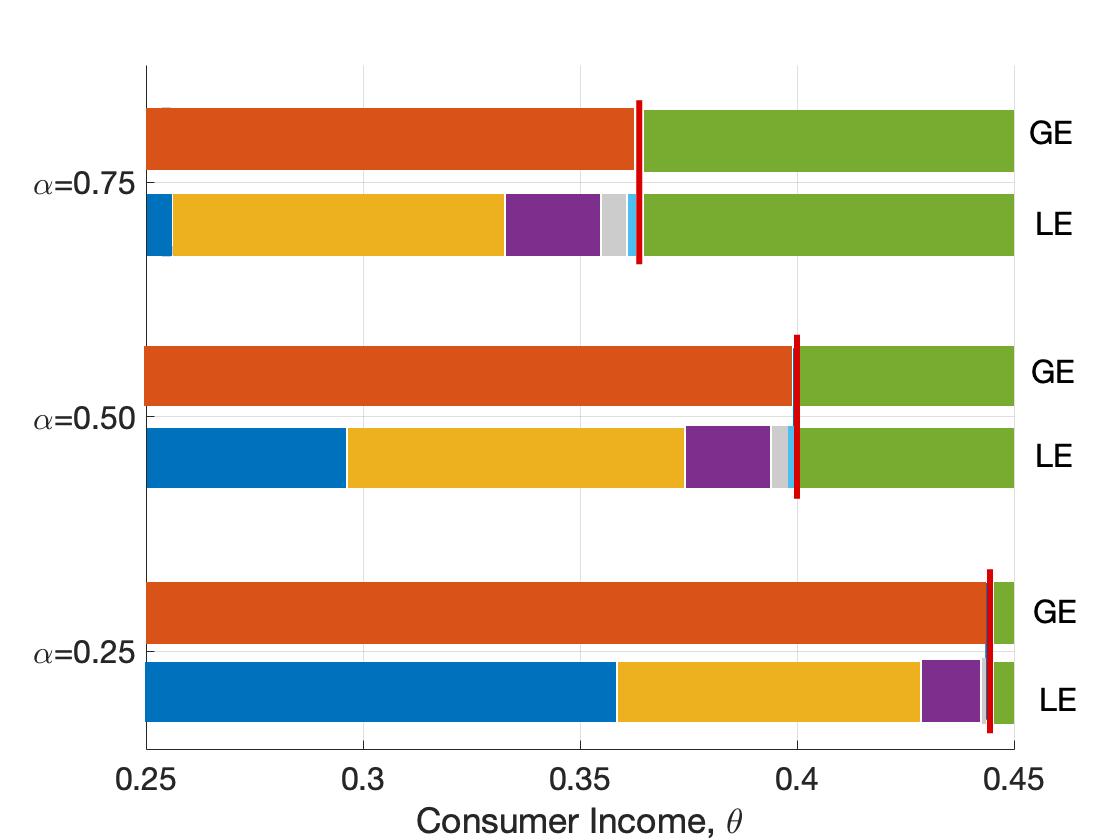}} 
\caption{Sufficient state space partitions under LE and GE for uniform income distribution with $\alpha=0.25,0.5$ and $0.75$ and $\rho=0.95$. The market discount rate $d$ under LE is set at the optimal rate under GE, $d^*$. } 
\label{fig:state_partition}
\end{figure}

Thus, under GE, non-creditworthy consumers (\(\theta<\bar{x}\)) are never profitable. Further, the NPV is strictly increasing in income, unlike the LE regime where non-creditworthy consumers may be profitable and the NPV need not be monotone in income. Figure \ref{dr_income_var} compares consumer segmentation under LE and GE for our uniform example. LE induces a fine-grained partition of the state space into segments \([y_{k-1},y_k)\) $(k=1,2,3, ..\ldots)$ shown by the colored stacked bars (the blue bar corresponds to Type-0 consumers, the orange to Type 1, the magenta to Type 2, the gray to Type  3 and so on). Under GE, the sufficient partition collapses to two segments (red, non-creditworthy to the left of the separating red line at $\bar{x}$ and creditworthy in green to its right). In both cases, $\bar{x}$ is the same, but the LE policy approaches $\bar{x}$ in small increments whereas GE first tests whether the consumer can repay $\bar{x}$ and then lends $(d^*\cdot\bar{x})$ in each subsequent period. 
This illustrates that with more actions to experiment with, the lender can partition the state space through a highly informative GE which is sufficient to determine whether the consumer is creditworthy or not. That is, the additional dimension of the action space allows the lender to concentrate all of its risk in the first experiment and, based on the results of that experiment, to proceed with riskless loans in all subsequent periods. This strategy is optimal when the demand elasticity is constant (as in Figure \ref{fig:state_partition}) or increasing, but not when the elasticity is decreasing. The reason is that with decreasing elasticity, the lender has an incentive to lend less early on (when market conditions are less favorable) and more later on (when they are more favorable), contrary to the pattern obtained under GE.   




\section{Profitability, Income Variability and the Value of Information}


\subsection{Impact of Income Variability}
\label{subsec_impact_income_var}

\begin{figure}[htb!]
\centering
{\includegraphics[width=0.55\textwidth]{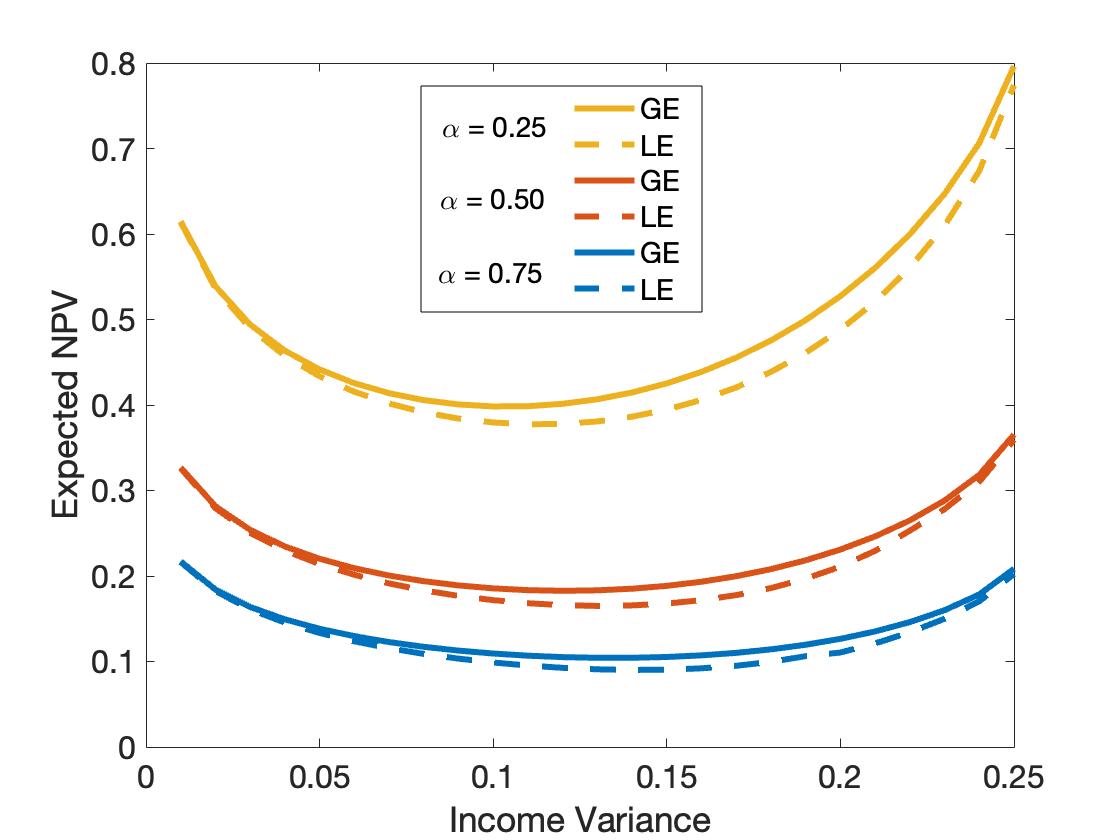}}\quad \qquad
\caption{Expected NPV for symmetric Beta distribution with shape parameter $a\in[0.02, 12]$. The corresponding income variances vary from $0.01$ to $0.24$. The demand elasticities are constant at $\alpha=0.3,\ 0.5$ and $0.9$ and $\rho = 0.95 $. For comparability, we set $d=d^*$ in the exogenous rate case.} 
\label{dr_income_var}
\end{figure}

How does the uncertainty about consumer incomes (i.e., the variability of the income distribution) affect the lender's expected profit? One might expect that lenders are averse to uncertainty. The results, however, are more nuanced. To examine the relationship between income variability and NPV, we consider symmetric Beta-distributed income with density $f(a,b,x)=\frac{x^{a-1}(1-x)^{b-1}}{B(a,b)}$, where $B(a,b)=\frac{\Gamma(a)\Gamma(b)}{\Gamma(a+b)}$ and $\Gamma(\cdot)$ is the Gamma function. We set $a=b$ so the mean income is $\frac{1}{2}$. Decreasing $a$ increases the variance of income given by $\frac{1}{4(2a+1)}$, without changing its mean. As in our earlier examples, we set $d=d^*$ in the exogenous interest rate case for comparability. Figure \ref{dr_income_var} shows the lender's NPV as a function of the income variance for different demand elasticities. First, the expected NPV is uniformly higher for GE, as might be expected since the lender has more degrees of freedom. The two NPVs converge when the variance is at its minimum ($0$) or maximum $(\frac{1}{4})$: When the variance is zero, income becomes deterministic at its mean value $\frac{1}{2}$. Then, both LE and GE give the perfect information solution, setting the repayment amount equal to income and choosing \(d=d^*\). When the variance is at the other extreme, $\frac{1}{4}$, the income distribution becomes Bernoulli with values $0$ or $1$ with probability $\frac{1}{2}$ each. Then, both LE and GE set the repayment amount to \(1\) with \(d=d^*\).
Also, as discussed earlier, the higher the demand elasticity, the lower the lender's profit in both cases.

NPV is a U-shaped function of the income variance: it first decreases and then increases. That is, 
profits are non-monotone in the income variance, and a higher variance may actually benefit the lender. In fact, the following Proposition shows that for the symmetric Beta distribution, the lender's NPV is U-shaped in the GE, constant-elasticity case.

\myprop{\label{prop_beta_ushape} For the symmetric Beta distribution with a fixed mean, in the constant-elasticity GE case the lender's expected NPV first decreases and then increases as the income variance increases. }

The common decreasing relationship between NPV and variance reflects the usual losses due to uncertainty. With no income variance, the lender has full information and can set the repayment amount equal to the consumer's income without any losses. As the income variance increases, the lender may lose money and to protect itself, it lends less. Then, its expected profits decline as a result its expected losses and its lower interest income.

The increasing part of the U-shape reflects the option value of experimentation. As the variance increases, the option value of experimentation increases as it increases the likelihood of identifying high-income borrowers. This also implies that the lender increases the lending threshold to better target high-income borrowers. This effect has to be balanced against the usual negative effect of uncertainty. The option value is particularly high for high values of the income variance, which leads to the U shape. 

We illustrate these general considerations with a simple example for the GE case. Consider a two-point income distribution with a fixed mean, $u$, where $\theta$ is $u+\delta$ or $u-\delta$ with probability $\frac{1}{2}$ each. In the special case where $u=\delta=\frac{1}{2},$ the income distribution is Beta with $a=b=0$; when $u=\frac{1}{2}$ and $\delta=0$, the income distribution is deterministic---a limiting form of the Beta distribution with $a$ and $b$ going to infinity. It is easy to see that one of following two policies must be optimal:  

\begin{enumerate}[label=(\Alph*)]

\item At $t=0$, lend $(u-\delta)+\epsilon$ at discount rate $d_0$ with $\epsilon$ being a small positive number, and proceed to offer $(u+\delta)$ in each period if the first-period loan is repaid. The corresponding expected NPV as $\epsilon \to 0$ is $\Pi^A(\delta) = -(d_0)^{a+1}(u-\delta)+\frac{1}{2}\rho (d_0)^{\alpha}\left((u-\delta)+ \left(\frac{1-(d^*)^{\alpha+1}}{1-\rho (d^*)^{\alpha}}-1\right)(u+\delta) \right)$. 
\item Lend $(u-\delta)$ with discount rate $d^*$ in each period unless the borrower defaults. The corresponding expected NPV is $\Pi^B(\delta) = \left(\frac{1-(d^*)^{\alpha+1}}{1-(d^*)^{\alpha}\rho}-1\right)(u-\delta)$.
\end{enumerate}

For each $\delta$, the lender's expected profit is $\max\left\{\Pi^A(\delta),\Pi^B(\delta)\right\}$, the maximum of two linear functions of $\delta$, where increasing $\delta$ increases the variance of the income distribution. Under Policy (A), the first experiment gives the lender the option to identify the high-income consumer. As is common in option valuation, the option value is an increasing function of the variance (parameterized here by $\delta$) -- if the consumer has the low income \(u-\delta\), she will default and won't get another loan anyway. Policy (B) lends to the entire market, including low-income consumers, and its profit decreases in $\delta$ (i.e., profits decrease with variance, as usual), as it does not allow the lender to identify the high-income consumer and as a result has no option value. Rather, under policy (B) the lender ``plays it safe." For large enough $\delta$ (and variance), policy (A), which targets the high income consumer, yields a higher expected profit. For small $\delta$, the income differences are small and the payoff of experimentation is low. As a result, for small $\delta$ the lender is better off with policy (B) which lends less to the entire market. The linear profit functions $\Pi^A(\delta)$ and $\Pi^B(\delta)$ intersect at some $\delta^*$, so the expected profit function $\max\left\{\Pi^A(\delta),\Pi^B(\delta)\right\}$ is U-shaped (in this case, V-shaped) with a minimum at the intersection point $\delta^*$. 

Similar considerations apply to the LE case.

\subsection{Value of Information}
\label{subsec_implications_unif}

We next compare our optimal policy to an oracle benchmark that knows the realization of $\theta$. The (relative) \emph{value of information} is measured by the (ex-ante) \emph{relative regret}:
\[
R^{I} \;:=\; \frac{\E(\Pi^{fb})}{\E(\Pi)}\,,
\]
which is the ratio of the oracle’s expected profit to that achieved by the optimal dynamic policy with the same parameters; we assume $\theta\sim U[0,1]$.

\paragraph{LE (exogenous interest rate).}
Under LE, if the lender knows $\theta$, it lends $(d\cdot\theta)$. The corresponding first-best profit is \(
\Pi^{fb}(\theta)=\frac{\rho-d}{1-\rho}\,\theta,
\)
hence
\[
\E(\Pi^{fb})=\int_0^1 \frac{\rho-d}{1-\rho}\,\theta\, d\theta \;=\; \frac{\rho-d}{2(1-\rho)}.
\]
We compare this to the expected optimal dynamic profit (Theorem~\ref{them_unif}):
\[
\E(\Pi)=\frac{1}{2(1-\rho)(2\rho-d-\rho d)^2}\Big[(\rho-d)^2\Big(2d-\rho-1 + d(1-\rho)\sqrt{\tfrac{\rho-d^2}{\rho d^2}}\Big)\Big].
\]
Trivially, $\E(\Pi^{fb})\ge \E(\Pi)$. The next result shows how large the gap can be.

\begin{proposition}
\label{prop_bounds_value_info}
Under an exogenous interest rate, $R^{I}\ge 2$, and for any $M>0$ there exist $(\rho,d)$ such that  $R^{I}>M$.
\end{proposition}

Although a creditworthy borrower generates arbitrarily many observations, Proposition~\ref{prop_bounds_value_info} shows that the online lender's NPV can never exceed half of the oracle benchmark, and that fraction can become arbitrarily small. The reason is that as discussed in Section \ref{subsec_lean_exp_strategy}, the lender is extremely conservative as it needs to protect itself against the risk of losing the entire principal amount. As a result, its experiments become infinitesimal and its repayment amount never exceeds $\bar{x}$ even when the consumer's income exceeds $\bar{x}$ with a probability that approaches 1. The opportunity cost of this extreme caution is half of the potential (oracle) profit or more. 

\paragraph{GE (endogenous interest rate with constant elasticity).}
Will relaxing the interest rate constraint and allowing the lender to set it in each period mitigate the $50\%$ information gap? The answer is no. With constant elasticity and $\theta\sim U[0,1]$, if the lender knows $\theta$, its first-best NPV is
\(
\Pi^{fb}(\theta)=\frac{1-s(d^*)d^*}{1-s(d^*)\rho}\,\theta
\), so
\[
\E(\Pi^{fb})=\int_0^1 \frac{1-s(d^*)d^*}{1-s(d^*)\rho}\,\theta\, d\theta \;=\; \frac{1-s(d^*)d^*}{2(1-s(d^*)\rho)}.
\]
By Theorem~\ref{thm_constant_elas}, the optimal profit is given by
\[
\E(\Pi)=\frac{\big((1-F(\bar{x}))\,d^*\big)^{\alpha+1}}{\alpha}\,\bar{x}
=\frac{\big((1-F(\tfrac{1}{\alpha+2}))\,d^*\big)^{\alpha+1}}{\alpha}\,\Big(\frac{1}{\alpha+2}\Big).
\]
We find again that $R^{I}=\frac{\E(\Pi^{fb})}{\E(\Pi)} \ge 2$.

\begin{proposition}
\label{prop_bounds_value_info_dr}
Under an endogenous interest rate with constant elasticity, $R^{I}\ge 2$. For any $M>0$, there exists a $\rho$ for which $R^{I}>M$.
\end{proposition}

Despite its additional degrees of freedom, the GE policy remains conservative. The ability to control interest rates accelerates convergence and allows the lender to discover with a single experiment whether the consumer is able to repay $\bar{x}$, but it subsequently lends $\bar{x}$ and no more. It follows that in either case, the lender can capture at most half of the oracle value, with the gap \(\E(\Pi^{fb})-\E(\Pi)\) representing the shadow cost of learning-by-doing, which is substantial under both LE and GE. The bottom line, then, is that online lending is not only extremely risky---it is also highly unprofitable compared to the full-information benchmark.
\section{Extension: Random Income Shocks}
\label{subsec_stochastic_income_fi}

We now consider a setting in which consumer income is stochastic. In each period $t$, income is given by $\theta + w_t$, where $w_t$ is a random shock which is not observable by the lender. Each shock $w_t$ has an arbitrary distribution with support $[-\epsilon,\epsilon]$, where $\epsilon>0$.  The distribution may vary over time or exhibit serial (inter-temporal) dependence. Here, we refer to $\theta$ as the consumer's type, with period-$t$ income given by $\theta + w_t$.

The lender's expected payoff under policy \myeq{\BFy=\{y_0,y_1,\dots\}\in\mathbb{R}^{\infty}} is 
\myeql{
\label{lender_expected_payoff_noise}
\Pi(\BFy)=\sum^{\infty}_{t=0}\ \rho^{t} \ [\Pi^{t-1}_{i=0}\Prob(\theta+w_i\geq y_i|\mathcal{H}_i)] \cdot [\rho\Prob(\theta+w_t\geq y_t|\mathcal{H}_t)y_t-d y_t] ,
} 

\noindent where we define  $\Pi^{-1}_{i=0}\Prob(\theta+w_i\geq y_i|\mathcal{H}_i)=1$.


The lender's problem can still be formulated as an infinite-horizon discounted dynamic programming problem. Denote the state at period \( t \) by \( x_t = \{y_0, \dots, y_{t-1}\} \), which summarizes all past repayment amounts. The state evolves according to \( x_{t+1} = x_t \cup \{y_t\} \). The corresponding Bellman equation is given by 
\begin{equation}
\label{eq:J_xt_noisy}
J(x_t) = \max_{y_t} \left\{ -d y_t + \rho \, \mathbb{P}(\theta + w_t \geq y_t \mid \theta + w_i \geq y_i,\, \textup{ for all } i < t) \cdot (J(x_{t+1}) + y_t) \right\},
\end{equation}
where the conditional probability reflects the belief update based on prior repayments. This formulation suffers from the curse of dimensionality: as the period \( t \) increases, the state space \( x_t \) grows in dimensionality, making direct computation intractable. However, the following lemma establishes that, when the income shock is relatively small, our results remain approximately valid in the random income setting.

\begin{proposition}\textbf{Random Income Shocks in the Exogenous Interest Rate Case}
\label{prop:bellman_bound}
\textit{Suppose $\theta \sim F$ has support on $[l,u]\subset[0,\infty)$ with strictly increasing CDF $F$, and assume $F$ is continuously differentiable and Lipschitz on $[l,u]$ with constant $L$. Let $w_t$ be random shocks supported on $[-\epsilon,\epsilon]$ and independent of $\theta$. Define the Bellman equations: 
\[
J(x_t)=\max_{y_t}\Big\{-d\,y_t+\rho\cdot\mathbb{P}\!\big(\theta\ge y_t \,\big|\, \theta\ge x_i,\ \textup{ for all } i<t\big)\cdot\big[y_t+J(x_t\cup\{y_t\})\big]\Big\},
\]
\[
J^{\epsilon}(x_t)=\max_{y_t}\Big\{-d\,y_t+\rho\cdot\mathbb{P}\!\big(\theta+w_t\ge y_t \,\big|\, \theta+w_i\ge x_i,\ \textup{ for all } s<t\big)\cdot\big[y_t+J^{\epsilon}(x_t\cup\{y_t\})\big]\Big\}.
\]
Then, for $\rho\in(0,1)$, each Bellman equation admits a unique solution. For $L\epsilon<\frac{1}{2}$, the lender's expected profits under the two models, $\Pi:=J(x_0=0)$ and $\Pi^{\epsilon}:=J^{\epsilon}(x_0=0)$, satisfy
\[
\big|\Pi-\Pi^{\epsilon}\big|\ \le\ \frac{2L\,(u+\epsilon)\,(1-d)}{1-\rho}\,\epsilon.
\]}
\end{proposition}

In spite of the complexity of the solution, Proposition \ref{prop:bellman_bound} allows us to bound the value difference between the two models. Here, $\frac{(u+\epsilon)(1-d)}{1-\rho}$ is a natural upper bound on any feasible NPV, as it is the lender's NPV with oracle knowledge that the consumer’s income attains its maximum $u+\epsilon$ each period. The inequality shows that the relative gap between the models is of the order of \(2L\epsilon\), where \(2\epsilon\) is the range of the random income shocks and the Lipschitz constant $L$ measures the sensitivity of the distribution function to income changes. For example, the Lipschitz constant of the standard uniform distribution is $L=1$, so the bound is proportional to $2\epsilon$. As the income shock shrinks (\(\epsilon\) goes to $0$) or \(F\) is smoother (smaller \(L\)), the gap between the two NPVs contracts (linearly in $L$ and \(\epsilon\)) and vanishes in the limit, so the two models deliver nearly identical expected profits. 



A similar bound applies to the case of endogenous interest rates. With the same random income shock model,
the following proposition bounds the difference in NPV between the two models.

\begin{proposition}\textbf{Random Income Shocks under Endogenous Interest Rates}
\label{prop:bellman_bound_endo}
\textit{ Suppose $\theta \sim F$ has support on $[l,u]\subset[0,\infty)$ with strictly increasing CDF $F$, and assume $F$ is continuously differentiable and Lipschitz on $[l,u]$ with constant $L$. Let $w_t$ be random shocks supported on $[-\epsilon,\epsilon]$ and independent of $\theta$. Define the Bellman equations: 
\[
J(x_t) = \max_{d_t,y_t} \left\{ -d_t y_t + \rho s(d_t) \cdot \mathbb{P}(\theta \ge y_t \mid \theta \ge x_i,\, \textup{ for all } i < t) \cdot \left[y_t + J(x_t \cup \{y_t\})\right] \right\} ,
\]
\[
J^{\epsilon}(x_t) = \max_{d_t, y_t} \left\{ -d_t y_t + \rho s(d_t)\cdot \mathbb{P}(\theta + w_t \ge y_t \mid \theta + w_i \ge x_i,\, \textup{ for all } i < t) \cdot \left[y_t + J^{\epsilon}(x_t \cup \{y_t\})\right] \right\} .
\]
Then, for $\rho\in(0,1)$, each Bellman equation admits a unique solution. Moreover, for $L\epsilon<\frac{1}{2}$, the lender's expected profits $\Pi:=J(x_0=0)$ and $\Pi^{\epsilon}:=J^{\epsilon}(x_0=0)$ satisfy
\[
|J^{\epsilon}(x_0) - J(x_0)| \le   \frac{2L(u+\epsilon)(1-s(d^*)d^*)}{(1-s(d^*)\rho)}\epsilon.
\]
}
\end{proposition}

Similar to Proposition \ref{prop:bellman_bound}, the bound scales as $2L\epsilon$ compared to the upper bound on any feasible NPV, suggesting that for relatively small shocks, our results approximate those of the random income shock model. 

\section{Concluding Remarks}
\label{sec_conclusion}

This paper studies optimal policies for online lending, an interesting area which has been transformed over the past thirty years by information and communication technologies and the increasing application of AI. Online lending is also an important example of sequential experimentation, or learning by doing, where the lender gradually learns about the borrower's creditworthiness by repeatedly lending to her. Our results provide online lending operators insights about the management of this activity and in particular, how to structure their loans, what the structure depends on, and how to think about the costs and benefits of experimentation. 

Online lending is also an important form of experimentation, and the canonical structures we discover---Lean Experimentation (LE) vs. a Grand Experiment (GE)---may transcend the lending domain. LE in particular is considered a best practice in the experimentation domain, and in the lending context it implies lending incrementally increasing amounts to gradually learn about a borrower's creditworthiness. We find that LE is indeed the optimal structure when the loan amount is the lender's only decision variable. However, we find that under certain conditions, a single multi-dimensional experiment (in our model, involving the initial interest rate and loan amount) can provide a sufficient partition of the state space, so decision-makers may use it to derive optimal results for all subsequent periods. Beyond the lending domain, our approach suggests that experiments may be structured so as to unravel an underlying partition of the state space which reduces the need for further costly experiments.

We derive optimal policies for models with exogenous and endogenous interest rates. In the exogenous rate scenario, the lender gradually increases the repayment amount, converging to a threshold; this solution belongs to the LE paradigm. Under endogenous rates, the shape of the demand elasticity function determines the structure of the optimal policy. When the demand elasticity decreases, LE is still optimal. When the demand elasticity is constant or increasing, the lender achieves optimality through a single GE without experimenting further in the subsequent periods. We derive the optimal policies explicitly and study the effects of income variability and the structure of consumer segmentation. We also bound the profit loss when our solution is applied to a setting where consumers are subject to random income shocks.  

In the experimentation context, we believe product developers (and in some cases, startups) should consider the design of a Grand Experiment as a potential alternative to the common LE approach. As we have seen in the lending domain, the GE approach does not always achieve optimality, but under a broad set of circumstances, it does. To apply it, the decision-maker needs to first identify a stable long-term policy for the problem at hand, and then identify a sufficient partition with respect to that policy. When such a partition exists, the GE maps the underlying state to the appropriate subset of the partition. If the sufficient partition has two subsets, ``desirable" vs. ``undesirable", there is hope that a GE policy may achieve optimality, and if it does not, it may achieve ``good-enough" results. Formalizing these ideas and identifying general conditions under which this approach works is an interesting area for future research. We hope researchers will follow this path when looking at other problems involving sequential experiments, and we hope to identify a more general structure that will guide the choice of an experimentation regime.

Our model is stylized and another avenue for future research is to subject it to empirical validation. The structure of lending policies and their impact on profitability may be tested directly using data from direct lenders. Our predictions may also be empirically tested using macroeconomic data. Online lending has been adopted in multiple countries, and cross-country data may be used to test some of our predictions, including the drivers of LE vs. GE strategies and the effects of the income distribution. For example, our models predict that highly regulated interest rate environments lead to the adoption of LE strategies whereas flexible rate environments lead (under the conditions studied in Section \ref{sec_endo}) to GE strategies. Thus, regulatory differences among states and countries may lead to sharply different lending strategies. Regarding income variability, our models predict that the relationship between income variance and profitability or ultimate lending amounts (represented in our model by the limit $\bar{x}$) is not monotone. Such relationships may be tested using macroeconomic income distribution data.

\newpage 
\bibliographystyle{pomsref}
 \let\oldbibliography\thebibliography
 \renewcommand{\thebibliography}[1]{%
    \oldbibliography{#1}%
    \baselineskip14pt 
    \setlength{\itemsep}{10pt}
 }
\bibliography{ref1}

\newpage
\begin{center}
\huge \textbf{E-Companion}
\end{center}

\setcounter{section}{0}
\def\thesection{\Alph{section}}

\section{Proofs} 

\label{sec:proofs}

\subsection{The Gamma, Weibull and Beta distributions satisfy Assumption \ref{assumpt1}}
\label{app_assump1_dist}

\subsubsection{The Gamma Distribution satisfies Assumption \ref{assumpt1}}
\hfill
\myproof{Consider the two-parameter Gamma distribution with shape parameter $a > 0$, scale parameter $b>0$, and probability density function (pdf) $f(x)=\frac{b^{a}}{\Gamma(a)}x^{a-1}e^{-b x}$. Then, \myeqln{G(x)=xH(x) = \frac{xf(x)}{1-F(x)}=\frac{x^{a}e^{-b x}}{\int^{\infty}_{x}t^{a-1}e^{-b t}dt}=\frac{1}{\int^{\infty}_{x}(t/x)^{a-1}x^{-1}e^{-b(t-x)}dt}.} It's straightforward that $\lim_{x\to0}G(x)=0$. For $\lim_{u\to\infty}G(u)\geq 1$, substituting $y=t-x\geq0$, we have  
\myeql{
\label{app_A_eq_gamma}G(x)=\frac{1}{\int^{\infty}_{0}e^{
-b y} (1+y/x)^{a-1}x^{-1} dy},}
hence $\lim_{x\to \infty} G(x)=\lim_{x\to\infty}\frac{x}{\int^{\infty}_{0}e^{-b y} dy}=\lim_{x\to\infty}b x>1$.

\noindent To show the (strict) monotonicity of $G(x)$, let $I(x,y)=(1+y/x)^{a-1}x^{-1}$. For any $y\geq0$,$\frac{\partial I(x,y)}{\partial x}=-\frac{(y/x + 1)^{a}(x + a y)}{x(x + y)^2}<0.$ Hence, for $x_1< x_2$ and $y\geq0$,  $I(x_1)>I(x_2)$ and by (\ref{app_A_eq_gamma}), \myeq{G(x_1)=\frac{1}{\int^{\infty}_{0}e^{-b y}I(x_1,y)dy}< \frac{1}{\int^{\infty}_{0}e^{-b y}I(x_2,y)dy}=G(x_2).}}

\subsubsection{The Weibull Distribution satisfies Assumption \ref{assumpt1}}
\hfill
\myproof{Consider the two-parameter Weibull distribution with shape parameter $k>0$, scale parameter $\lambda>0$, and pdf $f(x)=\frac{k}{\lambda}\left(\frac{x}{\lambda}\right)^{k-1}e^{-(x/\lambda)^k}$. Then, $G(x)=xH(x)$ is given by
\myeql{G(x)=\frac{xf(x)}{1-F(x)}=\frac{\frac{k}{\lambda}\left(\frac{x}{\lambda}\right)^{k-1}e^{-(x/\lambda)^k}}{e^{-(x/\lambda)^k}}=k\left(\frac{x}{\lambda}\right)^{k},\label{app_weibull}}
which implies that $\lim_{x\to0}G(x)=0$, $\lim_{x\to\infty}G(x)=\infty$, and for $k,x>0$, $G(x)$ is strictly increasing.
}

\subsubsection{The Beta Distribution satisfies Assumption \ref{assumpt1}}
\label{app_beta_gx}
\myproof{Consider the Beta distribution with positive parameters $a$ and $b$ and pdf  
\myeq{
f(x)=\frac{1}{B(a,b)}x^{a-1}(1-x)^{b-1},} 
where $B(a,b)=\frac{\Gamma(a)\Gamma(b)}{\Gamma(a+b)}$ and $\Gamma(\cdot)$ is the Gamma function. Proving that this distribution satisfies Assumption \ref{assumpt1} is more complex. We first show that for all $a,b>0$, $G(0)=0$ and $\lim_{x\to1}G(x)\geq1$. We then show that $G(x)$ is strictly increasing.

It's straightforward that $G(0)=0$ for all $a,b>0$. To show that $\lim_{x\to1}G(x)\geq1$, we consider the two cases \emph{(i)} $b\leq1$ and \emph{(ii)} $b>1$. 
\emph{(i)} When $b\leq 1$, $f(1)>0$ and as ${x\to1}$, $\frac{xf(x)}{1-F(x)}>1$ as it goes to infinity. \emph{(ii)} When $b>1$, 
\myeqmodeln{\lim_{x\to1}\frac{xf(x)}{1-F(x)}&=\lim_{x\to1}\frac{x^{a}(1-x)^{b-1}}{\int^{1}_{x}t^{a-1}(1-t)^{b-1}dt} =\lim_{x\to1}\frac{x^{a-1}(1-x)^{b-2}(a(1-x)+(1-b)x)}{-x^{a-1}(1-x)^{b-1}}\\
   & = \lim_{x\to1}\frac{(a+b)x-(a+x)}{1-x}=\lim_{x\to1}\frac{b-1}{1-x}>0.}
   
To show that $G(x)$ is increasing, we consider three cases: 
\begin{itemize}
    \item \textbf{Case 1}: $a\geq1$. It is well known that for $a\geq1$, the Beta distribution has an increasing failure rate, which implies that $G(x)$ is increasing. 
   
    \item \textbf{Case 2}: $a< 1$ and $b\leq1$. The derivative of $G(x)$ is 
    \myeq{G'(x) = \frac{(f(x)+xf'(x))(1-F(x))+xf(x)^2}{(1-F(x))^2}.}
    The sign of $G'(x)$ is the same as 
    \myeq{xf'(x)+f(x)=\frac{a(1-x)+(1-b)x}{B(a,b)x^{1-a}(1-x)^{2-b}}}.
    Since $b\leq1$, $xf'(x)+f(x)>0$ for all $x\in(0,1)$, hence $G'(x)>0$.
    
    \item \textbf{Case 3:} $a<1$ and $b>1$. Define the inverse hazard function $g(x)=\frac{1}{H(x)}=\frac{1-F(x)}{f(x)}$, which is bathtub-shaped in this case 
    (\cite{glaser1980bathtub}). Thus, there is a unique $x_0\in(0,1)$ such that
    $g'(x)>0$ for $x<x_0$, $g'(x_0)=0$ and $g'(x)<0$ for $x>x_0$. 
    Now, $G(x)=xH(x)$ is increasing if and only if its inverse $\frac{1}{G(x)}=\frac{1-F(x)}{f(x)x}=\frac{g(x)}{x}$ is decreasing, i.e., $\left[\frac{1}{G(x)}\right]'=\frac{g'(x)x-g(x)}{x^2}<0$, which holds if the numerator $\xi(x)=g'(x)x-g(x)$ is negative. We show this is the case by considering the two ranges: $x\in[x_0,1]$ and $x\in[0,x_0)$. 

    For $x\in[x_0,1]$, $g'(x)\leq 0$ and $g(x)>0$, hence $\xi(x)=xg'(x)-g(x)<0$. 
    
    For $x\in[0,x_0)$, define $\eta(x)=-\frac{f'(x)}{f(x)}=\frac{x(a+b-2)+(1-a)}{x-x^2},$ so
    \myeql{\label{beta_prove_g_x}g'(x)=g(x)\eta(x)-1.}
    Now,
    \myeq{
     \eta'(x)=\frac{a-1}{x^2}+\frac{b-1}{(1-x)^2} \ \textup{and}\ 
     \eta''(x)=2\left(\frac{1-a}{x^3}+\frac{b-1}{(1-x)^3}\right)
    .}  
    With $\eta''(x)>0$ for all $x\in(0, 1)$, $t_0=\frac{1}{1+\sqrt{(b-1)/(1-a)}}\in(0,1)$ is the unique global minimum of $\eta(x)$ satisfying $\eta'(x)=0$. Further, since $\eta(t_0)>0$, $\eta(x)$ is positive for all $x\in(0,1)$. 

    With these preliminaries, we now prove that $\xi(x)=xg'(x)-g(x)<0$ for $x\in(0,x_0)$. Rewrite $\xi(x)$ as
    \myeql{
    \xi(x)=xg'(x)-g(x)=xg'(x)-\frac{g'(x)+1}{\eta(x)}=\frac{g'(x)\left(\eta(x)x-1\right)-1}{\eta(x)},
    \label{beta_prove_g_x_numerator}}
    \noindent where $g(x)=\frac{g'(x)+1}{\eta(x)}$ by (\ref{beta_prove_g_x}). As $\eta(x)>0$, showing that $\xi(x)<0$ is equivalent to showing that the numerator $\gamma(x)$ of equation (\ref{beta_prove_g_x_numerator}), given by 
    \myeql{
    \gamma(x)=g'(x)(\eta(x)x-1)-1=g'(x)\left(\frac{(a+b-1)x-a}{1-x}\right)-1,
    \label{eq_eta1_positive}}
    is negative. To that end, we consider two ranges: \emph{(i)} $x\in(0,\min\{\frac{a}{a+b-1},x_0\}]$, and \emph{(ii)} $x\in(\min\{\frac{a}{a+b-1},x_0\}, x_0)$ (which may be empty). For \emph{(i)}, $\eta(x)x-1\leq 0$ and $g'(x)>0$, so $\gamma(x)$ is negative. For a nonempty \emph{(ii)}, substitute $g'(x)$ from (\ref{beta_prove_g_x}): 
    \myeql{\label{eq_beta_etax_1}
    \gamma(x)=\eta(x)(g(x)(\eta(x)x-1)-x)
    .} For $x>\frac{a}{a+b-1}$, $\eta(x)x-1=\frac{(a+b-1)x-a}{1-x}>0$. As $x_0$ is the maximizer of $g(x)$, \myeq{g(x)(\eta(x)x-1)<g(x_0)(\eta(x)x-1).} At $x_0$, $g'(x_0)=g(x_0)\eta(x_0)-1=0$, so $g(x_0)=\frac{1}{\eta(x_0)}$. From equation (\ref{eq_beta_etax_1}),
    \myeqln{
    \label{eq_beta_4}
    \gamma(x)<\eta(x)\left(g(x_0)(\eta(x)x-1)-x\right)=\eta(x)\left(\frac{x}{\eta(x_0)}[\eta(x)-\frac{1}{x}-\eta(x_0)]\right).
    }
    Let $\zeta(x)=\eta(x)-\frac{1}{x}$. For $a<1$ and $b>1$, 
    $\zeta(x)=\frac{(x(a+b-1)-a)}{x(1-x)}$ and $\zeta'(x)=\frac{a(1-x)^2+(b-1)x^2}{(x^2(1-x)^2)}>0,$ so $\zeta(x)$ is monotone increasing. Now, $\zeta(x)<\zeta(x_0)$ for $x\in(\frac{a}{a+b-1},x_0)$, hence $\xi(x)<\xi(x_0)<\eta(x_0)$ and
    \myeqln{\gamma(x)<\eta(x)\left(\frac{x}{\eta(x_0)}[\xi(x)-\eta(x_0)]\right)<\eta(x)\left(\frac{x}{\eta(x_0)}[\xi(x_0)-\eta(x_0)]\right)<0,\label{app_beta_mon_increasing}} 
     
\end{itemize}
\noindent\qquad  which completes the proof.
}

\subsection{Proof of Lemma \ref{lemma_increasing_control}}
\label{app_increasing_control}
\myproof{For any policy, let $t$ be the first time $y_t< y_{t-1}$, and let $x_t = \max_{ -1\leq i< t}  y_i=y_{t-1}$. We compare the expected NPV from period $t$ onwards for setting $y_{t}<y_{t-1}$ versus deviating to $\tilde{y}_t=y_{t-1}$. Under both policies, the state in period $t+1$ is $x_{t+1} =  \max_{ -1\leq i< t+1 } \ y_i= x_t$, so the future continuation value $J(x_t)$ is the same under both policies. Now, if the lender sets $y_t<x_t$, its expected NPV will be 
${-dy_t+ \rho \Prob(\theta\geq y_t|\theta\geq x_t)[y_t + J(x_t)] = (\rho-d)y_t+\rho J(x_t),}$
since $\Prob(\theta\geq y_t|\theta\geq x_t)=1$. 
The NPV with $\tilde{y}_t=y_{t-1}=x_t$ is 
${\left(\rho-d\right)x_t+\rho J(x_t).}$
The difference between the two NPV values 
is $\left(\rho-d\right)(x_t-y_t)>0$. Since this inequality holds for all $t$, it's never optimal to set $y_t<x_t$.}

\subsection{Proof of Lemma \ref{lemma_general_barx}}
\label{app_lemma_generalarx}

\myproof{
Define $T(x)=G(x)-\delta$ where $\delta=\frac{\rho-d}{\rho(1-d)}\in(0,1)$. By assumption \ref{assumpt1}, $T(0)<0$,  $\lim_{x\to u}T(x)>0$, and $T(\cdot)$ is  strictly increasing in the interval $[0,u)$. It follows from the continuity of $T(\cdot)$ that there exists a unique root $\bar{x}$ of $T(x)=0$ (which is equivalent to (\ref{eq_bar_x})). Since $G(\cdot)$ is increasing, its inverse $G^{-1}(\cdot)$ exists and is increasing. Now, $\bar{x} = G^{-1}(\delta)$, and the the proof follows from the fact that $\delta$ is increasing in $\rho$ and decreasing in $d$. 
}

\subsection{Proof of Theorem \ref{them_explore_exploit_1}}
\label{proof_them_explore_exploit_1}

\myproof{We first present the following Theorem, which generalizes Theorem \ref{them_explore_exploit_1} by assuming a general $x_0\in(0,u)$, and Theorem \ref{them_explore_exploit_1} is a special case with $x_0=0$, so it is sufficient to prove Theorem \ref{them_explore_exploit} : 

\myth{\label{them_explore_exploit} \textit{: There exists a unique optimal policy  $\BFy^*$ which depends on $x_0$ and $\bar{x}$ as follows:} 
 
\begin{itemize}
\item[] \textit{ (a) If $x_0<\bar{x}$, 
the lender keeps increasing the repayment amounts $y_t$ in each period $t$ as long as the consumer repays her loans, and stops lending if the consumer defaults. The infinite sequence of (potential) repayment amounts $\{y_t, t=0,1,2,\ldots\}$ is strictly increasing to the limit \myeq{\bar{x}}.  } 
\item[] \textit{(b) If $x_0\geq \bar{x}$, the optimal policy is to set the same repayment amount $y_t=x_0$ in each period. }
\end{itemize}
}

The proof is structured as follows. We first establish the existence and uniqueness of the solution to the Bellman equation (Lemma \ref{lemma_existence_uniquenss}). Next, Lemma \ref{cor_optimal_policy_structure} helps us establish that the optimal repayment amount must follow one of three structures: (1) constant amount, (2) strictly increasing amounts, or (3) initially strictly increasing followed by a constant amount. We then build the proof for the two parts of the Theorem, (a) $x_0 < \bar{x}$ and (b) $x_0 \geq \bar{x}$. 

\mylemma{
\label{lemma_existence_uniquenss}
There exists a unique solution to the Bellman equation (\ref{eq_J_xt}). 
}
\mylemmaproof{First, we can bound the repayment amounts $y_t$ for all $t$ since in any period $t$, there is an upper bound on $y_t$ above which the expected payoff becomes negative (in fact, as {$y_t\to\infty$}, the expected payoff $\lim_{y_t\to\infty}x_t-dy_t\to-\infty$). Thus, for some $M>0$ we can optimize (\ref{eq_J_xt}) within $y_t\in[0,M)$ for all $t$. With bounded controls and discount factor $\rho\in(0, 1)$, there exists a unique solution to the Bellman equation (see, e.g. \cite{bertsekas2012dynamic}). }

\noindent Building on the monotonicity result of Lemma \ref{lemma_increasing_control}, we next show that if the optimal control satisfies $ y_{t} = y_{t-1} = x_{t}$, then the optimal policy sets $y_{k} = x_{k} $ for all subsequent periods.

\begin{lemma}{\label{cor_optimal_policy_structure} For the optimal policy $\{\BFy^*_{t},t=1,2,\dots\}$, if for any $t$, $y^*_{t}=x_{t}$, then for all $k>t$, also $y^*_{k}=x_{t}$ .
}
\end{lemma}

\mylemmaproof{ If $y^*_{t}=x_t$, then $x_{t}$ maximizes the following Bellman equation:
\myeql{J(x_t) \ = \ \max_{y_t\in[x_t,u)} \ \ { x_t-dy_t+\rho\dfrac{1-F(y_t)}{1-F(x_t)}J(y_t) }.\label{eq_28}
}
By Lemma \ref{lemma_increasing_control}, the law of motion evolves as $x_{t+1}=y_{t}=x_{t}$, so the Bellman equation for period $t+1$ is the same as for period $t$: 
\myeql{J(x_{t+1})=J(x_t) \ = \ \max_{y_{t+1}\in[x_t,u)} \ \ { x_t-dy_{t+1}+\rho\dfrac{1-F(y_{t+1})}{1-F(x_t)}J(y_{t+1}) \label{eq_29}}.
}
As $x_{t}$ maximizes (\ref{eq_28}), it also maximizes (\ref{eq_29}), so $y^*_{t+1}=y^*_{t}=x_{t}$. The Lemma now follows by forward induction.}

It follows that there can only be three possible structures for the optimal policy: (1) $\{y_t,\ t\geq0\}$ are constant; (2)  $\{y_t,\ t\geq0\}$ are strictly increasing; or (3) $\{y_t,\ t\geq0\}$ first strictly increase and then become constant. We now prove Theorem \ref{them_explore_exploit} (a) ($x_0<\bar{x}$) by contradiction, showing (Lemma~\ref{lemma_constant_structure3_nopt}) that structures (1) and (3) are not optimal, so the optimal policy must follow (2). We complete the proof by showing that the repayment amounts go to the limit \(\bar{x}\).


\mylemma{\label{lemma_constant_structure3_nopt}
When $x_0<\bar{x}$, both a policy with constant repayment amounts and a policy with strictly increasing repayment amounts followed by constant repayment amounts are sub-optimal.}

\mylemmaproof{The proof is by contradiction.  
Under both structures (1) and (3), there exists a period $\tau$ where the policy stabilizes with \(y_{\tau} = x_{\tau}\). For structure (1), this may start at $\tau=0$ (when $y_0=x_0$) or at $\tau=1$ (when $y_0\neq x_0$ and $y_1=y_0=x_1$). For structure (3), $\tau>1$. For each of these cases, we construct an alternative policy with a strictly higher NPV. 

\emph{(i)} $\tau=0$. Here, the optimal policy has a constant repayment amount which is equal to $x_0$. The lender's NPV under this policy is $\Pi_1(x_0) = \frac{1-d}{1-\rho} x_0$. Consider an alternative policy with $y_t = y$ for all $t \geq 0$. The lender's NPV under the alternative policy is 
\begin{equation*}
\Pi_2(y|x_0) = x_0 - dy + \rho \frac{1-F(y)}{1-F(x_0)} \left(\frac{1-d}{1-\rho} y\right),
\end{equation*}
where $\Pi_1(x_0) = \Pi_2(x_0|x_0)$. The derivative of $\Pi_2(y|x_0)$ with respect to $y$ is given by
\begin{equation*}
\frac{d \Pi_2(y|x_0)}{dy} = d \frac{1-F(y)}{1-F(x_0)} \left(\frac{\rho (1-d)}{d(1-\rho)} - \frac{1-F(x_0)}{1-F(y)} - \frac{\rho (1-d)}{d(1-\rho)} G(y)\right).
\end{equation*}
Since $x_0 < \bar{x}$, evaluating this derivative at $y = x_0$ gives
\begin{equation*}
\frac{d \Pi_2(y|x_0)}{dy} \bigg|_{y = x_0, x_0 < \bar{x}} = \frac{\rho (1-d)}{d(1-\rho)} - 1 - \frac{\rho (1-d)}{d(1-\rho)} G(x_0) > 0,
\end{equation*}
so the assumed optimal policy is strictly improved, a contradiction.

\emph{(ii)} $\tau>0$ (this obviously covers both $\tau=1$ for policy structure (1) and $\tau>1$ for policy structure (3)). We distinguish between two cases: $x_{\tau}<\bar{x}$ and $x_{\tau}\geq\bar{x}$.  

\textbf{Case 1: $x_{\tau}<\bar{x}$}. The alternative policy directly generalizes the one for $\tau=0$ in that the repayment amount changes to $y$ starting from period $\tau$ (rather than $0$): $y_t=y$ for all $t\geq \tau$. We compare the lender's NPV from period $\tau$ onwards under the original policy, $\Pi_1(x_\tau)$, to that under the alternative policy, $\Pi_2(y|x_{\tau})$. Since in this case $x_{\tau}<\bar{x}$, the preceding proof (for $\tau=0$) carries over to this case starting from $\tau$, with the derivative of $\Pi_2(y|x_\tau)$ with respect to $y$ being positive, which shows that the alternative policy strictly improves on the assumed optimal policy, a contradiction.

\textbf{Case 2. } $x_{\tau}\geq\bar{x}$. Under the assumed optimal policy, $x_{\tau-1} < y_{\tau-1} = x_\tau = y_\tau =x_{\tau+1}= y_{\tau+1} = \dots$. We introduce an alternative policy with repayment amount $y$ for period \(\tau-1\) onwards and show that setting $y<y_{\tau-1}$ strictly improve the lender's NPV. Using similar notation to Case 1 but looking at period \(\tau-1\) and all subsequent periods, the lender's NPV with a repayment amount $y$ is: 
\myeql{\Pi_2(y|x_{\tau-1})= x_{\tau-1}- dy + \rho \frac{1-F(y)}{1-F(x_\tau-1)} \left(\frac{1-d}{1-\rho} y\right)
\label{eq_pi2},
}
and its derivative with respect to $y$, evaluated at $y\geq\bar{x}$ is 
\begin{equation*}
\frac{d \Pi_2(y|x_{\tau-1})}{dy}\bigg|_{y\geq\bar{x},y>x_{\tau-1}}  = d \frac{1-F(y)}{1-F(x_{\tau-1})} \left(\frac{\rho (1-d)}{d(1-\rho)} - \frac{\rho (1-d)}{d(1-\rho)} G(y) - \frac{1-F(x_{\tau-1})}{1-F(y)}\right)<0,
\end{equation*}
which contradicts the optimality of the assumed optimal policy.
}

Lemmas \ref{cor_optimal_policy_structure} and \ref{lemma_constant_structure3_nopt} together imply that when $x_0<\bar{x}$, the optimal policy $\{y_t, t\geq0\}$ is strictly increasing. Since the optimal $\{y_t, t\geq0\}$ are bounded (by Lemma \ref{lemma_existence_uniquenss}) and increasing, they have a finite limit. We prove by contradiction that that limit is $\bar{x}$. Suppose that limit is $\hat{x}\neq\bar{x}$. We distinguish between the two cases: \emph{(1)} $\hat{x} < \bar{x}$, and \emph{(2)} $\hat{x} > \bar{x}$. 

\textbf{Case 1: } $\hat{x}<\bar{x}$.  Let $\Pi_t$ denote the lender's NPV from period $t$ onwards. Given the limit $\hat{x}$, for any $\epsilon>0$, there exists an integer $M(\epsilon)$ such that 
\myeql{\hat{x}-x_{t}<\epsilon \quad\textup{and}\quad \hat{x}-y_{t}<\epsilon \quad\textup{for all}\  t\geq M(\epsilon)\label{eq_hat_x},}
and \myeql{|\Pi_{M(\epsilon)}-\frac{1-d}{1-\rho}\hat{x}|<\epsilon\label{eq_pi_hatx},} where $\frac{1-d}{1-\rho}\hat{x}$ is the lender's NPV from period $M(\epsilon)$ onwards if we set $y_{t}=\hat{x}$ for all $t\geq M(\epsilon)$. 

Now consider an alternative policy that sets $y_t=y'$ for $t\geq M(\epsilon)$. The lender's NPV from period $M(\epsilon)$ onwards under this policy is given by $\Pi_2(y'|x_{M(\epsilon)})$ given by (\ref{eq_pi2}), and $\Pi_2(x_{M(\epsilon)}|x_{M(\epsilon)})=\frac{1-d}{1-\rho}x_{M(\epsilon)}$. Moreover, evaluated at $y'=x_{M(\epsilon)}$, $\frac{d \Pi_2(y'|x_{M(\epsilon)}) }{d y}>0$, so there exist $\delta,\xi>0$, such that for $y'=x_{M(\epsilon)}+\delta$, $ \Pi_2(y'|x_{M(\epsilon)})-\frac{1-d}{1-\rho}x_{M(\epsilon)}>\xi$. Let $\epsilon=\min\{\frac{1}{3}\delta,\frac{1}{2}\frac{1-\rho}{1-d}\xi\}$. Then:

(1) $y'=x_{M(\epsilon)}+\delta>x_{M(\epsilon)}+2\epsilon>\hat{x}+\epsilon$, where the first inequality follows since $\epsilon\leq \frac{1}{3}\delta$, and the second inequality follows from (\ref{eq_hat_x}).

(2) $\Pi_2(y'|x_{M(\epsilon)})>\frac{1-d}{1-\rho}x_{M(\epsilon)}+\xi > \frac{1-d}{1-\rho}(\hat{x}-\epsilon)+\xi\geq \frac{1-d}{1-\rho}\hat{x}+\frac{1}{2}\xi>\Pi_{M(\epsilon)}-\epsilon + \frac{1}{2}\xi \geq \Pi_{M(\epsilon)}$, where the second inequality follow from equation (\ref{eq_hat_x}), the third inequality follows from $\epsilon\leq \frac{1}{2}\frac{1-\rho}{1-d}\xi$, the fourth inequality follows from (\ref{eq_pi_hatx}), and the last inequality follows from $\epsilon\leq\frac{1}{2}\frac{1-\rho}{1-d}\xi<\frac{1}{2}\xi$, as $\rho>d$. This is a contradiction as the alternative policy improves the lender's NPV.

\textbf{Case 2: } $\hat{x}>\bar{x}$. We first introduce the following lemma, which shows that whenever $x_t\geq\bar{x}$, the optimal policy must be constant from period $t$ onwards.

\mylemma{
\label{lemma_xtgeq_barx}
\textit{If $x_t\geq \bar{x}$ for some $t\geq0$, the optimal policy must set $y_{k}=x_t$ for all $k\geq t$. } 
}

\mylemmaproof{ We directly verify that the proposed policy satisfies the Bellman equation, where 
\myeql{\label{bellman_eq_36}J(x_t)=\frac{1-d}{1-\rho}x_t=\max_{y_t\in [x_t,u)} \ x_{t}-dy_{t}+\rho\frac{1-F(y_t)}{1-F(x_t)}J(y_t).}
Substituting of $J(y_t)=\frac{1-d}{1-\rho}y_t$ into equation (\ref{bellman_eq_36}), we obtain the maximization problem
\myeql{\max_{y_t\in [x_t,u)} \ x_{t}-dy_{t}+\rho\frac{1-F(y_t)}{1-F(x_t)} \frac{1-d}{1-\rho}y_t.\label{bellman_eq_37}}
The derivative of the objective function in (\ref{bellman_eq_37}) with respect to $y_t$ is $ d \frac{1-F(y)}{1-F(x_t)} \left(\frac{\rho (1-d)}{d(1-\rho)} - \frac{1-F(x_t)}{1-F(y)} - \frac{\rho (1-d)}{d(1-\rho)} G(y)\right)$, which is negative for any $x_t\geq\bar{x}$ and $y_t>x_t$. 
It follows that $y_{t}=x_t$ solves the Bellman equation, and equation (\ref{bellman_eq_36}) holds. The optimal policy is thus to set $y_k=x_t$ for all $k\geq t$, which yields $J(x_t)=\frac{1-d}{1-\rho}x_t$.}

Note that Lemma \ref{lemma_xtgeq_barx} holds in general and need not assume $x_0<\bar{x}$. However, we are assuming here that $x_0<\bar{x}$ and by Lemma \ref{lemma_constant_structure3_nopt}, a series of constant $y_t$ is not optimal, a contradiction. 

We have thus proven that the limit of the repayment amounts must be $\bar{x}$, which completes the proof of Theorem \ref{them_explore_exploit} (a). Lemma \ref{lemma_xtgeq_barx} in fact proves Theorem  \ref{them_explore_exploit} (b) by setting $t=0$: It shows that when $x_0\geq\bar{x}$, the optimal policy is to set $y_k=x_0$ for all $k\geq$0.  This completes the proof of Theorem \ref{them_explore_exploit}.  }

\subsection{Proof of Theorem \ref{them_unif}}
\label{app_unif_proof}

\myproof{
In the uniform case, the Bellman equation becomes 
\myeqln{J(x_t) = \max_{y_t} \ x_t-d y_t+\rho\dfrac{1-y_t}{1-x_t}J(y_t).}
We'll show that the proposed linear control satisfies the Bellman equation. Given the existence and uniqueness of the optimal solution (Lemma \ref{lemma_existence_uniquenss}), this will prove the Theorem.

With a linear control, $J(x)$ is of the form\footnote{For notational simplicity, we suppress the dependence on the time period.}

\myeqln{J(x)=
\begin{cases}
x + \dfrac{ax^2+bx+c}{1-x},& \textup{if  }x<\bar{x},\\
\dfrac{(1-d)}{(1-\rho)}x, &\textup{if  } x\geq \bar{x},
\end{cases}
}
where $\bar{x}$ is the unique solution to (\ref{eq_bar_x}), which is reduced to  
\myeq{\bar{x}=\frac{(\rho-d)}{2\rho - \rho d - d}
}
in the uniform case. We can now use the following conditions to solve for \myeq{\{a,b,c\}} (all for $x\in(0,\bar{x})$ ): 

\noindent \textbf{Value Matching}:
\myeql{J(\bar{x})=\bar{x}+\dfrac{a\bar{x}^2+b\bar{x}+c}{1-\bar{x}}=\dfrac{(1-d)}{(1-\rho)}\bar{x}
\label{c1};}
\textbf{Smoothness}:
\myeql{J'(\bar{x})=\dfrac{a\bar{x}^2+b\bar{x}+c}{(1-\bar{x})^2}+\dfrac{2a\bar{x}+b}{1-\bar{x}}+1=\dfrac{(1-d)}{(1-\rho)};
\label{c2}}
\textbf{Linear Control}:
\begin{align}
\begin{split}
\label{c3}
\dfrac{\partial J(x)}{\partial y}=&\dfrac{(x+( \rho (1+b)/d-1)) - 2\rho((1-a)/d)y }{(1 - x)/d}=0,\\
y=&mx + n,
\quad m=  \dfrac{d}{2\rho (1-a)},\quad n= \dfrac{(\rho - d + b\rho)}{2\rho(1-a)},\\
\bar{x}=&m\bar{x}+n;
\end{split}
\end{align}
\textbf{Consistency of Bellman equation at $x=0$}:
\myeql{J(0)=c=-dn+\rho\left(1-n\right)\left(n+\dfrac{an^2+bn+c}{1-n}\right).\label{c4}}
With some algebra, we find two sets \myeq{\{a_1,b_1,c_1\}} and \myeq{\{a_2,b_2,c_2\}} that satisfy (\ref{c1})-(\ref{c4}): 
\myeqmodeln{a_1= & \dfrac{1}{2}\left(1+d\sqrt{\dfrac{\rho-d^2}{\rho d^2}}\right), \\
b_1=& \dfrac{1}{2\rho - \rho d - d}\left((\rho -d)\left(-d\sqrt{\dfrac{\rho-d^2}{\rho d^2}}+d-1\right)\right),\\
c_1=& \dfrac{1}{2(1-\rho)(2\rho -d -\rho d)^2}\left((\rho-d)^2\left(1-2d+\rho+(1-\rho)\sqrt{1-\dfrac{d^2}{\rho}} \right) \right);  \\
a_2 = & \dfrac{1}{2}\left(1-d\sqrt{\dfrac{\rho-d^2}{\rho d^2}}\right),\\
b_2=&  \dfrac{1}{2\rho - \rho d - d}\left((\rho -d)\left(d\sqrt{\dfrac{\rho-d^2}{\rho d^2}}+d-1\right)\right),\\
c_2=& \dfrac{1}{2(1-\rho)(2\rho -d -\rho d)^2}\left((\rho-d)^2\left(1-2d+\rho-(1-\rho)\sqrt{1-\dfrac{d^2}{\rho}} \right) \right).
}
By the linear control requirement, $(1-m)\bar{x} = n$, and since $\bar{x}$ and $n$ are positive, we must have $m<1$. The first solution does not satisfy this requirement since $\rho>d$ implies $m(a_1)=\dfrac{d}{2\rho (1 - a_1 )}>1$.

We complete the proof by showing that $\{a_2,b_2,c_2\}$ satisfy the Bellman equation. For $x\in(0,\bar{x})$, \myeq{J(x_t)} should satisfy 
\myeql{
J(x_t)= x_t+\dfrac{ax_t^2+bx_t+c}{1-x_t}=  \max_{y_t\in[x_t,1)} x_t-dy_t+\rho\dfrac{1-y_t}{1-x_t}J(y_t) = \max_{y_t\in[x_t,1)} x_t-dy_t+\rho\dfrac{1-y_t}{1-x_t}\left(y_t+\dfrac{ay_t^2+by_t+c}{1-y_t}\right).
\label{app_correct_J}}
Plugging in $\{a_2,b_2,c_2\}$, equation (\ref{app_correct_J}) holds.}

\subsection{Proof of Proposition \ref{prop_mon_k}}
\label{app_lemma_mon_k}

\myproof{We first prove that if in  period $k^*$, $\Pi_{k^*}\geq0$, then $\Pi_{k+1}> \Pi_k\geq 0$ for all $k\geq k^*$. This implies that unprofitable consumers default in periods $k= 1,2,\ldots, k^*-1$, profitable consumers default in periods $k \ge k^*$, and creditworthy consumers never default; and that there exists a cutoff value $\theta^*(\rho,d)  = y_{k^*}$ which separates unprofitable from profitable consumers. 

As $\Pi_{k}=\sum^{k-1}_{i=0}\rho^{i}(\rho - d)y_i-\rho^{k}dy_{k}$, $\Pi_{k+1}-\Pi_k=y_k-dy_{k+1}$ and $y_k = mx_k+n$ for the uniform case. With some algebra, $\frac{1-d}{d}>\frac{m^{k+1}}{\sum^{k}_{i=0}m^i}$ implies that $y_k-dy_{k+1}>0$, which further implies $\Pi_{k+1}>\Pi_k$.  As \myeq{\frac{m^{k+1}}{\sum^{k}_{i=0}m^i}=\frac{1-m}{m^{-(k+1)}-1}} is decreasing in $k$, if $\frac{1-d}{d}>\frac{1-m}{m^{-k^*}-1}$ in period $k^*$, then $\frac{1-d}{d}>\frac{1-m}{m^{-(k+1)}-1}$ and therefore $\Pi_{k+1}>\Pi_k$ for all $k\geq k^*$.

We next prove that the cash flow received in period $k^*$, $y_{k^*}-dy_{k^*+1}$, is positive. As $dm<1$, $y_{k-1}-d y_{k}<y_{k}-d y_{k+1}$, so the cash flow is monotone increasing in the period index $k$. Let $\tilde{k}>0$ be the first period in which the lender receives a positive cash flow, i.e., $y_{\tilde{k}-1}-dy_{\tilde{k}}>0$. Since the sum of the discounted cash flows turns positive in period $k^*$, we have $k^*\geq \tilde{k}$. Hence, if a type-$k$ consumer is profitable, all type$-k'$ consumers with $k'\geq k$ are also profitable. 

Since $k^*$ is the smallest $k'$ such that $\Pi_{k'}\geq0$, $\theta^*(\rho,d)=y_{k^*}$. Consumers with $\theta\in[0,\theta^*(\rho,d))$ are unprofitable, and those with $\theta\in[\theta^*(\rho,d),\theta^{\infty}(\rho,d))$ are profitable. Consumers with $\theta \in[\theta^{\infty}(\rho,d),1)$ are creditworthy with $\theta^{\infty}(\rho,d)=\bar{x}$.

To complete the proof, we show that the three segments are nonempty. The unprofitable customers are from the interval $\theta\in[0,y_0)$, which is nonempty since $y_0=n>0$.
Also, since there always exists a $k^*$ such that $\Pi_{k^*}\geq0$ and $y_k^*<\bar{x}$, the set of profitable but non-creditworthy consumers is also nonempty. Finally, by Lemma \ref{lemma_general_barx}, $\bar{x}<u$, so the set of creditworthy consumers is notempty.}

\subsection{Proof of Lemma \ref{prop:dstar}}
\label{proof_dstar}

\myproof{
We rearrange (\ref{eq_d_star_dr_1}) as
$(\rho-d)s'(d)+\rho s(d)^2-s(d)=0$. The left hand side of this equation is monotone decreasing (with the derivative $(\rho-d)s''(d)+2s'(d)(\rho s(d)-1)<0$);   $\lim_{d\to 0}(\rho-d)s'(d)+\rho s(d)^2-s(d)>0$; and $\lim_{d\to \rho}(\rho-d)s'(d)+\rho s(d)^2-s(d)<0$. It follows that (\ref{eq_d_star_dr_1}) has a unique solution. }

\subsection{Proof of Proposition~\ref{prop:grand}}
\myproof{ We prove Proposition~\ref{prop:grand}(a) by contradiction. Proposition~\ref{prop:grand} (a) gives the first order conditions (FOCs) for the  continuously differentiable objective function, with $d_0,d^*\in[0,\rho]$ and $y_0\in[0,y^*]$, $y^*\in[y_0,u]$ (following Lemma~\ref{lemma_increasing_control_dr_1}). Suppose an optimal policy $(d'_0,y'_0;d^{*'},y^{*'})$ fails the FOCs.

At any interior maximizer the KKT conditions reduce to the FOCs, so if the FOCs fail then at least one component lies on the boundary of its feasible set. When $d=0$ or $d=\rho$ (the boundaries for $d$), the objective function value is not strictly positive, hence neither $d'_0$ nor $d^{*'}$ achieve a maximum. The remaining boundary possibilities are for the $y$s. However, both $y=0$ and $y=u$ yield negative objective function values, so they are not optimal either. The only remaining candidate is a flat policy with $y_0=y^*$.

By Lemma~A.5, under decreasing demand elasticity any flat policy $y_0=y^*$ is strictly dominated by a deviation with $y_0<y^*$, hence it cannot be optimal. This contradicts optimality of $(d'_0,y'_0;d^{*'},y^{*'})$. It follows that an optimal policy must be interior and satisfy the FOCs of Proposition~\ref{prop:grand} (a).

Theorems \ref{thm_constant_elas_A} and \ref{thm_increasing_elas_A} generalize Proposition \ref{prop:grand}(b) (since the two-stage problem is a constrained version of the general problem), and their proofs in Appendices \ref{app_thm_de} and \ref{app_thm_ie} follow the same steps as the proof of Proposition \ref{prop:grand}(b). Rather than present these steps here and then repeat them in more general form later, we avoid repetition and refer to Appendices \ref{app_thm_de} and \ref{app_thm_ie} for the proof of Proposition \ref{prop:grand}(b).
}

\subsection{Proof of Lemma \ref{lemma_increasing_control_dr_1}}
\myproof{Similar to the proof of Lemma \ref{lemma_increasing_control}, if the lender sets $y_t<x_t$ with discount rate $d_t$, its expected NPV will be ${-s(d_t)d_ty_t + \rho s(d_t)\Prob(\theta\geq y_t|\theta\geq x_t)[y_t+J(x_t)]
 = s(d_t)((\rho-d_t)y_t+\rho J(x_t)),}$
since $\Prob(\theta\geq y_t|\theta\geq x_t)=1$.
The NPV with $\tilde{y}_t=y_{t-1}=x_t$ under the same discount rate $d_t$ is 
${s(d_t)\left(\left(\rho-d_t\right)x_t+\rho J(x_t)\right)}$. The difference between the two NPV values is $s(d_t)\left(\rho - d_t\right)(x_t-y_t)>0$.  Since this inequality holds for all $t$, it's never optimal to set $y_t<x_t$.}

\subsection{Proof of Lemma \ref{lemma_general_barx_dr_1}}

\myproof{As $\rho>d$, $-\frac{(1-\rho s(d^*))}{(1-s(d^*)d^*)}\frac{d^*}{\rho}+1\in(0,1)$ for any $d^*\in(0,\rho)$, so  there is a unique $\bar{x}$ solving equation (\ref{eq_bar_x_dr_1}). 
}

\subsection{Proof of Theorem \ref{them_explore_exploit_dr_1}}
\label{app_thm_de}

\myproof{We prove Theorem \ref{them_explore_exploit_dr_1A} below; Theorem \ref{them_explore_exploit_dr_1} is a special case with $x_0=0$. 

\myth{
\label{them_explore_exploit_dr_1A}
\textit{Under monotone decreasing demand elasticity, there exists a unique optimal policy $(\BFy^*,\BFd^*)$ with the following structure:}
\begin{itemize}
\item[] \textit{(a) If $x_0<\bar{x}$, 
the lender keeps increasing the repayment amounts $y_t$ in each period $t$ as long as the consumer repays her loans, and stops lending if the consumer defaults. The infinite sequence of (potential) repayment amounts $\{y_t, t=0,1,2,\ldots\}$ is strictly increasing to a limit $\bar{x}$ defined by  equation (\ref{eq_bar_x_dr_1}). }
\item[] \textit{(b) If $x_0\geq \bar{x}$, the optimal policy is to set the repayment amount at $y_t=x_0$ in each period with a fixed discount rate $d^*$.}
\end{itemize}
}

The proof is similar to the proof of Theorem \ref{them_explore_exploit} (Appendix \ref{proof_them_explore_exploit_1}). We first establish the existence and uniqueness of the solution to the Bellman equation as in Lemma \ref{lemma_existence_uniquenss}: since both controls are bounded and the discount factor $\rho\in(0,1)$, there exists a unique solution to the Bellman equation. We next show that if the repayment amounts $y_t$ satisfy $ y_{t} = y_{t-1} = x_{t}$, then the optimal policy sets $y_{k} = x_{k}$ for all subsequent periods with $d_k=d^*$. If $y^*_{t}=x_t$, then $x_{t}$ maximizes the Bellman equation in period $t$:
\myeq{J(x_t) \ = \ \max_{y_t\in[x_t,u),d_t\in(0,\rho)} \ \ { x_t-s(d_t) d_t y_t+\rho s(d_t) \frac{1-F(y_t)}{1-F(x_t)}J(y_t) }.} Now, the Bellman equation for period $t+1$ is the same as for period $t$. Hence, $y=x_t$ also maximizes the Bellman equation for period $t+1$, so $y^*_{t+1}=y^*_{t}=x_{t}$. Further, when the repayment amount stabilizes at $x_t$, the lender's expected NPV is given by $\frac{1-s(d_t)d_t}{1-s(d_t)\rho}x_t$, with $d^*$ being the unique maximizer. The result now follows by forward induction.

It follows that the optimal repayment amounts have to follow one of the same three structures ((1)-(3)) we identified in Appendix \ref{proof_them_explore_exploit_1}. The following Lemma, analogous to Lemma \ref{lemma_constant_structure3_nopt} in Appendix \ref{proof_them_explore_exploit_1}, shows that the optimal repayment amounts must be strictly increasing. We then show that the limit of the repayment amounts must be $\bar{x}$.

\mylemma{\label{lemma_constant_structure3_nopt_de}
When $x_0<\bar{x}$, under an endogenous interest rate and decreasing demand elasticity, both a policy with constant repayment amounts and a policy with strictly increasing repayment amounts followed by constant repayment amounts are sub-optimal.
}

\mylemmaproof{The proof is by contradiction. Under both structures, there exists a period $\tau$ where the policy stabilizes with $y_{\tau}=x_{\tau}$. We construct alternative policies with strictly higher NPVs for $\tau=0$ and $\tau>0$.

\emph{(i)} $\tau=0$. Here, the optimal policy has a constant repayment amount $y_t=x_0$, with $d_t=d^*$, and the lender's NPV is 
\myeq{\Pi_1(d^*,x_0) =\frac{1-s(d^*)d^*}{1-s(d^*)\rho}x_0 = x_0-s(d^*)d^*x+\rho s(d^*)\frac{1-s(d^*)d^*}{1-s(d^*)\rho}x_0}.
We now consider an alternative policy which sets the repayment amounts at $y$ for all $t \geq 0$ with  the same discount rate $d^*$. The lender's NPV under the alternative policy is 
\myeqln{\Pi_2(d^*,y|x_0) =x_0-s(d^*)d^*y +\rho s(d^*)\frac{1-F(y)}{1-F(x_t)}\frac{1-s(d^*)d^*}{1-s(d^*)\rho}y ,}
where $\Pi_1(d^*,x_0) = \Pi_2(d^*,x_0|x_0)$. The derivative of $\Pi_2(d^*,y|x_0)$ with respect to $y$ is given by
\myeqln{ \frac{\partial \Pi_2(d^*,y|x_0) }{\partial y}=-s(d^*)d^*+\frac{\rho s(d^*)(1-s(d^*)d^*)}{1-s(d^*)\rho}\left(\frac{1-F(y)}{1-F(x_t)}-\frac{f(y)}{1-F(x_t)}y\right). }
If $x_0<\bar{x}$, evaluating this derivative at $y = x_0$ gives, 
\myeqln{\frac{d \Pi_2(d^*,y|x_0)}{dy} \bigg|_{y = x_0, x_0 < \bar{x}}=\frac{\rho s(d^*)(1-s(d^*)d^*)}{1-s(d^*)\rho}\left(1-\frac{d^*(1-s(d^*)\rho)}{\rho (1-s(d^*)d^*)}-G(x)\right)>0.}
This implies that when $x_0 < \bar{x}$, the assumed optimal policy is strictly improved by deviating to a larger constant value $y>x_0$ for all $t \geq 0$ with the same discount rate $d^*$, a contradiction.

\emph{(ii)} $\tau>0$. We distinguish between two cases : $x_{\tau}<\bar{x}$ and $x_{\tau}\geq\bar{x}$.

\textbf{Case 1. }  $x_{\tau}<\bar{x}$. Here, the alternative policy directly generalizes the one for $\tau=0$ in that the repayment amount changes to $y$ starting from period $\tau$ rather than $0$, and the above proof follows through.

\textbf{Case 2. } $x_{\tau}\geq\bar{x}$. Looking at period $\tau-1$ and all subsequent periods, under the assumed policy where $x_{\tau-1} < y_{\tau-1} = x_\tau = y_\tau =x_{\tau+1}= y_{\tau+1} = \dots$, the lender's NPV with repayment amount $y_{\tau-1}=y$ and discount rate $d_{\tau}$ is 
\myeql{\Pi_2(d_{\tau-1},y|x_{\tau-1})= x_{\tau-1}- s(d_{\tau-1})d_{\tau-1}y_{\tau-1} + \rho s(d_{\tau-1})\frac{1-F(y)}{1-F(x_{\tau-1})} \left(\frac{1-s(d^*)d^*}{1-s(d^*)\rho} y\right)\label{eq_pi2_d}.}
The discount rate $d_{\tau-1}$ that maximizes the lender's NPV satisfies
\myeql{\frac{1-F(x_{\tau-1})}{1-F(y)}=\frac{s(d^*)+s'(d^*)d^*}{\rho s'(d^*)} \frac{\rho s'(d_2)}{ s(d_{\tau-1})+s'(d_{\tau-1})d_{\tau-1}}.\label{proof_d2_B11}}
Further, $d_{\tau-1}<d^*$. To see this, let $\eta(d)=\frac{s(d)+s'(d)d}{\rho s'(d)}$. Equation (\ref{proof_d2_B11}) becomes $\eta(d_{\tau-1})=\frac{1-F(x_{\tau-1}+\epsilon)}{1-F(x_{\tau-1})}\eta(d^*)$. Moreover, $\eta(d)$ is monotone increasing as $\eta'(d)=\frac{2\rho s'(d)-\rho s''(d)s(d)}{(\rho s'(d))^2}>0$. Hence $d_{\tau-1}<d^*$. 

We now take the derivative of $\Pi_2$ with respect to $y$. By the evenlope theorem, $\frac{d \Pi_2(d_{\tau-1},y|x_{\tau-1})}{d y}=\frac{\partial \Pi_2(d_{\tau-1},y|x_{\tau-1})}{\partial y}+\frac{\partial \Pi_2(d_{\tau-1},y|x_{\tau-1})}{\partial d_{\tau-1}}\frac{\partial d_{\tau-1}}{\partial y} $, and $\frac{\partial \Pi_2(d_{\tau-1},y|x_{\tau-1})}{\partial d_{\tau-1}}=0$. The derivative $\frac{d \Pi_2(d_{\tau-1},y|x_{\tau-1})}{d y}$ is thus given by
\myeql{\label{eq_derivative_d2}
\frac{d \Pi_2(d_{\tau-1},y|x_{\tau-1})}{d y} = \rho s(d_{\tau-1})\frac{1-F(y)}{1-F(x_{\tau-1})}\frac{1-s(d^*)d^*}{1-s(d^*)\rho}\left[1-\frac{1-F(x_{\tau-1})}{1-F(y)}\frac{1-s(d^*)\rho}{1-s(d^*)d^*}\frac{d_{\tau-1}}{\rho}-G(y)\right].
}
Plugging (\ref{proof_d2_B11}) into the derivative (\ref{eq_derivative_d2}), we have 
\myeqln{
\frac{d \Pi_2(d_{\tau-1},y|x_{\tau-1})}{d y} = \rho s(d_2)\frac{1-F(y)}{1-F(x_{\tau-1})}\frac{1-s(d^*)d^*}{1-s(d^*)\rho}\left[1-\left(\frac{s(d^*)+s'(d^*)d^*}{s'(d^*)}\frac{s'(d_{\tau-1})}{s(d_{\tau-1})+s'(d_{\tau-1})d_{\tau-1}}\frac{d_{\tau-1}}{d^*}\right)\frac{1-s(d^*)\rho}{1-s(d^*)d}\frac{d^*}{\rho}-G(y)\right].
}
However, evaluating at $y\geq\bar{x}$, 
\myeql{\frac{d \Pi_2(d_{\tau-1},y|x_{\tau-1})}{d y}  \bigg|_{y \geq \bar{x} } <0. \label{eq_dv_neg_d2}}
This is because as the elasticity is decreasing and $d_{\tau-1}<d^*$,
\myeq{\frac{s(d^*)+s'(d^*)d^*}{s'(d^*)}\frac{s'(d_{\tau-1})}{s(d_{\tau-1})+s'(d_{\tau-1})d_{\tau-1}}\frac{d_{\tau-1}}{d^*}>1}. Following the definition of $G(\bar{x})$ (equation (\ref{eq_d_star_dr_1}), Lemma \ref{lemma_general_barx_dr_1}), equation (\ref{eq_dv_neg_d2}) holds. This implies that the assumed optimal policy can be strictly improved (by decreasing the repayment amount $y_{\tau-1}$), a contradiction.
}

The previous results imply that when $x_0<\bar{x}$, the optimal repayment amounts $y_t$ are strictly increasing. We next show that the limit of the repayment amount sequence is $\bar{x}$. Suppose by contradiction that the limit is $\hat{x}\neq\bar{x}$. We then distinguish between two cases, $\hat{x}<\bar{x}$ and $\hat{x}>\bar{x}$.

\textbf{Case 1: $\hat{x}<\bar{x}$}. The proof follows that of Appendix \ref{proof_them_explore_exploit_1}, replacing $|\Pi_{M(\epsilon)}-\frac{1-d}{1-\rho}\hat{x}|<\epsilon|$ by $|\Pi_{M(\epsilon)}-\frac{1-s(d^*)d^*}{1-s(d^*)\rho}\hat{x}|<\epsilon|$, where $\frac{1-s(d^*)d^*}{1-s(d^*)\rho}\hat{x}$ is the lender's NPV from period $M(\epsilon)$ onwards with $y_{t}=\hat{x}$ and $d_t=d^*$. Further, we replace $\Pi_2(y'|x_{M(\epsilon)})$ with $\Pi_2(d^*, y'|x_{M(\epsilon)})$ (equation (\ref{eq_pi2_d}), which is the counterpart of equation (\ref{eq_pi2}) in Appendix \ref{proof_them_explore_exploit_1}).

\textbf{Case 2: $\hat{x}\geq\bar{x}$}. We first introduce the following Lemma, which states that whenever $x_t\geq\bar{x}$, the optimal policy becomes constant with repayment amount and discount rate $(x_t,d^*)$ from period $t$ onwards.

\mylemma{
\label{lemma_xtgeq_barx_de}
If $x_t\geq \bar{x}$ for some $t\geq0$, then the optimal policy is to set $(y_{k}, d_k)=(x_t,d^*)$ for all $k\geq t$. 
}
\mylemmaproof{ We directly verify the optimality of the Bellman equation given by  
\myeql{\label{bellman_eq_36_de}J(x_t)=\frac{1-s(d^*)d^*}{1-s(d^*)\rho}x_t=\max_{y_t\in [x_t,u),d_t\in(0,\rho)} \ x_{t}-s(d)d y_{t}+\rho s(d)\frac{1-F(y_t)}{1-F(x_t)}J(y_t).}
Substituting $J(y_t)=\frac{1-s(d^*)d^*}{1-s(d^*)\rho}y_t$, the right hand side of (\ref{bellman_eq_36_de}) becomes
\myeql{\max_{y_t\in [x_t,u), d_{t}\in(0,\rho)} \ x_{t}-ds(d)y_{t}+\rho s(d)\frac{1-F(y_t)}{1-F(x_t)} \frac{1-s(d^*)d^*}{1-s(d^*)\rho}y_t.\label{max_eq_d2_xtgeqbarx}}
The optimal $d_t$ under any $y_t$ should satisfy (\ref{proof_d2_B11}) (by changing $d_{\tau-1}$ in equation (\ref{proof_d2_B11}) to $d_t$). By (\ref{eq_dv_neg_d2}), $\frac{d \Pi_2(d_t,y|x_t)}{d y}  \bigg|_{y > x_t \geq  \bar{x}}<0$. 
Any $y>x_t$ is not an optimal solution of the problem (\ref{max_eq_d2_xtgeqbarx}), hence $y_{t}=x_t$ and $d_t=d^*$ is the unique pair that solves the Bellman equation when $x_t\geq\bar{x}$, and equation (\ref{bellman_eq_36_de}) holds. The optimal policy is to set $y_k=x_t$, $d_k=d^*$ for all $k\geq t$, and under this policy, $J(x_t)=\frac{1-s(d^*)d^*}{1-s(d^*)\rho}x_t$.
}

Lemma \ref{lemma_xtgeq_barx_de} shows that under our assumptions, the repayment amount becomes constant, but  Lemma \ref{lemma_constant_structure3_nopt_de} shows that when $x_0<\bar{x}$, this solution cannot be optimal, a contradiction. 

We have thus proved that the limit of the repayment amounts is $\bar{x}$, which completes the proof of Theorem \ref{them_explore_exploit_dr_1A} (a). Lemma \ref{lemma_xtgeq_barx_de} also proves Theorem \ref{them_explore_exploit_dr_1A} (b) by setting $t=0$. This completes the proof of Theorem \ref{them_explore_exploit_dr_1A}, hence also Theorem \ref{them_explore_exploit_dr_1}. }

\subsection{Proof of Theorem \ref{thm_constant_elas}}
\label{app_thm_cons}
\myproof{Theorem \ref{thm_constant_elas} is a special case of Theorem \ref{thm_constant_elas_A} below.
\myth{
\label{thm_constant_elas_A} \textit{Under constant demand elasticity, there exists a unique solution to the Bellman equation}
\myeql{
\label{eq_Jx_const}
J(x_t)=\begin{cases}
x_t-(d_{x_t})^{\alpha+1}\bar{x}+\rho (d_{x_t})^{\alpha}\dfrac{1-F(\bar{x})}{1-F(x_t)}\dfrac{1-(d^{*})^{\alpha+1}}{1-(d^{*})^{\alpha}\rho} \bar{x}, \quad x_t<\bar{x},\\
\dfrac{1-(d^{*})^{\alpha+1}}{1-(d^{*})^{\alpha}\rho} x_t, \quad x_t\geq\bar{x},
\end{cases}}
where 
\myeql{\label{eq_barx_const}d_{x_t}=\frac{1-F(\bar{x})}{1-F(x_t)}d^* and \quad \bar{x}=G^{-1}\left(\frac{1}{\alpha+1}\right).}
The optimal policy $(\BFy^*,\BFd^*)=(\{y_0,d_0\},\{y_1,d_1\},\dots,\{y_t,d_t\})$ is given by
\begin{itemize}
    \item[]  \textit{(a) if $x_0<\bar{x}$, the lender sets the repayment amount at $y_0=\bar{x}$ with the corresponding discount rate $d_0=\frac{1-F(\bar{x})}{1-F(x_0)}d^*$. If the consumer repays $y_0$, the lender fixes the repayment amount $y_t=\bar{x}$ and discount rate $d_t=d^*$ for all future periods.}
    \item[] \textit{ (b) if $x_0\geq \bar{x}$, the optimal policy is to set $y_t=x_0$ in each period with a fixed discount rate $d^*$.}
\end{itemize}
}

We prove Theorem \ref{thm_constant_elas_A} by directly verifying that the proposed solution satisfies the Bellman equation. For notational simplicity, we omit here the dependence on $t$ and just write $x$ instead of $x_t$. The results hold for all $t$.
We also define $\beta = \frac{1-(d^{*})^{\alpha+1}}{1-(d^{*})^{\alpha}\rho}$, so equation (\ref{eq_d_star_dr_1}) becomes $\beta =\frac{1-(d^*)^{\alpha+1}}{1-(d^*)^{\alpha}\rho}=\frac{1+\alpha }{\rho \alpha }d^*$. 

We consider separately $x<\bar{x}$ and $x\geq\bar{x}$.

\textbf{Case 1: $x<\bar{x}$}. We verify that the following equation holds,
\myeqmodeln{
J(x)|_{x<\bar{x}}&= x-(d_x)^{\alpha+1}\bar{x}+\rho (d_x)^{\alpha}\frac{1-F(\bar{x})}{1-F(x)}\frac{1-(d^{*})^{\alpha+1}}{1-(d^{*})^{\alpha}\rho} \bar{x}\\
&=\max\left\{
\max_{d, y\in[x,\bar{x})} \ x-d^{\alpha+1}y+\rho d^{\alpha}\frac{1-F(y)}{1-F(x)}\left(y-d^{\alpha+1}_y\bar{x}+\rho d^{\alpha}_y\frac{1-F(\bar{x})}{1-F(y)}\beta \bar{x}\right) , \max_{d, y\in[\bar{x},u)}\  x-d^{\alpha+1}y+\rho d^{\alpha} \frac{1-F(y)}{1-F(x)}\beta y
\right \},
}
where
\myeqln{d_x = \frac{1-F(\bar{x})}{1-F(x)}d^* \quad\textup{and }\quad d_y=\frac{1-F(\bar{x})}{1-F(y)}d^*.}
\noindent This involves two cases: evaluating $\max_{d, y\in[x,\bar{x})} \ x-d^{\alpha+1}y+\rho d^{\alpha}\frac{1-F(y)}{1-F(x)}\left(y-d^{\alpha+1}_y\bar{x}+\rho d^{\alpha}_y\frac{1-F(\bar{x})}{1-F(y)}\beta \bar{x}\right)$ for $y\in[x,\bar{x})$ (Case 1.1 below), and evaluating $\max_{d, y\in[\bar{x},u)}\  x-d^{\alpha+1}y+\rho d^{\alpha} \frac{1-F(y)}{1-F(x)}\beta y$ for $y\in[\bar{x},u)$ (Case 1.2 below). 

\textbf{Case 1.1: $y\in[x,\bar{x})$.} We prove that
\myeql{
\max_{d, y\in[x,\bar{x})} \ x-d^{\alpha+1}y+\rho d^{\alpha}\frac{1-F(y)}{1-F(x)}\left(y-d^{\alpha+1}_y\bar{x}+\rho d^{\alpha}_y\frac{1-F(\bar{x})}{1-F(y)}\beta \bar{x}\right)=x-d_x^{\alpha+1}\bar{x}+\rho d_x^{\alpha}\frac{1-F(\bar{x})}{1-F(x)}\frac{1-(d^{*})^{\alpha+1}}{1-(d^{*})^{\alpha}\rho} \bar{x}.
\label{prop_1_pi1}
}
Consider the optimization problem
\myeqln{\max_{d, y\in[x,\bar{x})} \Pi_1(d, y) \ := x-d^{\alpha+1}y+\rho d^{\alpha}\frac{1-F(y)}{1-F(x)}\left(y-d^{\alpha+1}_y\bar{x}+\rho d^{\alpha}_y\frac{1-F(\bar{x})}{1-F(y)}\beta \bar{x}\right).}
We first optimize over $d$. The optimal $d$ is always in the interior of $(0,\rho)$ and it satisfies the first order condition
\myeq{-(\alpha+1)d^{\alpha}y+\rho \alpha d^{\alpha-1}\frac{1-F(y)}{1-F(x)}\left(y-d^{\alpha+1}_y\bar{x}+\rho d^{\alpha}_y\frac{1-F(\bar{x})}{1-F(y)}\beta \bar{x}\right) = 0 .}
With some algebra,
\myeql{d=\frac{\rho \alpha}{\alpha+1}\frac{1-F(y)}{1-F(x)}\left(y-d^{\alpha+1}_y\bar{x}+\rho d^{\alpha}_y\frac{1-F(\bar{x})}{1-F(y)}\beta \bar{x}\right)\frac{1}{y}=\frac{\rho\alpha}{\alpha+1}\frac{1-F(y)}{1-F(x)}\frac{C(y)}{y},
\label{eq_d_pi1}}
where
\myeqln{C(y)=\left(y-d^{\alpha+1}_y\bar{x}+\rho d^{\alpha}_y\frac{1-F(\bar{x})}{1-F(y)}\beta \bar{x}\right).}

We next optimize over $y$, taking the derivative of  
$\Pi_1$ with respect to $y$:  
\myeq{\frac{d\Pi_1}{d y}=\frac{\partial \Pi_1}{\partial y}+\frac{\partial \Pi_1}{\partial d }\frac{\partial d}{\partial y}.}
By the envelope theorem, $\frac{\partial \Pi_1}{\partial d }=0$, and
\begin{align}\frac{d\Pi_1}{d y}=\frac{\partial \Pi_1}{\partial y}&=-d^{\alpha+1}+\rho d^{\alpha}\frac{1-F(y)}{1-F(x)}\frac{d C(y)}{d y}-\rho d^{\alpha}\frac{f(y)}{1-F(x)}C(y).
\label{eq_73}
\end{align}
Substituting (\ref{eq_d_pi1}) into (\ref{eq_73}), we obtain
\myeqln{\frac{d\Pi_1}{d y}=
\frac{\rho d^{\alpha}}{y}\frac{1-F(y)}{1-F(x)}\left(-\frac{\alpha}{\alpha+1}C(y)+ y\frac{dC(y)}{dy}-G(y)C(y)\right).
}
By the definition of $d_y$,
$\frac{\partial C(y)}{\partial d_y}=0$, which implies,
\myeqln{\frac{d C(y)}{d y}=\frac{\partial C(y)}{\partial y}+\frac{\partial C(y)}{\partial d_y}\frac{\partial d_y}{\partial y}=\frac{\partial C(y)}{\partial y}= 1+\rho d_y^{\alpha}\frac{1-F(\bar{x})}{1-F(y)}\frac{f(y)}{1-F(y)}\beta\bar{x}.}
Further, 
\begin{align*}\frac{d\Pi_1}{d y}&=
\frac{d^{\alpha}\rho}{y}\frac{1-F(y)}{1-F(x)}\left(-\frac{\alpha}{\alpha+1}C(y)+ y\left(1+\rho d_y^{\alpha}\frac{1-F(\bar{x})}{1-F(y)}\frac{f(y)}{1-F(y)}\beta\bar{x}\right)-G(y)C(y)\right)\\
& = \frac{d^{\alpha}\rho}{y}\frac{1-F(y)}{1-F(x)}\left(-\frac{\alpha}{\alpha+1}C(y)+ y + G(y)\left(\rho d_y^{\alpha}\frac{1-F(\bar{x})}{1-F(y)}\beta\bar{x}\right)-G(y)C(y)\right)\\
&=\frac{d^{\alpha}\rho}{y}\frac{1-F(y)}{1-F(x)}\left(-\left(\frac{\alpha}{\alpha+1}+G(y)\right)\left(y-d^{\alpha+1}_y\bar{x}\right)+y-\frac{\alpha}{\alpha+1}\left(\rho d_y^{\alpha}\frac{1-F(\bar{x})}{1-F(y)}\beta\bar{x}\right)\right).
\end{align*}
By definition, $d_y=\frac{1-F(\bar{x})}{1-F(y)}d^*=\frac{\alpha\rho}{\alpha+1}\frac{1-F(\bar{x})}{1-F(y)}\beta$, and
\begin{align*}\frac{d\Pi_1}{d y} &=\frac{d^{\alpha}\rho}{y}\frac{1-F(y)}{1-F(x)}\left(-\left(\frac{\alpha}{\alpha+1}+G(y)\right)\left(y-d^{\alpha+1}_y\bar{x}\right)+y-d^{\alpha+1}_y \bar{x}\right)\\
&=\frac{d^{\alpha}\rho}{y}\frac{1-F(y)}{1-F(x)}\left(1-\left(\frac{\alpha}{\alpha+1}+G(y)\right)\right)\left(y-d^{\alpha+1}_y\bar{x}\right).
\end{align*}
Since $y<\bar{x}$, \myeq{G(y)<-\frac{\alpha}{\alpha+1}+1,}
hence $\left(1-\left(\frac{\alpha}{\alpha+1}+G(y)\right)\right)>0$.

Let $q(y)=y-d^{\alpha+1}_y\bar{x}$, then
\myeqln{\frac{d\Pi_1}{d y}=\frac{d^{\alpha}\rho}{y}\frac{1-F(y)}{1-F(x)}\left(1-\left(\frac{\alpha}{\alpha+1}+G(y)\right)\right)q(y)}
for $y\in[x,\bar{x})$. At $y=\bar{x}$, $q(y)|_{y=\bar{x}}=y-d^{\alpha+1}_y\bar{x}=(1-d^{\alpha+1}_{\bar{x}})\bar{x}>0$. Moreover, $q'(y)= 1-d^{\alpha}_y\alpha \rho\frac{1-F(\bar{x})}{1-F(y)}\beta \bar{x}$ is monotone decreasing in $y$. Together with the fact that $q(0)<0$, $q(\bar{x})>0$ and $q'(y)$ is monotone decreasing, $q(y)$ crosses the $x-$axis only once for $x\in[0,\bar{x})$. This implies the single-crossing property for $\frac{d\Pi_1}{d y}(y)$: there exists a $y'\in(0,\bar{x})$ such that for $y<y'$,  $\frac{d\Pi_1}{d y}<0$, and for $y>y'$,  $\frac{d\Pi_1}{d y}>0$

We further prove that $\frac{d\Pi_1}{dy}|_{y=x}>0$ for all $x<\bar{x}$:
\myeqln{\frac{d\Pi_1}{dy}\bigg|_{y=x}=(d^*)^{\alpha}\rho\beta\left(-\frac{\alpha}{\alpha+1}+1- G(x)\right)>0, \quad \textup{for }x<\bar{x}.}

Finally, by the single-crossing property of $\frac{d\Pi_1}{dy}$, $\frac{d\Pi_1}{d y}>0$ for $y\in(x,\bar{x}]$. This implies that $y=\bar{x}$ optimizes $\Pi_1$, and equation (\ref{prop_1_pi1}) holds.

\textbf{Case 1.2: $y\in[\bar{x},u)$}. We prove that 
\myeql{
\max_{d, y\in[\bar{x},u)}\  x-d^{\alpha+1}y+\rho d^{\alpha} \frac{1-F(y)}{1-F(x)}\beta y =x-d_x^{\alpha+1}\bar{x}+\rho d_x^{\alpha}\frac{1-F(\bar{x})}{1-F(x)}\beta \bar{x}.
\label{prop_1_pi2}
}
Consider the optimization problem 
\myeql{\max_{d, y\in[\bar{x},u)}\ \Pi_2(d,y) := x-d^{\alpha+1}y+\rho d^{\alpha} \frac{1-F(y)}{1-F(x)}\beta y .}
We first optimize over $d$, obtaining \myeq{d=\frac{\alpha\rho}{\alpha+1}\frac{1-F(y)}{1-F(x)}\beta=\frac{1-F(y)}{1-F(x)}d^*.}

We next optimize over $y$. The  derivative of $\Pi_2$ with respect to $y$ is \myeq{\frac{d\Pi_2}{d y}=\frac{\partial \Pi_2}{\partial y}+\frac{\partial \Pi_2}{\partial d }\frac{\partial d}{\partial y}.} By the envelope theorem, $\frac{\partial \Pi_2}{\partial d }=0$, and
\myeqln{\frac{d\Pi_2}{d y}=\frac{\partial \Pi_2}{\partial y}=d^{\alpha}\left(-d+\rho \frac{1-F(y)}{1-F(x)}\beta-\rho\beta \frac{f(y)}{1-F(x)} y\right)=d^{\alpha}\rho\beta\frac{1-F(y)}{1-F(x)}\left(-\frac{\alpha}{\alpha+1}+1- G(y)\right).
}
Since $G(y)> 1-\frac{d^*}{\rho\beta}=1-\frac{\alpha}{\alpha+1}$ for all $y>\bar{x}$ and $G(y)=1-\frac{\alpha}{\alpha+1}$ for $y=\bar{x}$, it follows that  
\myeq{\frac{d\Pi_2}{d y}|_{y>\bar{x}}<0} and \myeq{\frac{d\Pi_2}{d y}|_{y=\bar{x}}=0.}
Hence, equation (\ref{prop_1_pi2}) holds, and for $x<\bar{x}$,
\begin{align*}J(x)|_{x<\bar{x}}=&x-d_x^{\alpha+1}\bar{x}+\rho d_x^{\alpha}\frac{1-F(\bar{x})}{1-F(x)}\frac{1-(d^{*})^{\alpha+1}}{1-(d^{*})^{\alpha}\rho} \bar{x}\\
=&\max\left\{
\max_{d, y\in(x,\bar{x})} \ x-d^{\alpha+1}y+\rho d^{\alpha}\frac{1-F(y)}{1-F(x)}\left(y-d^{\alpha+1}_y\bar{x}+\rho d^{\alpha}_y\frac{1-F(\bar{x})}{1-F(y)}\beta \bar{x}\right) ,\quad 
\max_{d, y\in[\bar{x},u)}\  x-d^{\alpha+1}y+\rho d^{\alpha} \frac{1-F(y)}{1-F(x)}\beta y
\right\}
\end{align*}
with corresponding 
\myeq{
d_x= \frac{1-F(\bar{x})}{1-F(x)}d^*,\quad y_x=\bar{x}.
}

\textbf{Case 2: $x\geq\bar{x}$.}\\
We verify that the following equation holds:
\myeql{\label{eq_xgeqbarx_proof}
J(x)|_{x\geq \bar{x}}= \beta x =
\max_{d, y\in[x,u)}\  x-d^{\alpha+1}y+\rho d^{\alpha} \frac{1-F(y)}{1-F(x)}\beta y.
}
The proof is similar to that of Case 1.2. Consider the optimization problem 
\myeql{\max_{d, y\in[x,u)}\Pi_2(d,y) = x-d^{\alpha+1}y+\rho d^{\alpha} \frac{1-F(y)}{1-F(x)}\beta y .}
The optimal $d=\frac{1-F(y)}{1-F(x)}d^*$, and the derivative of $\Pi_2$ with respect to $y$ is  
\myeql{\frac{d\Pi_2}{d y}=d^{\alpha}\rho\beta\frac{1-F(y)}{1-F(x)}\left(-\frac{\alpha}{\alpha+1}+1- G(y)\right)
.}
For $y>x>\bar{x}$, $\frac{d\Pi_2}{d y}|_{y>\bar{x}}<0$. Hence, the optimal $y$ is attained at $y=x$, with $d_x=d^*$, and equation (\ref{eq_xgeqbarx_proof}) holds. 

We have thus verified the optimality of the proposed solution for Cases 1 and 2, which completes the proof.}

\subsection{Proof of Proposition \ref{prop_d_increase_in_alpha}}
\label{app_d_increase_in_alpha}

\proof{ We first prove that $d^*$ increases in $\alpha$. From the implicit function theorem, 
\[\dfrac{\partial d^*}{\partial \alpha}=\frac{\rho (d^*)^{\alpha+1}log(d^*)+\rho-d^*}{(\alpha+1)(1-\rho (d^*)^{\alpha})}.\]
In order to show that $\frac{\partial d^*}{\partial \alpha}>0$, it's sufficient to show that 
$\rho (d^*)^{\alpha+1}log(d^*)+\rho-d^*>0$. By the definition of $d^*$, $\rho=\dfrac{\alpha+1}{\alpha + (d^*)^{\alpha+1}}d^*,$
hence $ \rho (d^*)^{\alpha+1}log(d^*)+\rho-d^* = \dfrac{d^*}{\alpha+(d^*)^{\alpha}}\left(1-(d^*)^{\alpha+1}(1-log(d^*)(\alpha+1))\right),$
so it's sufficient to show that $1-(d^*)^{\alpha+1}(1-log(d^*)(\alpha+1))>0$.

Let $\delta(d)=d^{\alpha+1}(1-log(d)(\alpha+1))$, with the derivative $\delta'(d)=-(\alpha+1)^2d^{\alpha}log(d)>0$.
Since $d\in(0,\rho)$, $\delta(1)=1$ and $1-(d^*)^{\alpha+1}(1-log(d^*)(\alpha+1))=\delta(1)-\delta(d^*)>0$, hence, $\frac{\partial d^*}{\partial \alpha}>0$.

Finally, equation (\ref{eq_barx_const}) shows that $\bar{x}$ decreases in $\alpha$. 
}

\subsection{Proof of Theorem \ref{thm_increasing_elas}}
\label{app_thm_ie}
\myproof{As before, we state and prove Theorem \ref{thm_increasing_elas} for the generalized $x_0$. 

\myth{\label{thm_increasing_elas_A} \textit{When both the demand elasticity and the hazard rate of the consumer's income distribution are increasing, there exists a unique optimal policy  $(\BFy^*,\BFd^*)=((y_0, d_0),(y_1, d_1),\dots)$ satisfying:}
\begin{itemize}
    \item[]  (a) if $x_0<\bar{x}$, in period $0$ the lender sets the repayment amount and discount rate $(y_0,d_0)$ that jointly solve the equations
    \myeqln{G(y_0)+\frac{1-F(x_0)}{1-F(y_0)}\frac{d_0}{\rho}\frac{1-s(d^*)\rho}{1-s(d^*)d^*}=1,\quad \frac{1-F(y_0)}{1-F(x_0)}\frac{1-s(d^*)d^*}{1-s(d^*)\rho}=\frac{s(d_0)+s'(d_0)d_0}{\rho s'(d_0)},}
    which have a unique solution with $d_0\in(0,d^*)$. If the consumer repays $y_0$, the lender fixes the repayment amount at $y_t=y_0$ and the discount rate at $d_t=d^*$ for all subsequent periods.
    \item[]  (b) if $x_0\geq \bar{x}$, the optimal policy is to set the repayment amount in each period at $y_t=x_0$ with a fixed discount rate $d^*$.
\end{itemize}
} 

As in the preceding proof, we omit the dependence on the time period $t$. We define $\beta = \frac{1-s(d^*)d^*}{1-s(d^*)\rho}$. We have $G(\bar{x})=1-\frac{d^*}{\rho\beta}$.  

Along the lines of the proof of Theorem \ref{thm_constant_elas_A}, our proof proceeds by showing that the proposed solution satisfies the Bellman equations for $x<\bar{x}$ and $x\geq\bar{x}$ separately. We evaluate these equations for different states, and the following Lemma addresses the existence and uniqueness of solutions to first order conditions of the form 
\myeql{\label{eq_1_ie}G(\Upsilon)+\frac{1-F(X)}{1-F(\Upsilon)}\frac{\delta}{\rho}\frac{1-s(d^*)\rho}{1-s(d^*)d^*}=1,}
and 
\myeql{\label{eq_2_ie}\frac{1-F(\Upsilon)}{1-F(X)}\frac{1-s(d^*)d^*}{1-s(d^*)\rho}=\frac{s(\delta)+s'(\delta)\delta}{\rho s'(\delta.}}

\begin{lemma}\label{lemma_existence_equation_sol}
Let $X\in(0,\bar{x})$. There exists a unique $(\Upsilon,\delta)$ satisfying equations (\ref{eq_1_ie})-(\ref{eq_2_ie}) for $\delta\in(0,d^*]$.
\end{lemma}

\mylemmaproof{Define $\eta(\delta)=\delta+\frac{s(\delta)}{s'(\delta)}$. With some algebra, we can rewrite equations (\ref{eq_1_ie})-(\ref{eq_2_ie}) with $\Upsilon$ as functions of $\delta$: $\Upsilon=G^{-1}\left(1-\frac{\delta}{\eta(\delta)}\right)$ and $\Upsilon=F^{-1}\left(1-\frac{1-F(x_0)}{\rho\beta}\eta(\delta)\right)$, and  
it is sufficient to show that these functions intersect at a single $\delta$. Defining $M(\delta)=G^{-1}\left(1-\frac{\delta}{\eta(\delta)}\right)-F^{-1}\left(1-\frac{1-F(x_0)}{\rho\beta}\eta(\delta)\right)$, we thus need to prove that for $\delta\in(0,d^*]$, $M(\delta)$ crosses the $\delta$-axis exactly once. Since the function $M(\cdot)$ is differentiable, it is sufficient to show that \emph{(i)} $\lim_{\delta\to0}M(\delta)<0$; \emph{(ii)} $\lim_{\delta\to d^*}M(\delta)>0$; and \emph{(iii)} at any point $\delta\in(0,d^*]$ where $M(\delta)=0$, $M'(\delta)>0$. 

\textbf{Proof of \emph{(i)}}: As $\delta\to0$, $\lim_{\delta\to0}M(\delta)=G^{-1}(1)-\lim_{\delta\to0}F^{-1}\left(1-\frac{1-F(x_0)}{\rho\beta}\eta(\delta)\right)$. Since $G(\cdot)$ is monotone increasing with $\lim_{x\to u}G(x)\to\infty$, there exists a $\hat{y}<u$ such that $G(\hat{y})=1$, or $\hat{y}=G^{-1}(1)$. For the second term, $\lim_{\delta\to0}F^{-1}\left(1-\frac{1-F(x_0)}{\rho\beta}\eta(\delta)\right)=u$, hence $\lim_{\delta\to0}M(\delta)=\hat{y}-u<0$.

\textbf{Proof of \emph{(ii)}}: $\lim_{\delta\to d^*}M(\delta)=G^{-1}(1-\frac{d^*}{\eta(d^*)})-F^{-1}\left(1-\frac{1-F(x_0)}{\rho\beta}\eta(d^*)\right)=G^{-1}(1-\frac{d^*}{\eta(d^*)})-x_0=\bar{x}-x_0 > 0$. The fact that the first expression is reduced to $\bar{x}$ follows from the definition of $\bar{x}$.
As for the second expression, when $\delta=d^*$, let $\Upsilon=F^{-1}\left(1-\frac{1-F(x_0)}{\rho\beta}\eta(d^*)\right)$. This implies $\left(1-F(\Upsilon)\right)\rho\beta=\left(1-F(x_0)\right)\eta(d^*)$, and by the definition of $d^*$,  $\eta(d^*)=\rho\beta$, hence $\Upsilon=F^{-1}\left(1-\frac{1-F(x_0)}{\rho\beta}\eta(d^*)\right)=x_0$. 

\textbf{Proof of \emph{(iii)}}: Let $l(\delta)=\frac{\delta}{\eta(\delta)}$, then 
\myeql{\label{eq_Mpd}M'(\delta)=-\frac{l'(\delta)}{G'\left(G^{-1}\left(1-\frac{\delta}{\eta(\delta)}\right)\right)}+\frac{\frac{1-F(x_0)}{\rho\beta}\eta'(\delta)}{f\left(F^{-1}\left(1-\frac{1-F(x_0)}{\rho\beta}\eta(\delta)\right)\right)}.}
If $M(\delta)=0$, then $G^{-1}\left(1-\frac{\delta}{\eta(\delta)}\right)=F^{-1}\left(1-\frac{1-F(x_0)}{\rho\beta}\eta(\delta)\right)$ and the corresponding $\Upsilon$ is given by $\Upsilon=G^{-1}\left(1-\frac{\delta}{\eta(\delta)}\right)=F^{-1}\left(1-\frac{1-F(x_0)}{\rho\beta}\eta(\delta)\right)$. With that, (\ref{eq_Mpd}) is reduced to 
\myeq{
M'(\delta)=-\frac{l'(\delta)}{G'(\Upsilon)}+\frac{\frac{1-F(x_0)}{\rho\beta}\eta'(\delta)}{f(\Upsilon)}
}. As $G'(\Upsilon)=H(\Upsilon)+H'(\Upsilon)\Upsilon$ where $H(\cdot)$ is the monotone increasing hazard function, $G'(\Upsilon)>H(\Upsilon)$ and
\myeqln{
M'(\delta)>-\frac{l'(\delta)}{H(\Upsilon)}+\frac{\frac{1-F(x_0)}{\rho\beta}\eta'(\delta)}{f(\Upsilon)}=\frac{1-F(\Upsilon)}{f(\Upsilon)}\left(\frac{1-F(x_0)}{1-F(\Upsilon)}\rho\beta\eta'(\delta)-l'(\delta)\right).}

As $\Upsilon=F^{-1}(1-\frac{1-F(x_0)}{\rho\beta}\eta(\delta))$, $\frac{1-F(x_0)}{1-F(\Upsilon)}\rho\beta=\eta(\delta)$, and 
$M'(\delta)>\frac{1-F(\Upsilon)}{f(\Upsilon)}\left(\eta(\delta)\eta'(\delta)-l'(\delta)\right)$. With some algebra, $\eta(\delta)\eta'(\delta)-l'(\delta)=\frac{\eta^3(\delta)\eta'(\delta)-\eta(\delta)+\delta\eta'(\delta)}{\eta^2(\delta)}$. Since $\eta'(\delta)=2-\frac{s''(\delta)}{[s'(\delta)]^2}>2$, $\eta(\delta)\eta'(\delta)-l'(\delta)>\frac{2\eta^{3}(\delta)+2\delta-\eta(\delta)}{\eta^2(\delta)}=\frac{2\delta^3(\xi(\delta)+1)^3+\delta(\xi(\delta)-1)\xi^2(\delta)}{\eta^2(\delta)\xi^3(\delta)}$, where $\xi(\delta)$ is the demand elasticity. Since $\xi(\delta)>0$ and $\delta\in(0,1)$, $2\delta^3(\xi(\delta)+1)^3+\delta(\xi(\delta)-1)\xi^2(\delta)>0$. Hence, $M(\delta)=0$ implies $M'(\delta)>0$.}

Proceeding with the proof of Theorem \ref{thm_increasing_elas_A}, we verify the Bellman equation for $x<\bar{x}$ (Case 1) and $x\geq\bar{x}$ (Case 2).

\textbf{Case 1: $x<\bar{x}$}. We define ($y_x,d_x)$ as the solution to (\ref{eq_1_ie})-(\ref{eq_2_ie}) with $\Upsilon=y_x,\delta=d_x$ and $X=x$. Lemma \ref{lemma_existence_equation_sol} guarantees the existence and uniqueness of the solution. We verify that the following equation holds: 
\begin{align}J(x)|_{x<\bar{x}}=&x-s(d_x)d_x y_x+\rho s(d_{x})\frac{1-F(y_x)}{1-F(x)}\frac{1-s(d^*)d^*}{1-s(d^*)\rho}y_x\\
=&\max\{\max_{d,y\in[x,\bar{x})}x-s(d) d y +\rho s(d)\frac{1-F(y)}{1-F(x)}\left(y-s(d_y)d_y y_y+\rho s(d_y)\frac{1-F(y_y)}{1-F(y)}
\beta y_y \right), \\
&\qquad \ \max_{d,y\in[\bar{x},u)} x-s(d)dy+\rho s(d)\frac{1-F(y)}{1-F(x)}\beta y \}\label{eq_ie_case1},
\end{align}
Here, ($y_y,d_y)$ is the solution to (\ref{eq_1_ie})-(\ref{eq_2_ie}) with $\Upsilon=y_y,\delta=d_y$ and $X=y$ (where $\Upsilon=y_y$ and $\delta=d_y$ satisfy the first order conditions for maximizing $\Pi_y(\Upsilon,\delta)=\Upsilon-s(\delta)\delta \Upsilon+\rho s(\delta)\frac{1-F(\Upsilon)}{1-F(y)}\beta\Upsilon$, with the current state being $y$). We further define $C(y)=\Pi_y(y_y,d_y)=\left(y-s(d_y)d_y y_y+\rho s(d_y)\frac{1-F(y_y)}{1-F(y)} \beta y_y \right)$.
The existence and uniqueness of $(y_y,d_y)$ follow from Lemma \ref{lemma_existence_equation_sol}. 

The future continuation value in the Bellman equation depends on the range that $y$ is in. We consider two cases: $y\in[x,\bar{x})$ (Case 1.1) and $y\in[\bar{x},u)$ (Case 1.2). 

\textbf{ Case 1.1: $y\in[x,\bar{x})$}. We prove that 
\myeqln{ \max_{d,y\in[x,\bar{x})}x-s(d) d y +\rho s(d)\frac{1-F(y)}{1-F(x)}C(y) \leq x-s(d_x)d_x y_x+\rho s(d_x)\frac{1-F(y_x)}{1-F(x)}\beta y_x.}

Consider the optimization problem 
\myeql{\max_{d,y\in[x,\bar{x})} \Pi_{1}(d,y):= x-s(d) d y +\rho s(d)\frac{1-F(y)}{1-F(x)}C(y).}  For all $y\leq\bar{x}$, $C(y)\leq \beta y_y$, as
\myeqln{C(y)=y-s(d_y)d_y y_y+\rho s(d_y)\frac{1-F(y_y)}{1-F(y)}\beta y_y = y+[\eta(d_y)-d_y]s(d_y)y_y  \leq y_y + [\eta(d^*)-d^*]s(d^*)y_y =  \beta y_y, }
where the inequality follows from the fact that the derivative of $\eta(d)-d$ is $1-\frac{s''(d)}{[s'(d)]^2}>0$. Since  $d_y=\eta^{-1}\left(\rho\beta \frac{1-F(y_y)}{1-F(y)}\right)$ and $d^*=\eta^{-1}(\rho \beta)$, $d^*>d_y$. Hence, for $y\in[x,\bar{x})$,
\myeqln{
\Pi_1(y,d)=x-s(d)dy+\rho s(d)\frac{1-F(y)}{1-F(x)}C(y)<x-s(d)dy+\rho s(d)\frac{1-F(y)}{1-F(x)}\beta y_y.
}
Extending the set of feasible $y$ value to $y\in[x,u)$ can only improve the objective function value. Therefore, 
\myeqln{
\max_{d, y\in[x,\bar{x})}  \ x-s(d)dy+\rho s(d)\frac{1-F(y)}{1-F(x)}\beta y_y
\leq \max_{d, y\in[x,u)}  \ x-s(d)dy+\rho s(d)\frac{1-F(y)}{1-F(x)}\beta y_y.
}
As $d_x$ and $y_x$ are the maximizers in $\max_{d, y\in[x,u)}x-s(d)dy+\rho s(d)\frac{1-F(y)}{1-F(x)}\beta y$, we have 
\myeqln{
\max_{d, y\in[x,u)}  \ x-s(d)dy+\rho s(d)\frac{1-F(y)}{1-F(x)}\beta y_y\leq x-s(d_x)d_x y_x+\rho s(d_x)\frac{1-F(y_x)}{1-F(x)}\beta y_x.
}

\textbf{Case 1.2: $y\in[\bar{x}, u)$}. We prove  that $(d_x, y_x)$ jointly maximize   
\myeq{
\max_{d,y\in[\bar{x},u)}\Pi_2(d,y)=x-s(d)dy+\rho s(d)\frac{1-F(y)}{1-F(x)}\beta y
}, and that $y_x\geq\bar{x}$. We first optimize over $d$. The optimal $d$, denoted by $d_x$, satisfies $\eta(d)=\rho\beta\frac{1-F(y)}{1-F(x)}$. We then optimize over $y$. Let $\Pi_2(d,y) = x-s(d)dy+\rho s(d)\frac{1-F(y)}{1-F(x)}\beta y.$ We first prove that at $y=\bar{x}$, $\frac{d\Pi_2(d_x,y)}{dy}>0$. This together with $\lim_{y\to u}\Pi_2(d_x,y)<0$ implies that for $y\in[\bar{x},u)$, the solution that maximizes $\Pi_2$ is an interior solution that satisfies the first order condition.

We next prove that at $y=\bar{x}$, $\frac{d\Pi_2(d_x,y)}{dy}>0$. This is because 
\myeqln{
\frac{d\Pi_2(d_x,y)}{dy}\bigg|_{y=\bar{x}}=s(d_x)\rho\beta\frac{1-F(y)}{1-F(x)}\left(-G(\bar{x})-\frac{d_x}{\rho\beta}\frac{1-F(x)}{1-F(\bar{x})}+1\right)=s(d_x)\rho\beta\frac{1-F(y)}{1-F(x)}\left(\frac{d^*}{\rho\beta}-\frac{d_x}{\rho\beta}\frac{1-F(x)}{1-F(\bar{x})}\right),
}
where the second equality follows from the definition of $\bar{x}$. Since 
\myeq{\frac{d_x}{\rho\beta}\frac{1-F(x)}{1-F(\bar{x})}=\frac{d_x}{\eta(d_x)},\ \frac{d^*}{\rho\beta}=\frac{d^*}{\eta(d^*)}}, 
\myeqln{
\frac{d\Pi_2(d_x,y)}{dy}\bigg|_{y=\bar{x}}=s(d_x)\rho\beta\frac{1-F(y)}{1-F(x)}\left(\frac{d^*}{\eta(d^*)}-\frac{d_x}{\eta(d_x)}\right).
}
For $x<\bar{x}$, $d_x<d^*$, as
\myeq{\rho\beta\frac{1-F(\bar{x})}{1-F(x)}=\eta(d_x)} and \myeq{\rho\beta=\eta(d^*).}
Since $\eta(\cdot)$ is monotone increasing, $d_x<d^*$. Also, $\frac{d}{\eta(d)}=\frac{1}{1+\frac{1}{\xi(d)}}$ is monotone increasing when $\xi(\cdot)$ (the demand elasticity) is monotone increasing. Hence, $\frac{d\Pi}{dy}|_{\bar{x}}=s(d_x)\rho\beta\frac{1-F(y)}{1-F(x)}\left(\frac{d^*}{\eta(d^*)}-\frac{d_x}{\eta(d_x)}\right)>0$,
while the same derivative is negative as $y\to u$. 

It follows that if $y_x$ maximizes $\Pi_2$, it must satisfy the first order condition, so there exist $y_x\in(\bar{x},u)$ and $d_x\in(0,d^*]$, such that $(y_x, d_x)$ satisfies the first order conditions $(\ref{eq_1_ie})-(\ref{eq_2_ie})$ with $\Upsilon=y_x, \delta=d_x$, and state $X=x$. As $x<\bar{x}$, Lemma \ref{lemma_existence_equation_sol} implies that $(y_x,d_x)$ is the unique solution to $(\ref{eq_1_ie})-(\ref{eq_2_ie})$.

This completes the proof for Case 1 ($x<\bar{x}$).

\textbf{Case 2: $x\geq\bar{x}$}. We verify that the following equation holds:
\myeqln{
J(x)|_{x\geq \bar{x}} =\frac{1-s(d^*)d^*}{1-s(d^*)\rho}x=\max_{d,y\in[x,u)} \ x-s(d)d y + \rho s(d)\frac{1-F(y)}{1-F(x)}\beta y.
}
We first optimize over $d$. The optimal $d$, denoted by $d_x$, should satisfy $\rho \beta \frac{1-F(y)}{1-F(x)}=\eta(d)$, so $d_x=\eta^{-1}\left(\rho\beta\frac{1-F(y)}{1-F(x)}\right)$.
We next optimize over $y$. Let $\Pi_2(d_x,y)=x-s(d_x)d_x y+\rho s(d_x)\frac{1-F(y)}{1-F(x)}\beta y$, then
\myeqln{
\frac{d\Pi_2(d_x,y)}{dy}
=s(d)\rho\beta\frac{1-F(y)}{1-F(x)}\left(-G(y)-\frac{d_x}{\rho \beta}\frac{1-F(x)}{1-F(y)}+1\right).
}
Now, $\frac{d_x}{\rho \beta}\frac{1-F(x)}{1-F(y)} =\frac{d_x}{\eta(d_x)}=\frac{1}{1+1/\xi(d_x)}$,
where $\xi(d)$ is the demand elasticity. Let 
$\lambda(x)=\frac{1}{1+1/x}$, $\Lambda(y)=G(y)+\frac{d_x}{\rho \beta}\frac{1-F(x)}{1-F(y)}$,  $\lambda(\xi(d_x))=\frac{d_x}{\rho\beta}\frac{1-F(x)}{1-F(y)}$, and $\Lambda(y) = G(y)+\lambda\left(\xi\left(\eta^{-1}\left(\rho\beta\frac{1-F(y)}{1-F(x)}\right)\right)\right)$. With these definitions,  
\myeql{\label{eq_dpi2_ie}
\frac{d\Pi_2(d_x,y)}{dy}=s(d_x)\rho\beta\frac{1-F(y)}{1-F(x)}\left(1-\Lambda(y)\right).
}
By the inverse function theorem,
\myeqln{
\frac{d \ \lambda\left(\xi\left(\eta^{-1}\left(\rho\beta\frac{1-F(y)}{1-F(x)}\right)\right)\right) }{d y} = -\frac{\xi'(d_x)}{(1+\xi(d_x))^2\eta'(d_x)}\frac{\rho\beta f(y)}{(1-F(x))}.
}
Recall that $H(y)=\frac{f(y)}{1-F(y)}$ is the hazard rate of the income distribution evaluated at $y$. Now,  
\begin{align*}
\frac{d\Lambda(y)}{dy}&=\frac{f(y)}{1-F(y)}+H'(y)y-\frac{\xi'(d_x)}{(1+\xi(d_x))^2\eta'(d_x)}\frac{\rho\beta f(y)}{(1-F(x))}\\
&\geq \frac{f(y)}{1-F(y)}-\frac{\xi'(d_x)}{(1+\xi(d_x))^2\eta'(d_x)}\frac{\rho\beta f(y)}{(1-F(x))}\\
& \geq \frac{f(y)}{1-F(y)}\left(1-\frac{\xi'(d_x)\eta(d_x)}{(1+\xi(d_x))^2\eta'(d_x)}\right),
\end{align*}
where the first inequality follows from $H'(y)\geq0$ and the second inequality follows from the fact that $x\leq y$, hence $1-F(x)\geq 1-F(y)$.

We next show that for all $d\in(0,1)$, $\frac{d\Lambda(y)}{dy}\geq0$. For all $d\in(0,1)$, $\xi'(d)\eta(d)>0$, $(1+\xi(d))^2\eta'(d)>0$, and $s'(d)>0$, $s''(d)<0$. Thus, 
\begin{align*}
(1+\xi(d))^2\eta'(d)-\xi'(d)\eta(d)=\left(1+\frac{ds'(d)}{s(d)}\right)^2\left(2-\frac{s''(d)}{(s'(d))^2}\right)-\left(d+\frac{s(d)}{s'(d)}\right)\frac{\left(s(d)(s'(d)+s''(d)d)-d(s'(d))^2\right)}{(s(d))^2}\\
=\frac{s(d)+ds'(d)}{(s(d)s'(d))^2}\left((s'(d))^2(s(d)+3ds'(d))-s''(d)(s(d)+ds'(d)(1+s(d)))\right)>0,
\end{align*}
which implies $\frac{\xi'(d_x)\eta(d_x)}{(1+\xi(d_x))^2\eta'(d_x)}<1$.

It follows that $\frac{d\Lambda(y)}{dy}>0$, so $\Lambda(y)$ is monotone increasing and at $y=x\geq\bar{x}$, 
\myeqln{
\frac{d\Pi_2(d_x,y)}{dy}\bigg|_{y=x}=s(d^*)\rho\beta\left(-G(x)+\frac{d^*}{\rho \beta}+1\right)\leq 0.
}

To complete the proof, we show that for $y>x$, $\frac{d\Pi_2}{dy}$ given by equation (\ref{eq_dpi2_ie}), is negative, which implies that the maximum is achieved at $y=x$. This follows from the fact that as shown above, $\Lambda(y)$ is strictly increasing so $(1-\Lambda(y))<0$ for $y>x$ while the other terms in (\ref{eq_dpi2_ie}) are positive. We thus conclude that when $x\geq\bar{x}$, $(x,d^*)$ maximizes $\Pi_2(d,y)$, which completes the proof of Case 2. }

\subsection{Proof of Proposition \ref{prop_beta_ushape}}

\label{app_prop_51}
\myproof{We start by proving a general property of the lender's expected NPV, $\E(\Pi(\lambda))=\frac{(d^*)^{\alpha+1}}{\alpha}[1-F_{\lambda}(\bar{x}(\lambda))]^{\alpha+1}\bar{x}(\lambda)$, when the income distribution $F_{\lambda}(\cdot)$ and the NPV are parameterized by an arbitrary parameter $\lambda$: if for all $x$, $1-F_{\lambda}(x)$ is quasi-convex in $\lambda$, then $\E(\Pi(\lambda))$ is quasi-convex in $\lambda$. In particular, this is satisfied by the Beta distribution.

To prove this property, first note that quasi-convexity of $1-F_{\lambda}(x)$ means that for all $\lambda_1<\lambda_2$ and $\gamma\in[0,1]$, if we define $\lambda=\gamma\lambda_1+(1-\gamma)\lambda_2$, then for all $x$, $1-F_{\lambda}(x)\leq \max\{1-F_{\lambda_1}(x), 1-F_{\lambda_2}(x)\}$.  

If $1-F_{\lambda_1}(x)\geq 1-F_{\lambda_2}(x)$, then $1-F_{\lambda}(x)\leq 1-F_{\lambda_1}(x)$. Moreover,
\myeqln{\E(\Pi(\lambda))= \frac{(d^*)^{\alpha+1}}{\alpha}[1-F_{\lambda}(\bar{x}(\lambda))]^{\alpha+1}\bar{x}(\lambda)
\leq \frac{(d^*)^{\alpha+1}}{\alpha}[1-F_{\lambda_1}(\bar{x}(\lambda))]^{\alpha+1}\bar{x}(\lambda)
\leq \frac{(d^*)^{\alpha+1}}{\alpha}[1-F_{\lambda_1}(\bar{x}(\lambda_1))]^{\alpha+1}\bar{x}(\lambda_1)= \E(\Pi(\lambda_1)),
}
where the second inequality follows from $\bar{x}(\lambda_1)$ being the maximizer of the profit function under distribution $F_{\lambda_1}$. Hence, $\E(\Pi(\lambda))\leq \E(\Pi(\lambda_1))$. Similarly, if $1-F_{\lambda_1}(x)<1-F_{\lambda_2}(x)$, by quasi convexity, $1-F_{\lambda}(x)\leq 1-F_{\lambda_2}(x)$, and following the same argument, $\E(\Pi(\lambda))\leq \E(\Pi(\lambda_2))$. Thus, if $1-F_{\lambda}(x)\leq \max\{1-F_{\lambda_1}(x), 1-F_{\lambda_2}(x)\}$, then
$\E(\Pi(\lambda))\leq \max\{\E(\Pi(\lambda_1)), \E(\Pi(\lambda_2))\}$. It follows that quasi-convexity of $1-F_{\lambda}(x)$ for all $x$ implies quasi-convexity of $E(\Pi(\lambda))$. 

For the Beta distribution with fixed mean $\lambda=a=b$, let $\lambda=\gamma\lambda_1+(1-\gamma)\lambda_2$ for $\gamma\in[0,1]$. Then, $F_{\lambda}(x)\geq \min\{F_{\lambda_1}(x), F_{\lambda_2}(x)\}$ because decreasing $\lambda$ is a mean-preserving spread (\cite{rothschild1978increasing}). Hence, $1-F_{\lambda}(x)\leq  \max\{1-F_{\lambda_1}(x), 1-F_{\lambda_2}(x)\}$, and as shown above, the expected NPV is quasi-convex in $\lambda$. 

We are now ready to show that the NPV is U-shaped. As $\lambda\to0$, the distribution becomes a standard Bernoulli distribution and $\E(\Pi(\lambda=0))=\frac{2^{-1 - \alpha}}{\alpha}{ \left( \frac{{\alpha (-1 + d^{1 + \alpha}) \rho}}{{(1 + \alpha) (-1 + d^\alpha \rho)}} \right)^{1 + \alpha}}$. As $\lambda\to\infty$, the distribution becomes degenerate with $P(\theta=0.5)=1$ and $\lim_{\lambda\to\infty}\E(\Pi(\lambda)) =\frac{1-(d^*)^{\alpha+1}}{1-(d^*)^{\alpha}\rho}\frac{1}{2}$. When $\lambda=1$, the distribution is uniform and $\E(\Pi(\lambda=1))=\frac{(d^*)^{\alpha+1}}{\alpha}(1 + \alpha)^{1 + \alpha}(2 + \alpha)^{-(2+\alpha)}<\min\{\E(\Pi(\lambda=0)),\lim_{\lambda\to\infty}\E(\Pi(\lambda))$ for $\alpha\in(0,1)$. Since $\E(\Pi(\lambda))$ is continuous and quasi-convex, it follows that $\E(\Pi(\lambda))$ first decreases and then increases in $\lambda$ or, equivalently, it first decreases and then increases in the income variance $\frac{1}{4(2a+1)}$. }

\subsection{Proof of Proposition \ref{prop_bounds_value_info} }

To prove Proposition \ref{prop_bounds_value_info}, we first introduce Lemmas \ref{lemma_monotone_r} and \ref{lemma_monotone_rho} to show that the relative regret is monotone increasing in $d$ and monotone decreasing in $\rho$. Hence, the relative regret attains its minimum when $\rho$ goes to $1$ and $d$ goes to $0$. We then show that at the limit, the relative regret goes to is $2$. 

\mylemma{\label{lemma_monotone_r}
For any fixed $\rho\in(0,1)$, $R^{I}(\rho, d)$ is monotone increasing in $d$, and $\lim_{d\to \rho}R^{I}(\rho,r)=\infty$ and  $\lim_{d\to 0}R^{I}(\rho,r)=2$.
}

\myproof{\textbf{Proof.} Let $t=\frac{\rho}{d}-1$, so $d=\frac{\rho}{1+t}$. For fixed $\rho$, it's sufficient to show that 
$f(\rho,t)=\frac{1}{R^I(\rho,t)}$ is monotone increasing in $t$, where 
\myeql{f(\rho,t)=\dfrac{1}{R^I(\rho,t)}=\dfrac{t(1-\rho + t(1+\rho) - (1-\rho)\sqrt{(1+t)^2-\rho})}{\rho(2t - \rho + 1)^2.}.
}
We need to show that the derivative  
\myeql{
\dfrac{\partial f(\rho, t) }{\partial t}=
\dfrac{(1-\rho)(2\rho + t\rho + 2\rho t^2 - t  - \rho^2 - 1 + (1+ 2\rho t - \rho)\sqrt{(1+t)^2 - \rho})}{\rho (2t + 1-\rho)^3\sqrt{((1+t)^2 - \rho )}}
}
is nonnegative, which is equivalent to $g(\rho,t)\geq0$, where
\myeql{g(\rho,t)=2\rho + t\rho + 2\rho t^2 - t  - \rho^2 - 1 + (1+ 2\rho t - \rho)\sqrt{(1+t)^2 - \rho}\geq0 .}
Now, $g(\rho,t)\ge0$ is equivalent to 
\myeql{\label{eq_mond_1} (1+ 2\rho t - \rho)\sqrt{(1+t)^2 - \rho}\geq 1+ t +\rho^2 -2\rho -\rho t - 2\rho t^2.}
If the right hand side of ***(71)*** is negative, the inequality is trivial. If the right hand side is positive, equation (\ref{eq_mond_1}) is equivalent to 
\myeql{ \sqrt{(1+t)^2 - \rho}\geq \dfrac{1+ t +\rho^2 -2\rho -\rho t - 2\rho t^2}{(1+ 2\rho t - \rho)}.}
With some algebra, 
\myeql{ (1+t)^2 - \rho- \left(\dfrac{1+ t +\rho^2 -2\rho -\rho t - 2\rho t^2}{(1+ 2\rho t - \rho)}\right)^2 = \dfrac{\rho(1-\rho+2t)^3}{(2\rho t - \rho + 1)^2}
\geq 0,}
which implies $g(\rho,t)\geq0$. It follows that $f(\rho, t)$ is monotone increasing in $t$, hence $R^{I}$ is monotone decreasing in $t$ and $R^{I}(\rho,d)$ is monotone increasing in $d$. 

Finally, for any fixed $\rho\in(0,1)$, $\lim_{t\to0} R^I(\rho, t) \to \infty$ and $\lim_{t\to\infty} R^I(\rho, t) \to 2$. As $t=\frac{\rho}{d}-1$, $\lim_{d\to \rho}R^{I}(\rho,r)=\infty$, and $\lim_{d\to 0}R^{I}(\rho,r)=2$
}

Lemma \ref{lemma_monotone_rho} proves the monotonicity of the relative regret in $\rho$.

\mylemma{\label{lemma_monotone_rho}
For any fixed $d\in(0,1)$, $R^{I}(\rho, d)$ is monotone decreasing in $\rho$, and $\lim_{\rho\to 1}R^{I}(\rho,d)=2$
}

\myproof{\textbf{Proof.} 
Set $t=\frac{\rho}{d}-1$, so $\rho=d(1+t)$. It's sufficient to show that $\frac{1}{R^{I}(d, t)}$ is monotone decreasing in $t$. Let
\myeql{\phi(d,t)=\dfrac{1}{R^I(d,t)}=\dfrac{t\left((dt - d + 1)/d+ (\sqrt{(t - d + 1)/(d^2 (t + 1))}(d + dt - 1))\right)}{(2t+1 -d - dt)^2}. }
We need to prove that $\dfrac{\partial \phi(d, t) }{\partial t}\geq0$. With some algebra,
\myeql{
\dfrac{\partial \phi(d, t) }{\partial t} = \dfrac{\psi(d,t)}{(2(t + 1)^2\sqrt{(t - d + 1)/(d^2(t + 1))}(2t+1 -d - dt)^3}}
where $\psi(d,t)$ is given by $\psi(d,t)=t^3 + t(1/d - 1)^2(10t^2 + 12t + 3) + t^2(4t + 3)(2/d - 2) + 2(1/d - 1)^2(1-2t)(t + 1)^2(q(d,t) - 1/d + 1) + 2q(d,t)t(1/d - 1)(t + 1)^2$, and $q(d,t)=\sqrt{ (t - d + 1)/(d^2(t + 1))}$. As $2t+1-d(1+t)>0$, it's sufficient to show that $\psi(d,t)\geq0$. 

When $t\leq \frac{1}{2}$, the monotonicity is trivial. Consider case where $t\geq\frac{1}{2}$. As $q(d,t)\in[\frac{1}{d}-1, \frac{1}{d}]$, 
\myeql{
\psi(d,t)\geq (1/d - 1)^2((10t^3 + 12t^2 + 3t)-2(2t-1)(t + 1)^2)=6t^3 + 6t^2 + 3t + 2\geq0,
}
so $\phi(r, t)$ is monotone increasing in $t$ and $R^{I}(d,t)$ is monotone decreasing in $t$. As $t$ is monotone increasing in $\rho$, $R^{I}(\rho, r)$ is monotone decreasing in $\rho$. 

Finally, as $\rho$ goes to $1$, for any fixed $d$, $\lim_{\rho\to1} R^I(\rho, t) = 2.$ 

Combining Lemmas \ref{lemma_monotone_r} and \ref{lemma_monotone_rho} gines us the proof of Proposition \ref{prop_bounds_value_info}.}

\subsection{Proof of Proposition \ref{prop_bounds_value_info_dr}}

With some algebra, $\frac{1-s(d^*)d^*}{(1-s(d^*)\rho)}=\frac{d^*(\alpha+1)}{\rho\alpha}$, so
\myeql{
R^{I}(\rho) = \dfrac{\alpha+1}{2\rho (d^*)^{\alpha}}\frac{1}{(1-F(\bar{x}))^{\alpha+1}\bar{x}}\geq \dfrac{\alpha+1}{2\rho (d^*)^{\alpha}}\frac{1}{(1-F(\bar{x}))\bar{x}}=\frac{(\alpha+2)^{2}}{2\rho (d^*)^{\alpha}}.}
As $d^*\leq \rho<1$ and $\alpha\in(0,1)$, it follows that $R^{I}(\rho)\geq 2$. Finally, it is straightforward to verify that as $\rho$ goes to zero, the relative regret goes to infinity.

\subsection{Proof of Proposition \ref{prop:bellman_bound}}

We first introduce the following lemma.  

\begin{lemma}\textbf{Sandwich Bound on Noisy Conditional Survival Probability}
\label{lem:noisy_conditional_bound}
\textit{Let $\theta \sim F$ be a continuous random variable supported on $[\ell, u] \subset (0, \infty)$ with strictly increasing CDF $F$. Assume $w_0, w_1, \dots, w_t$ are random shocks independent of $\theta$, each supported on $[-\epsilon, \epsilon]$ for some $\epsilon > 0$. Define any history $x_0, \dots, x_{t-1} \in [0, u]$ and set $\bar{x}_t := \max_{i < t} x_i$. For any $t$ and $y_t$, define $\delta^{\epsilon+}(x):=\min\{\max\{x+\epsilon,0\},u\}$ and $\delta^{\epsilon-}(x):=\min\{\max\{x-\epsilon,0\},u\}$. 
The following inequality holds:}
\[
\frac{1 - F(\delta^{\epsilon+}(\max\{\bar{x}_t,y_t\})}{1 - F(\delta^{\epsilon-}(\bar{x}_t))}
\le 
\mathbb{P}(\theta + w_t \ge y_t \mid \theta + w_i \ge x_i,\; \textup{ for all } i < t)
\le 
\min\{\frac{1 - F(\delta^{\epsilon-}(\max\{\bar{x}_t,y_t\})}{1 - F(\delta^{\epsilon+}(\bar{x}_t))},1\}.
\]
\end{lemma}

\myproof{
We start by expressing the conditional probability using the definition:
\[
\mathbb{P}(\theta + w_t \ge y_t \mid \theta + w_i \ge x_i,\; \textup{ for all } i < t)
= \frac{\mathbb{P}(\theta + w_t \ge y_t,\; \theta + w_i \ge x_i\ \textup{ for all } i < t)}{\mathbb{P}(\theta + w_i \ge x_i\ \textup{ for all } i < t)}.
\]

The event $\{\theta + w_i \ge x_i, \textup{ for all } \ i<t\}$ implies  \(
\theta \ge \min\{\max\{\bar{x}_t - \epsilon,0\},u\}\), and the event $\{\theta\geq \min\{\max\{\bar{x}_t+\epsilon,0\},u\}$ implies that $\{\theta + w_i \ge x_i , \textup{ for all } \ i<t \}$.  Therefore, the denominator satisfies
\(
\mathbb{P}(\theta + w_i \ge x_i,\ \textup{ for all } i < t) \in [1 - F(\delta^{\epsilon+}(\bar{x}_t)),\; 1 - F(\delta^{\epsilon-}(\bar{x}_t) )].
\)

For the numerator, observe that the joint event \(\{\theta + w_t \ge y_t\}\) and  \(\{\theta + w_i \ge x_i\ \textup{ for all } i < t \} \) implies
\( \theta \ge \delta^{\epsilon-}(\max\{\bar{x}_t,y_t\}) \)
and is implied by \(\theta \ge \delta^{\epsilon+}(\max\{\bar{x}_t,y_t\}) \). 
Therefore, the numerator satisfies:
\(
\mathbb{P}(\theta+w_t\geq y_t, \theta + w_i \ge x_i,\ \textup{ for all } i < t) \in [1 - F(\delta^{\epsilon+}(\max\{\bar{x}_t,y_t\}) ),\; 1 - F( \delta^{\epsilon-}(\max\{\bar{x}_t,y_t\}))]
\)
and the full ratio satisfies:
\[
\frac{1 - F(\delta^{\epsilon+}(\max\{\bar{x}_t,y_t\})}{1 - F(\delta^{\epsilon-}(\bar{x}_t))}
\le 
\mathbb{P}(\theta + w_t \ge y_t \mid \theta + w_i \ge x_i,\; \textup{ for all } i < t)
\le 
\frac{1 - F(\delta^{\epsilon-}(\max\{\bar{x}_t,y_t\})}{1 - F(\delta^{\epsilon+}(\bar{x}_t))}.
\]
Finally, $\mathbb{P}(\theta + w_t \ge y_t \mid \theta + w_i \ge x_i,\; \textup{ for all } i < t)\leq 1$ as is the case for any probability.
}

With Lemma \ref{lem:noisy_conditional_bound}, we are ready to prove Proposition \ref{prop:bellman_bound}.

Let  $J(x_t)$ be the Bellman value function with no random shocks and $J^{\epsilon}(x_t)$ the Bellman value function with random income shocks: 
\[
J(x_t) = \max_{y_t\in[0,u)} \left\{ -d y_t + \rho \cdot \mathbb{P}(\theta \ge y_t \mid \theta \ge x_i,\, \textup{ for all } i < t) \cdot \left[y_t + J(x_t \cup \{y_t\})\right] \right\},
\]
\[
J^{\epsilon}(x_t) = \max_{y_t\in[0,u)} \left\{ -d y_t + \rho \cdot \mathbb{P}(\theta + w_t \ge y_t \mid \theta + w_i \ge x_i,\, \textup{ for all } i < t) \cdot \left[y_t + J^{\epsilon}(x_t \cup \{y_t\})\right] \right\}.
\]
For $\rho\in(0,1)$ and $y\in[0,u]$ for bounded $u$, both Bellman equations satisfy the contraction mapping property, so there exist unique solutions for both $J(x_t)$ and $J^{\epsilon}(x_t)$.  We further present the following Lemma.

\begin{lemma}
\label{lemma:sandwich_value}
Let $J^\epsilon$, $J^\epsilon_\ell$, and $J^\epsilon_u$ be the fixed points of the following Bellman equations, respectively:
\begin{align*}
J^\epsilon_\ell(x_t) &= \max_{y \in \mathcal{Y}} \left\{ -d y + \rho \cdot \mathbb{P}_\ell(y \mid x_t) \cdot \left[ y + J^\epsilon_\ell(x_t \cup \{y\}) \right] \right\}, \\
J^\epsilon(x_t) &= \max_{y \in \mathcal{Y}} \left\{ -d y + \rho \cdot \mathbb{P}_\epsilon(y \mid x_t) \cdot \left[ y + J^\epsilon(x_t \cup \{y\}) \right] \right\}, \\
J^\epsilon_u(x_t) &= \max_{y \in \mathcal{Y}} \left\{ -d y + \rho \cdot \mathbb{P}_u(y \mid x_t) \cdot \left[ y + J^\epsilon_u(x_t \cup \{y\}) \right] \right\},
\end{align*}

where $\mathbb{P}_\epsilon(y \mid x_t) := \mathbb{P}(\theta + w_t \ge y \mid \theta + w_i \ge x_i,\ \textup{ for all } i < t)$, $\mathbb{P}_\ell(y \mid x_t) := \frac{1 - F(\delta^{\epsilon+}(\max\{\bar{x}_t,y_t\})}{1 - F(\delta^{\epsilon-}(\bar{x}_t))}$, $\mathbb{P}_u(y \mid x_t) := \min\{\frac{1 - F(\delta^{\epsilon-}(\max\{\bar{x}_t,y_t\})}{1 - F(\delta^{\epsilon+}(\bar{x}_t))},1\}$, $\bar{x}_t := \max_{i < t} x_i$, $\mathcal{Y}:=[0,u]$, and $\rho \in (0,1)$ is the discount factor. Then:
\[
J^\epsilon_\ell(x_t) \le J^\epsilon(x_t) \le J^\epsilon_u(x_t) \quad \text{for all } x_t.
\]
\end{lemma}

\myproof{
Let $\mathcal{T}_\ell$, $\mathcal{T}_{\epsilon}$, and $\mathcal{T}_u$ denote the Bellman operators corresponding to $J^\epsilon_\ell$, $J^\epsilon$, and $J^\epsilon_u$ respectively :
\[
(\mathcal{T}_i J)(x_t) := \max_{y \in \mathcal{Y}} \left\{ -d y + \rho \cdot \mathbb{P}_i(y \mid x_t) \cdot \left[ y + J(x_t \cup \{y\}) \right] \right\}, \quad \text{for } i \in \{\ell, \epsilon, u\}.
\]

From Lemma~\ref{lem:noisy_conditional_bound}, we have for all $x_t$ and $y$:
\(\mathbb{P}_\ell(y \mid x_t) \le \mathbb{P}_\epsilon(y \mid x_t) \le \mathbb{P}_u(y \mid x_t)\), hence for any fixed value function $J$ and any $x_t$, the following inequality holds:
\begin{equation}
\mathcal{T}_\ell J(x_t) \le \mathcal{T}_\epsilon J(x_t) \le \mathcal{T}_u J(x_t). \label{ineq_Je_mon}
\end{equation}
Then, by the monotonicity of the Bellman operator, for each $\mathcal{T}_i$, if $J_1 \le J_2$ pointwise, then $\mathcal{T}_i J_1 \le \mathcal{T}_i J_2$ pointwise. Consider the following sequences: $J^{(k+1)}_\ell = \mathcal{T}_\ell J^{(k)}_\ell,\ 
J^{(k+1)} = \mathcal{T}_\epsilon J^{(k)}, \ 
J^{(k+1)}_u = \mathcal{T}_u J^{(k)}_u$, with $J^{(k)}_\ell=J^{(k)}=J^{(k)}_u=0$ for $k=0$. Using the monotonicity in $J$ and inequality (\ref{ineq_Je_mon}), we get by induction:
\[
J^{(k)}_\ell \le J^{(k)} \le J^{(k)}_u \quad \text{for all } k.
\]

Finally, for $\rho\in(0,1)$ and $y\in[0,u]$, following standard textbook results (see e.g. \cite{bertsekas1995dynamic}), $\mathcal{T}_i$ are contraction mappings, and $\lim_{k\to\infty} J^{(k)}_l(x_t):=J^{\epsilon}_l(x_t)$, $\lim_{k\to\infty} J^{(k)}(x_t):=J^{\epsilon}(x_t)$, $\lim_{k\to\infty} J^{(k)}_u(x_t):=J^{\epsilon}_u(x_t)$. Hence \[
J^\epsilon_\ell(x_t) \le J^\epsilon(x_t) \le J^\epsilon_u(x_t) \quad \text{for all } x_t.
\]
}


We are now ready to prove the bound between $J^{\epsilon}(x_t)$ and $J(x_t)$ by bounding the difference between $J(x_t)$ and $J^{\epsilon}_l(x_t)$, $J^{\epsilon}_u(x_t)$ respectively. First, under $J^{\epsilon}_l(x_t)$ and $J^{\epsilon}_u(x_t)$, the optimal policies are weakly increasing following a similar argument as in Lemma \ref{lemma_increasing_control}. Let $x_{t+1}=y_t$. We can rewrite the Bellman equation as \begin{align*}J^\epsilon_\ell(x_t) &= \max_{y_t \in [x_t,u)} \left\{x_t -d y_t + \rho \cdot \mathbb{P}_\ell(y_t \mid x_t) \cdot J^\epsilon_\ell(y_t)  \right\}, \\
J^\epsilon_u(x_t) &= \max_{y_t \in [x_t,u)} \left\{x_t - d y_t + \rho \cdot \mathbb{P}_u(y \mid x_t) \cdot  J^\epsilon_u(y_t) \right\},
\end{align*}
with state evolution $x_{t+1}=y_{t}$. Similarly, 
\begin{equation*}
J(x_t)=\max_{y_t \in [x_t,u)} \left\{x_t -d y_t + \rho \cdot \frac{1-F(y)}{1-F(x_t)}\cdot J(y_t), \right\}
\end{equation*}
and 
\begin{align*}
J^{\epsilon}_l(x_0)&\geq \max_{y_t \in [x_t,u)} \left\{x_t -d y_t + \rho \cdot \frac{1-F(y)-L\epsilon}{1-F(x_0)+L\epsilon} \cdot J^\epsilon_\ell(y_t)  \right\}\\
&\geq \max_{y_t \in [x_t,u)} \left\{x_t -d y_t + \rho \cdot \frac{1-F(y)}{1-F(x_0)} \cdot J^\epsilon_\ell(y_t)  - \frac{2L\epsilon}{1-F(x_0)}J(y_t)\right\}\\
&\geq \max_{y_t \in [x_t,u)} \left\{x_t -d y_t + \rho \cdot \frac{1-F(y)}{1-F(x_0)} \cdot J^\epsilon_\ell(y_t)\right\}  - \frac{2L\epsilon}{1-F(x_0)}\frac{(1-d)(u+\epsilon)}{1-\rho}\\
&\geq J(x_0)- \frac{2L\epsilon}{1-F(x_0)}\frac{(1-s(d^*)d^*)(u+\epsilon)}{1-s(d^*)\rho}.
\end{align*}

The first inequality comes from the Lipschitz continuity of $F(x)$; the second inequality comes from the fact that $\frac{1-F(y)}{1-F(x_0)}\leq 1$; the third inequality comes from the fact that $J(y_t)$ is upper bounded by $\frac{1-s(d^*)d^*}{1-s(d^*)\rho}(\mu+\epsilon)$, which is the expected return under the optimal policy with full information where the consumer has the maximum possible income; and the last inequality comes from the definition of $J(x_0)$.

Similarly,
\begin{align*}
J^{\epsilon}_u(x_0)&\leq \max_{y_t \in [x_t,u)} \left\{x_t -d y_t + \rho \cdot \frac{1-F(y)+L\epsilon}{1-F(x_0)-L\epsilon} \cdot J^\epsilon_\ell(y_t)  \right\}\\
&\leq \max_{y_t \in [x_t,u)} \left\{x_t -d y_t + \rho \cdot \frac{1-F(y)}{1-F(x_0)} \cdot J^\epsilon_\ell(y_t)  + \frac{L\epsilon}{1-F(x_0)-L\epsilon}J(y_t)\right\}\\
&\leq \max_{y_t \in [x_t,u)} \left\{x_t -d y_t + \rho \cdot \frac{1-F(y)}{1-F(x_0)} \cdot J^\epsilon_\ell(y_t)\right\}  + \frac{L\epsilon}{1-F(x_0)-L\epsilon}\frac{(1-d)(u+\epsilon)}{1-\rho}\\
&\leq J(x_0)+ \frac{L\epsilon}{1-F(x_0)-L\epsilon}\frac{(1-d)(u+\epsilon)}{1-\rho},
\end{align*}

hence
$J(x_0)- \frac{2L\epsilon}{1-F(x_0)}\frac{(1-d)(u+\epsilon)}{1-\rho} \leq J^{\epsilon}(x_0)\leq J(x_0)+ \frac{L\epsilon}{1-F(x_0)-L\epsilon}\frac{(1-d)(u+\epsilon)}{1-\rho}$, and 
\[
|J^{\epsilon}(x_0) - J(x_0)| \le  \max\left\{ \frac{2L(u+\epsilon)(1-d)}{(1-F(x_0))(1-\rho)},\frac{L\epsilon(1-d)}{(1-F(x_0)-L\epsilon)(1-\rho)}\right\}\epsilon.
\]

Finally, when $L\epsilon<\frac{1}{2}$, with $x_0=0$, 
\[ \max\left\{ \frac{2L(u+\epsilon)(1-d)}{(1-F(x_0))(1-\rho)},\frac{L\epsilon(1-d)}{(1-F(x_0)-L\epsilon)(1-\rho)}\right\}\epsilon =\frac{2L\,(u+\epsilon)\,(1-d)}{1-\rho}\,\epsilon.\]

\subsection{Proof of Proposition \ref{prop:bellman_bound_endo}}

We first introduce the following lemma. 

\begin{lemma}
\label{lemma:sandwich_value_endo}
Let $J^\epsilon$, $J^\epsilon_\ell$, and $J^\epsilon_u$ be the fixed points of the following Bellman equations, respectively:
\begin{align*}
J^\epsilon_\ell(x_t) &= \max_{y \in \mathcal{Y},d\in[0,\rho)} \left\{ -d y + s(d)\rho \cdot \mathbb{P}_\ell(y \mid x_t) \cdot \left[ y + J^\epsilon_\ell(x_t \cup \{y\}) \right] \right\}, \\
J^\epsilon(x_t) &= \max_{y \in \mathcal{Y},d\in[0,\rho)} \left\{ -d y + s(d) \rho \cdot \mathbb{P}_\epsilon(y \mid x_t) \cdot \left[ y + J^\epsilon(x_t \cup \{y\}) \right] \right\}, \\
J^\epsilon_u(x_t) &= \max_{y \in \mathcal{Y},d\in[0,\rho)} \left\{ -d y + s(d)\rho \cdot \mathbb{P}_u(y \mid x_t) \cdot \left[ y + J^\epsilon_u(x_t \cup \{y\}) \right] \right\},
\end{align*}
Then:
\[
J^\epsilon_\ell(x_t) \le J^\epsilon(x_t) \le J^\epsilon_u(x_t) \quad \text{for all } x_t.
\]
\end{lemma}
\myproof{ The proof resembles that of Lemma \ref{lemma:sandwich_value}. For $\rho\in(0,1)$, $y\in[0,u]$ $d\in[0,\rho)$, and $s(d)\in(0,1)$, each period return is bounded, so the Bellman operators are contraction mappings. It follows that \(
J^\epsilon_\ell(x_t) \le J^\epsilon(x_t) \le J^\epsilon_u(x_t) \ \text{for all } x_t.\)
}
Similarly, we define the Bellman equations as \begin{align*}
J^\epsilon_\ell(x_t) &= \max_{y_t \in [x_t,u), d_t\in[0,\rho)} \left\{x_t -d y_t + s(d_t)\rho \cdot \mathbb{P}_\ell(y_t \mid x_t) \cdot J^\epsilon_\ell(y_t)  \right\}, \\
J^\epsilon_u(x_t) &= \max_{y_t \in [x_t,u),d_t\in[0,\rho)} \left\{x_t - d y_t + s(d_t)\rho \cdot \mathbb{P}_u(y \mid x_t) \cdot  J^\epsilon_u(y_t) \right\},
\end{align*}
with state evolution $x_{t+1}=y_{t}$. Now,  
\begin{equation*}
J(x_t)=\max_{y_t \in [x_t,u), d_t\in[0,\rho)} \left\{x_t -d_t y_t + s(d_t)\rho \cdot \frac{1-F(y)}{1-F(x_t)}\cdot J(y_t). \right\}
\end{equation*}
We have
\begin{align*}
J^{\epsilon}_l(x_0)&\geq \max_{y_t \in [x_t,u),d_t\in(0,\rho)} \left\{x_t -d_t y_t + s(d_t)\rho \cdot \frac{1-F(y)-L\epsilon}{1-F(x_0)+L\epsilon} \cdot J^\epsilon_\ell(y_t)  \right\}\\
&\geq \max_{y_t \in [x_t,u),d_t\in[0,\rho)} \left\{x_t -d_t y_t + s(d_t)\rho \cdot \frac{1-F(y)}{1-F(x_0)} \cdot J^\epsilon_\ell(y_t)  - \frac{2L\epsilon}{1-F(x_0)}J(y_t)\right\}\\
&\geq \max_{y_t \in [x_t,u),d_t\in[0,\rho)} \left\{x_t -d y_t + s(d_t) \rho \cdot \frac{1-F(y)}{1-F(x_0)} \cdot J^\epsilon_\ell(y_t)\right\}  - \frac{2L\epsilon}{1-F(x_0)}\frac{(1-s(d^*)d^*)(u+\epsilon)}{1-s(d^*)\rho}\\
&\geq J(x_0)- \frac{2L\epsilon}{1-F(x_0)}\frac{(1-s(d^*)d^*)(u+\epsilon)}{1-s(d^*)\rho}.
\end{align*}

The first inequality comes from the Lipschitz continuity of $F(x)$. The second comes from the fact that $
\frac{1-F(y)}{1-F(x_0)}\leq 1$. The third inequality follows from the fact that $J(y_t)$ is upper bounded by $\frac{1-s(d^*)d^*}{1-s(d^*)\rho}(\mu+\epsilon)$, which is the expected return under the optimal policy with full information where the consumer has the maximum possible income, and the last inequality comes from the definition of $J(x_0)$.

Similarly,
\begin{align*}
J^{\epsilon}_u(x_0)&\leq \max_{y_t \in [x_t,u),d_t\in[0,\rho)} \left\{x_t -d_t y_t + s(d_t)\rho \cdot \frac{1-F(y)+L\epsilon}{1-F(x_0)-L\epsilon} \cdot J^\epsilon_\ell(y_t)  \right\}\\
&\leq \max_{y_t \in [x_t,u),d_t\in[0,\rho)} \left\{x_t -d_t y_t + s(d_t)\rho \cdot \frac{1-F(y)}{1-F(x_0)} \cdot J^\epsilon_\ell(y_t)  + \frac{L\epsilon}{1-F(x_0)-L\epsilon}J(y_t)\right\}\\
&\leq \max_{y_t \in [x_t,u),d_t\in[0,\rho)} \left\{x_t -d_t y_t + s(d_t)\rho \cdot \frac{1-F(y)}{1-F(x_0)} \cdot J^\epsilon_\ell(y_t)\right\}  + \frac{L\epsilon}{1-F(x_0)-L\epsilon}\frac{(1-s(d^*)d^*)(u+\epsilon)}{1-s(d^*)\rho}\\
&\leq J(x_0)+ \frac{L\epsilon}{1-F(x_0)-L\epsilon}\frac{(1-s(d^*)d^*)(u+\epsilon)}{1-s(d^*)\rho}.
\end{align*}

Hence
\[
|J^{\epsilon}(x_0) - J(x_0)| \le  \max\left\{ \frac{2L(u+\epsilon)(1-s(d^*)d^*)}{(1-F(x_0))(1-s(d^*)\rho)},\frac{L\epsilon(1-s(d^*)d^*)}{(1-F(x_0)-L\epsilon)(1-s(d^*)\rho)}\right\}\epsilon.
\]

Finally, when $L\epsilon<\frac{1}{2}$ and $x_0=0$, 
\[\max\left\{ \frac{2L(u+\epsilon)(1-s(d^*)d^*)}{(1-F(x_0))(1-s(d^*)\rho)},\frac{L\epsilon(1-s(d^*)d^*)}{(1-F(x_0)-L\epsilon)(1-s(d^*)\rho)}\right\}\epsilon = \frac{2L(u+\epsilon)(1-s(d^*)d^*)}{(1-s(d^*)\rho)}\epsilon.\]



\end{document}